\documentclass[fleqn]{annalen}
\usepackage{graphicx}
\usepackage{subfigure}
\usepackage{epsf}
\usepackage{epic}
\usepackage{psfig}
\usepackage[dvips]{color}
\begin{document}
\newcommand{\volume}{11}             
\newcommand{\xyear}{2002}            
\newcommand{\issue}{9}               
\newcommand{\recdate}{9 July 2002}   
\newcommand{\revdate}{dd.mm.yyyy}    
\newcommand{\revnum}{0}              
\newcommand{\accdate}{30 July 2002}  
\newcommand{\coeditor}{B.\ Kramer}   
\newcommand{\firstpage}{ }           
\newcommand{\lastpage}{  }           
\setcounter{page}{1}                 
\newcommand{\keywords}{transition-metal oxides, metal-insulator transition, 
                       electronic structure} 
\newcommand{\PACS}{71.20.-b, 71.30.+h, 72.15.Nj}
\newcommand{\shorttitle}{V.\ Eyert, The metal-insulator transitions of 
                         $ {\rm VO_2} $} 
\title{The metal-insulator transitions of $ {\rm \bf VO_2} $: \\
       A band theoretical approach}
\author{Volker Eyert$ ^{\ast} $ 
        \footnotetext{$ ^{\ast} $Corresponding author: 
                      eyert@physik.uni-augsburg.de}} 
\newcommand{\address}
  {Institut f\"ur Physik, Universit\"at Augsburg, 86135 Augsburg, Germany}
\newcommand{\email}{\tt }
\maketitle
\begin{abstract}
The results of first principles electronic structure calculations for 
the metallic rutile and the insulating monoclinic $ {\rm M_1} $ phase 
of vanadium dioxide are presented. In addition, the insulating 
$ {\rm M_2} $ phase is investigated for the first time. 
The density functional calculations allow for a consistent understanding 
of all three phases. In the rutile phase metallic conductivity is carried 
by metal $ t_{2g } $ orbitals, which fall into the one-dimensional 
$ d_{\parallel} $ band, and the isotropically dispersing $ e_{g}^{\pi} $ 
bands. Hybridization of both types of bands is weak. In the $ {\rm M_1} $ 
phase splitting of the $ d_{\parallel} $ band due to metal-metal 
dimerization and upshift of the $ e_{g}^{\pi} $ bands due to increased 
$ p $--$ d $ overlap lead to an effective separation of both types of bands. 
Despite incomplete opening of the optical band gap due to the shortcomings 
of the local density approximation, the metal-insulator transition can be 
understood as a Peierls-like instability of the $ d_{\parallel} $ band in an 
embedding background of $ e_{g}^{\pi} $ electrons. In the $ {\rm M_2} $ 
phase, the metal-insulator transition arises as a combined embedded 
Peierls-like and antiferromagnetic instability. 
The results for $ {\rm VO_2} $ fit into the general scenario of an 
instability of the rutile-type transition-metal dioxides at the 
beginning of the $ d $ series towards dimerization or antiferromagnetic 
ordering within the characteristic metal chains. This scenario was 
successfully applied before to $ {\rm MoO_2} $ and $ {\rm NbO_2} $. In 
the $ d^1 $ compounds, the $ d_{\parallel} $ and $ e_{g}^{\pi} $ bands 
can be completely separated, which leads to the observed metal-insulator 
transitions. 
\end{abstract}

\renewcommand{\topfraction}{1.0}
\renewcommand{\bottomfraction}{1.0}
\renewcommand{\textfraction}{0.0}

\section{Introduction}
\label{intro}

At a temperature of 340\,K and ambient pressure, stoichiometric $ {\rm VO_2} $ 
undergoes a metal-insulator transition, which is accompanied by a structural 
transition from a high-temperature rutile phase to a low-temperature 
monoclinic phase \cite{morin59,brueckner83}. Since its discovery more than 
fourty years ago, this transition has been attracting considerable interest 
for fundamental reasons, and for possible applications \cite{brueckner83}. 
The latter result from the abrupt change in the resistivity over several 
orders of magnitude, as well as from the fact that the transition occurs 
near room temperature \cite{morin59,brueckner83}. Despite this strong 
motivation, our understanding of the origin of the phase transition is 
still far from complete. Several models have been proposed ranging from 
Peierls- \cite{goodenough60,adler67a,adler67b,goodenough71a,goodenough71b,wentz94a,wentz94b} 
to Mott-Hubbard-type \cite{mott61,zylbersztejn75,paquet80,rice94} scenarios.  
They stress, to a different degree, the role of lattice instabilities, 
electron-phonon interaction and electron-electron correlations. So far, 
neither of these approaches has been successful in explaining the broad 
range of phenomena occurring in vanadium dioxide. In particular, a complete 
and generally accepted picture of the physics of this material has not yet 
been developed. More generally, a comprehensive understanding of the 
rutile-related transition-metal dioxides is not available. 

This class of materials has been intensively studied for a long time, 
due to the complexity of the physical properties \cite{rogers69}. As 
is obvious from Tab.\ \ref{tab:intro1}  
\begin{table}[ht]
\begin{center}
\caption{Properties of transition-metal dioxides with rutile-related 
         structure (from Ref.\ \protect \cite{mattheiss76}).}
\label{tab:intro1}
{\large 
\begin{tabular}{llllllll}
\\[-3mm] \hline \\[-3mm]
       & $ d^0 $ & $ d^1 $ & $ d^2 $ & $ d^3 $ & $ d^4 $ & $ d^5 $ & $ d^6 $ \\
\\[-3mm] \hline \\[-3mm]
\textcolor{green}{$ 3d $} & 
\textcolor{green}{$ {\rm TiO_2}       $} & 
\textcolor{green}{$ {\rm VO_2^{\ast}} $} & 
\textcolor{green}{$ {\rm CrO_2}       $} & 
\textcolor{green}{$ {\rm MnO_2}       $} &            
                                               &         & \\ 
       & 
\textcolor{green}{ (S)                     }   & 
\textcolor{green}{ (M--S)                  }   & 
\textcolor{green}{ (F--M)                  }   & 
\textcolor{green}{ (AF--S)                 }   &      
                                    &                            & \\ 
\multicolumn{7}{l}{ } \\
\textcolor{blue}{$ 4d $} & 
                                    & 
\textcolor{blue}{$ {\rm NbO_2^{\ast}} $} & 
\textcolor{blue}{$ {\rm MoO_2^{\ast}} $} & 
\textcolor{blue}{$ {\rm TcO_2^{\ast}} $} & 
\textcolor{blue}{$ {\rm RuO_2}        $} & 
\textcolor{blue}{$ {\rm RhO_2}        $} & \\ 
       &                            & 
\textcolor{blue}{ (M--S)                  } & 
\textcolor{blue}{ (M)                     } & 
\textcolor{blue}{ (M)                     } & 
\textcolor{blue}{ (M)                     } & 
\textcolor{blue}{ (M)                     } & \\ 
\multicolumn{7}{l}{ } \\
\textcolor{magenta}{ $ {\bf 5d} $ } & 
                                    & 
\textcolor{magenta}{$ {\rm TaO_2}        $} & 
\textcolor{magenta}{$ {\rm WO_2^{\ast}}  $} & 
\textcolor{magenta}{$ {\rm ReO_2^{\ast}} $} & 
\textcolor{magenta}{$ {\rm OsO_2}        $} & 
\textcolor{magenta}{$ {\rm IrO_2}        $} & 
\textcolor{magenta}{$ {\rm PtO_2^{\ast}} $} \\ 
       &                            & 
\textcolor{magenta}{ (?)                  } & 
\textcolor{magenta}{ (M)                  } & 
\textcolor{magenta}{ (M)                  } & 
\textcolor{magenta}{ (M)                  } & 
\textcolor{magenta}{ (M)                  } & 
\textcolor{magenta}{ (M)                  } \\ 
\\[-3mm] \hline \\[-3mm]
\multicolumn{8}{l}{$ {\bf ^\ast} $ deviations from rutile,  
                   M = metal, S = semiconductor} \\ 
\multicolumn{8}{l}{F/AF = ferro-/antiferromagnet} \\ 
\end{tabular} 
}
\end{center} 
\end{table} 
this diversity shows up mainly in the $ 3d $ series, which comprises, in 
addition to $ {\rm VO_2} $, the large-gap semiconductor $ {\rm TiO_2} $, 
the half-metallic ferromagnet $ {\rm CrO_2} $, and the antiferromagnetic 
semiconductor $ {\rm MnO_2} $ \cite{mattheiss76,bookdft}. In contrast, 
except for $ {\rm NbO_2} $, which, like $ {\rm VO_2} $, undergoes a 
metal-insulator transition accompanied by a structural transition, the 
$ 4d $ and $ 5d $ compounds are neither semiconducting nor magnetic. 
Nevertheless, there exist several members in each group, which display 
small but characteristic deviations from the rutile structure 
\cite{rogers69,mattheiss76,bolzan97}.  

\begin{figure}
\centering
\includegraphics[width=0.8\textwidth]{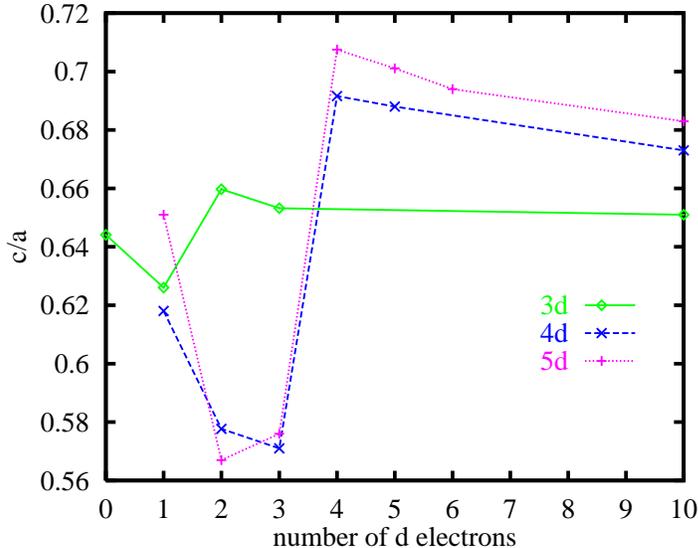}
\caption{Axial ratios of rutile-related transition-metal dioxides.} 
\label{fig:intro1}
\end{figure}

Interestingly, the distortions of the rutile structure found at the phase 
transitions of $ {\rm VO_2} $ and $ {\rm NbO_2} $ are very similar for all 
the transition-metal dioxides which display such deviations, i.e.\ for 
$ {\rm VO_2} $, $ {\rm NbO_2} $, $ {\rm MoO_2} $, $ {\rm WO_2} $, 
$ {\rm TcO_2} $, and $ \alpha $-$ {\rm ReO_2} $ \cite{rogers69}. They are 
characterized by two distinct modes, namely a pairing of the metal atoms 
along characteristic chains parallel to the tetragonal $ c $ axis, and a 
lateral zigzag-like displacement 
\cite{andersson56,rogers69,longo70,marezio72}. This leads to the socalled 
$ {\rm M_1} $ structure, which has a simple monoclinic lattice. The unit 
cell is twice as large as that of the rutile structure. 
$ {\rm NbO_2} $ is an exception: below the transition temperature 
this compound has a body-centered tetragonal lattice with a unit cell, which 
comprises 16 rutile cells \cite{pynn76}. Due to the strong metal-metal 
bonding within the chains parallel to the rutile $ c $ axis, which causes 
the aforementioned pairing, the above six dioxides have the smallest ratios 
of the symmetrized lattice constants $ c_R/a_R $ (Fig.\ \ref{fig:intro1}). 

The structural similarities of the six compounds mentioned above are 
in contrast to their completely different electronic properties: while 
the $ d^1 $ compounds $ {\rm VO_2} $ and $ {\rm NbO_2} $ undergo a 
metal-insulator transition, their four $ d^2 $ and $ d^3 $ neighbours 
like all other $ 4d $ and $ 5d $ compounds listed in Tab.\ \ref{tab:intro1}  
remain metallic at all temperatures. An explanation for this diversity 
is still lacking. The different trends in the structural and electronic 
properties within the transition-metal dioxides at the beginning of 
the $ d $ series led Goodenough to rule out the Jahn-Teller effect as a 
major source for the crystal structure characteristics \cite{goodenough60}. 
Mechanisms based on Fermi surface nesting could be excluded due to the 
change in $ d $ electron occupation. A comprehensive theory, explaining 
both the structural similarities of all the above transition-metal 
dioxides and the diversity in their electronic properties on an equal 
footing, is still missing. 

The discovery of two additional insulating low-temperature phases of 
$ {\rm VO_2} $ led to further complications. These phases, denoted 
$ {\rm T} $ and $ {\rm M_2} $, appear on doping or applying uniaxial 
pressure. They are generally regarded as metastable forms of stoichiometric 
$ {\rm VO_2} $ at ambient pressures. Their crystallographic distortions 
and magnetic properties add to the wealth of phenomena observed in vanadium 
dioxide, and demonstrate even more our lack in understanding. 

We have studied the transition-metal dioxides at the beginning of the 
$ d $ series using density functional theory based electronic structure 
calculations \cite{habil}. In particular, our investigations so far 
include metallic $ {\rm MoO_2} $ as well as the other metal-insulator 
transition system besides $ {\rm VO_2} $, namely $ {\rm NbO_2} $ 
\cite{moo2pap,nbo2pap}. In the present paper we report on results of the 
different phases of $ {\rm VO_2} $. In doing so we complement previous 
work on the rutile and $ {\rm M_1} $ phases by a more detailed analysis 
of their electronic properties, and we present new results for the 
$ {\rm M_2} $ phase \cite{habil}. 

The purpose of this work is twofold. First, by investigating a large class 
of related compounds within the same theoretical approach, we are able to 
develop a more comprehensive understanding of the relevant mechanisms.  
Second, our density functional calculations benefit from the fact that 
they do not depend on adjustable parameters. Hence, despite possible 
limitations arising from the local density approximation, they establish 
a reference, which can be directly compared to experimental data. Finally, 
our results can be used as the starting point for more elaborate treatments, 
as e.g.\ the LDA+DMFT approach, which only recently has proven very 
successful in modelling the symmetry conserving metal-insulator transition 
of $ {\rm V_2O_3} $ \cite{held01,held02}.

The paper is organized as follows: In Sec.\ \ref{survey}, an overview of 
previous results for $ {\rm VO_2} $ is presented. In Sec.\ \ref{cryst}, 
the crystal structures of the different phases are described. Sec.\ 
\ref{method} summarizes the gross features of our calculational method. 
In Secs.\ \ref{resrut}, \ref{resm1}, and \ref{resm2} we discuss the 
results obtained for the rutile, $ {\rm M_1} $, and $ {\rm M_2} $ phases, 
respectively.  Finally, a summary is given in Sec.\ \ref{concl}.

\section{Summary of previous results for $ {\rm \bf VO_2} $}
\label{survey}

\subsection{Experimental results}
\label{experiment}

The important role of the lattice degrees of freedom for the stabilization 
of the different phases of $ {\rm VO_2} $ has been deduced from several 
experiments, which revealed striking differences between the metallic and 
the insulating state. Together with the symmetry change at the metal-insulator 
transition these findings were taken as evidence for strong electron-phonon 
coupling and a lattice softening at the transition \cite{mcwhan74}. Of 
course, this became most obvious from the fact that crystals could even 
break at the transition \cite{prieur77,brueckner83}. Furthermore, X-ray 
measurements revealed large thermal displacements in the rutile phase of both 
stoichiometric and Cr-doped $ {\rm VO_2} $, which exceed the values measured 
in the respective low-temperature phases as well as in the neighbouring rutile 
compounds $ {\rm TiO_2} $ and $ {\rm CrO_2} $ considerably 
\cite{marezio72,mcwhan74}. Rao {\em et al.}\ observed a large 
anisotropy of the thermal expansion in the rutile phase 
($ \alpha_{\parallel} \approx 5-6 \times \alpha_{\perp} $) 
\cite{rao67,kucharczyk79}, which causes a strong temperature dependence of the 
electric field gradient at the vanadium site \cite{pouget76}. In addition, an 
anisotropic change of the lattice constants at the phase transition was 
reported \cite{minomura64,kucharczyk79}: While the rutile $ c $ axis increases 
by $ \approx 1 $\% the $ a $ axis decreases by $ \approx 0.6 $\%. At the same 
time, the volume changes discontinuously by -0.044\% and the thermal expansion 
coefficient drops by a factor of three \cite{kawakubo64,kucharczyk79}. 
Ultrasonic microscopy revealed a strong elastic anisotropy in the metallic 
phase, which disappears almost completely in the insulating state 
\cite{maurer99}. X-ray diffuse scattering measurements performed just 
above the metal-insulator transition by Terauchi and Cohen \cite{terauchi78} 
indicated a lattice 
instability at the tetragonal R-point with a wave vector parallel to 
$ {\rm \Gamma} $-R and a polarization vector parallel to the $ c $ axis. 
This result confirmed the symmetry analysis by Brews, who showed the 
rutile-to-monoclinic transition to be compatible with a phonon instability 
at the R-point \cite{brews70}. Indeed, using a shell model to calculate 
the phonon dispersion of several rutile-type compounds, Gervais and Kress 
found a phonon softening at the tetragonal R-point for $ {\rm VO_2} $ 
\cite{gervais85}. Furthermore, the analysis of the eigenvectors of the R-point
soft mode resulted in a displacement pattern of the metal atoms equivalent 
to that of the monoclinic phase \cite{gervais85}. Finally, an oxygen isotope 
effect was observed for the transition temperature \cite{terukov78}. 

The electronic structure of metallic $ {\rm VO_2} $ has been probed by optical 
and photo\-emission experiments. According to optical measurements by Verleur 
{\em et al.}\ the lowest unfilled V $ 3d $ levels lie about 2.5\,eV above the 
top of the O $ 2p $ bands \cite{verleur68}. This result was supported by 
photoemission measurements by Powell {\em et al.}\ \cite{powell69}. UPS and 
XPS experiments revealed an $ \approx 8.5 $\,eV wide occupied band falling into 
the low binding, approximately 1.5\,eV wide V $ 3d $ band and an $ \approx 6 $ 
eV wide group of O $ 2p $ bands at higher binding energies 
\cite{shin90,bermudez92,goering96,goering97a,goering97b}. According to 
O K-edge XAS experiments, the unoccupied V $ 3d $ bands extend from 
$ {\rm E_F} $ to 1.7\,eV and from 2.2 to 5.2\,eV 
\cite{blaauw75,shin90,abbate91,bermudez92,goering97b,mueller97}. 
The nearly isotropic electrical conductivity, which is only twice as large 
parallel to the rutile $ c $ axis as compared to its in-plane value 
\cite{bongers65,kosuge67}, has been attributed to the existence of at least 
two different bands near the Fermi energy and a rather complex Fermi surface 
\cite{berglund69a}. The same conclusion was drawn from optical, Seebeck 
coefficient and magnetization measurements 
\cite{verleur68,berglund69a,brown78}. 
For the insulating phase, photoemission studies revealed a sharpening and 
downshift of the occupied V $ 3d $ bands 
\cite{blaauw75,shin90,bermudez92,goering97a,goering97b} relative to the 
metallic phase. XAS measurements showed a splitting of the 
lowest unoccupied $ 3d $ state \cite{abbate91,mueller97}. 
Finally, the vanadium atom pairing is reflected by the magnetic 
susceptibility, which shows Curie-Weiss behaviour above $ {\rm T_c} $ and 
a nearly constant van Vleck-like contribution below $ {\rm T_c} $ 
\cite{kosuge67,berglund69a}.

\subsection{Theoretical work}
\label{theory}

Theoretical approaches to the insulating ground state of $ {\rm VO_2} $ 
as well as to the metal-insulator transition range from phenomenological 
models to first principles studies. 
In a survey over oxides with octahedrally coordinated transition metals,  
Goodenough pointed out the important role of direct cation-cation 
interactions, which add to and often supersede the cation-anion-cation 
interactions \cite{goodenough60}. If strong enough, the cation-cation 
interactions may give rise to a covalent-type bonding at low temperatures. 
This bonding would go along with a displacement of the metal atom away from 
the center of symmetry of the anion interstice as well as with a spin 
pairing of the bonding electrons. In the case of $ {\rm VO_2} $, the 
latter would explain the observed Curie-Weiss behaviour above $ {\rm T_c} $ 
and the small temperature-independent van Vleck susceptibility below 
$ {\rm T_c} $ \cite{kosuge67}. Since those $ d $ electrons, which are 
responsible for the metal-metal bonding, do not contribute to metallic 
conductivity, the phase transition may be from the metallic to the 
semiconducting state \cite{goodenough60}. Subsequently, 
Kawakubo proposed that the metal-insulator transition might be due to 
electronic excitations from the bonding to the antibonding state 
\cite{kawakubo65}. Giving arguments for a linear dependence of the energy 
gap on the number of excited carriers Adler and Brooks demonstrated that 
a system with a half-filled conduction band will lower its total energy 
and enter an insulating state by developing either antiferromagnetic order 
or a crystalline distortion \cite{adler67a,adler67b}. 

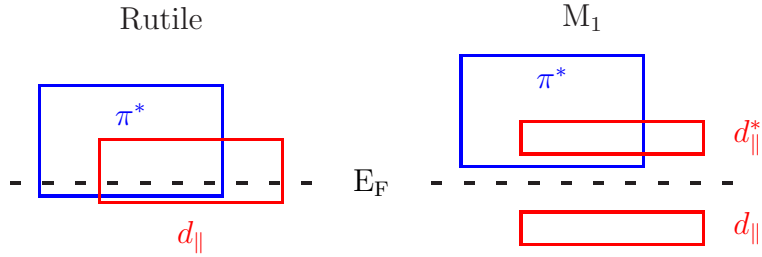
\begin{figure}
\begin{center}
\unitlength0.8mm
\begin{picture}(140,55)
\thicklines
\put( 10.0, 18.0){\textcolor{blue}{\framebox(30.0,18.0){ }}}
\put( 20.0, 17.0){\textcolor{red}{\framebox(30.0,10.0){ }}}
\put( 80.0, 23.0){\textcolor{blue}{\framebox(30.0,18.0){ }}}
\put( 90.0, 25.0){\textcolor{red}{\framebox(30.0, 5.0){ }}}
\put( 90.0, 10.0){\textcolor{red}{\framebox(30.0, 5.0){ }}}
\dashline{2}(  5.0, 20.0)( 55.0, 20.0)
\dashline{2}( 75.0, 20.0)(125.0, 20.0)
\put(  5.0, 40.0){\makebox(50.0,15.0){\large Rutile}}
\put( 75.0, 40.0){\makebox(50.0,15.0){\large $ {\rm M_1} $}}
\put( 55.0, 12.0){\makebox(20.0,16.0){\textcolor{black}{\large $ {\rm E_F} $}}}
\put( 10.0, 27.0){\makebox(30.0, 9.0){\textcolor{blue}{\large $ \pi^{\ast} $}}}
\put( 80.0, 32.0){\makebox(30.0,11.0){\textcolor{blue}{\large $ \pi^{\ast} $}}}
\put( 20.0,  5.0){\makebox(30.0,12.0){\textcolor{red}{\large $ d_{\parallel} $}}}
\put(120.0,  7.0){\makebox(15.0,11.0){\textcolor{red}{\large $ d_{\parallel} $}}}
\put(120.0, 22.0){\makebox(15.0,11.0){\textcolor{red}{\large $ d_{\parallel}^{\ast} $}}}
\end{picture}
\unitlength1pt
\caption{Schematic band diagrams for $ {\rm VO_2} $.}
\label{fig:intro2}   
\end{center}
\end{figure}

Lateron, Goodenough presented a schematic energy band diagram for both the 
metallic and the insulating phase \cite{goodenough71a,goodenough71b}. Using 
electrostatic considerations for the effective ionic charges he placed the 
oxygen $ 2p $ levels well below the vanadium $ 3d $ states. The latter 
are subject to octahedral crystal field splitting into lower $ t_{2g} $ and 
higher $ e_g $ levels. Furthermore, the $ t_{2g} $ states, which are located 
near the Fermi energy, are split into a $ d_{\parallel} $ state, which is 
directed along the rutile $ c $ axis, and the remaining $ \pi^{\ast} $ 
states, see Fig.\ \ref{fig:intro2}. 
In the monoclinic phase, metal-metal pairing within the vanadium chains 
parallel to the rutile $ c $ axis causes splitting of the $ d_{\parallel} $ 
band into filled bonding and empty antibonding states. In contrast, the 
$ \pi^{\ast} $ bands move to higher energies due to the antiferroelectric 
zigzag-type displacement of the vanadium atoms. As a result, a band gap 
opens between the bonding $ d_{\parallel} $ band and the other $ t_{2g} $ 
bands. According to Shin {\em et al.}\ the $ d_{\parallel} $ band splitting 
amounts to $ \approx 2.5 $\,eV, while the $ \pi^{\ast} $ bands raise by 
$ \approx 0.5 $\,eV \cite{shin90}. Nevertheless, in proposing this scheme 
Goodenough questioned the covalent-type bonding arising from metal-metal 
pairing as the major source of the phase transition. Instead, he suggested 
that the metal-insulator transition results predominantly from the increased 
$ p $--$ d $ overlap coming with the antiferroelectric distortion of the 
$ {\rm VO_6} $ octahedra \cite{goodenough71a,goodenough71b}. It would 
lower and raise the bonding and antibonding $ \pi $ and $ \pi^{\ast} $ 
levels. The $ d_{\parallel} $ states would be split as a consequence of the 
symmetry change. Eventually this would open the optical band gap. 
Using the general band scheme sketched in Fig.\ \ref{fig:intro2}, Hearn 
performed numerical calculations for one-dimensional vanadium chains and 
demonstrated the validity of the scenario outlined by Goodenough 
\cite{hearn72}. 

A mechanism differing from the molecular orbital picture of Goodenough 
was proposed by Zylbersztejn and Mott. These authors attributed the coupled 
metal-insulator and structural transition to the presence of strong 
electron-electron correlations especially in the $ d_{\parallel} $ band 
rather than to electron-lattice interaction \cite{zylbersztejn75}. 
However, as Zylbersztejn and Mott argue, these correlations are efficiently 
screened by the $ \pi^{\ast} $ bands in the metallic phase. In the 
insulating phase, the screening of the $ d_{\parallel} $ electrons 
is diminished since the $ \pi^{\ast} $ bands experience energetical 
upshift due to the antiferroelectric displacement of the V atoms. As 
a consequence, the narrow $ d_{\parallel} $ bands at the Fermi energy 
are susceptible to strong Coulomb correlations and undergo a Mott 
transition. This opens the optical band gap. 

Lateron, Paquet and Leroux-Hugon presented a band model for $ {\rm VO_2} $,  
which aimed at incorporating both the electron-electron interactions and 
the electron-lattice coupling on an equal footing \cite{paquet80}. To do 
so, the authors started from a tight-binding representation for the 
$ d_{\parallel} $ and $ \pi^{\ast} $ bands. In addition, they included 
Hubbard-type interactions for both bands as well as a contribution due 
to the phonons. Using experimentally derived parameters Paquet and 
Leroux-Hugon concluded that the metal-insulator transition is primarily 
driven by electron-electron correlations. However, these correlations 
would be strongly renormalized by the lattice distortion and the 
electrostatic interaction between the $ d_{\parallel} $ and the 
$ \pi^{\ast} $ electrons. 

First principles studies for rutile $ {\rm VO_2} $ range from calculations 
using semiempirical or overlapping free atom potentials 
\cite{caruthers73a,gupta77} to self-consistent state-of-the-art 
investigations \cite{nikolaev92,allen93,wentz94a,wentz94b,habil,kurmaev98}. 
Work on the insulating and monoclinic ($ {\rm M_1} $) phase includes a 
semiempirical calculation \cite{caruthers73b}, the molecular dynamics study  
by Wentzcovitch {\em et al.}\ \cite{wentz94a,wentz94b}, state-of-the-art 
investigations using the LDA and LDA+U \cite{habil,yangdiss}, and a recent 
model GW calculation \cite{continenza99}. Most of these calculations 
confirmed the general molecular-orbital scheme sketched by Goodenough.  
They found the same order of bands for the metallic phase and related 
the change in electronic 
structure occurring at the phase transition to the simultaneous 
symmetry-lowering distortions of the crystal structure. From a maximum 
in the generalized susceptibility occurring at the tetragonal R-point, 
which could be traced back exclusively to the $ d_{\parallel} $ bands,  
Gupta {\em et al.}\ concluded that the phase transition could be due to 
formation of a charge-density wave accompanied by a periodic lattice 
distortion leading to the monoclinic $ {\rm M_1} $ structure 
\cite{gupta77}. Support for the predominant influence of the lattice 
degrees of freedom on the transition came from first principles molecular 
dynamics calculations by Wentzcovitch {\em et al.} \cite{wentz94a}. Their 
variable cell shape approach allowed for simultaneously relaxing the 
atomic positions and the primitive translations. As a result, starting 
from different intermediate structures these authors obtained the 
monoclinic $ {\rm M_1} $ structure as the most stable one. In addition, 
a slightly metastable rutile structure was found, which was about 54 meV 
per f.u.\ higher in energy as compared to the $ {\rm M_1} $ structure. 
Calculated crystal structure parameters for both cells were in good 
agreement with the experimental data (see Sec.\ \ref{cryst}). However, 
the band structure obtained for monoclinic $ {\rm VO_2} $ contrasted 
the experimental finding of an 0.6\,eV wide insulating gap 
\cite{verleur68,rosevear73,blaauw75,shin90}. Instead, semimetallic 
behaviour with a band overlap of 0.04\,eV was found \cite{wentz94a}. This 
result was attributed to the typical failure of the local density 
approximation to correctly reproduce measured optical band gaps. The 
situation is thus not unlike that in small band gap 
semiconductors as Ge, where use of the local density approximation likewise 
leads to a metallic ground state. Wentzcovitch {\em et al.}\ suspected that, 
by further strengthening the V--V bonds, the gap should eventually open and, 
hence, $ {\rm VO_2} $ could be regarded as a band insulator. This was 
interpreted as assertion of the energy band scheme by Goodenough 
\cite{goodenough71a,goodenough71b}. More recent studies using the local 
density approximation confirmed the results presented by Wentzcovitch 
{\em et al.}\ in that they also found semimetallic rather than 
semiconducting behaviour \cite{habil,kurmaev98}. Nevertheless, a recent 
LDA+U approach and a quasiparticle band structure calculation as based on 
a model GW scheme produced the correct optical band gap 
\cite{yangdiss,continenza99}. 

Despite the huge amount of theoretical investigations there still remain 
fundamental questions. First of all, none of the previous studies has 
presented a detailed analysis of the band structure or the density of 
states in terms of the orbitals discussed by Goodenough. For this reason, 
the molecular orbital considerations as well as the more recent model 
approaches lack a profound justification. In addition, most of the 
previous studies, including those of Goodenough as well as Zylbersztejn 
and Mott, emphasize the role of the $ \pi^{\ast} $ bands for the transition. 
Yet, as will be discussed in more detail in Sec.\ \ref{cryst}, the role of 
the zigzag-like displacement of the metal atoms for the metal-insulator 
transition is still unclear. This is related to the different response of 
the surrounding oxygen atoms. In $ {\rm VO_2} $, the oxygen atoms stay 
essentially at their original positions and the vanadium atom shifts lead 
to an an increased metal-oxygen bonding. This causes an energetical upshift 
of the $ \pi^{\ast} $ bands. In contrast, in $ {\rm MoO_2} $ the metal atom 
displacements are to a large part compensated by the accompanying shift of 
the oxygen octahedron. As a consequence, the zigzag-like displacement of 
the metal atoms does not lead to an increased metal-oxygen bonding. As 
a consequence, it does not generally act as an antiferroelectric mode. 
Hence, the zigzag-like displacement cannot be regarded as a common origin 
for the destabilization of the rutile structure.

\subsection{The $ M_2 $ phase}
\label{m2phase}

\begin{figure}[htp]
\centering
\includegraphics[width=0.8\textwidth]{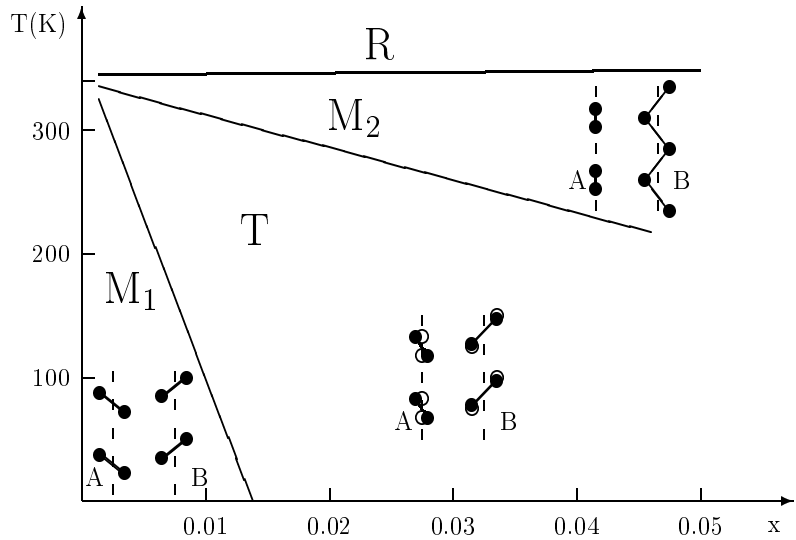}
\caption{Phase diagram of $ {\rm V_{1-x}Cr_xO_2} $ according to Pouget and 
         Launois \protect \cite{pouget76}. The distortion patterns of the 
         metallic chains are given in the insets (open circles in the T 
         phase refer to the positions of the $ {\rm M_2} $ phase).}
\label{fig:phasdg}   
\end{figure}

The understanding of $ {\rm VO_2} $ and the metal-insulator transition 
is even more complicated due to the presence of two additional phases of 
this material. They appear on application of uniaxial stress or on 
doping of $ {\rm VO_2} $ with small amounts of Cr, Fe, Al, or Ga of 
the order of few percent \cite{marezio72,pouget76,brueckner83}. The 
phase diagram of Cr doped $ {\rm VO_2} $ as given by Pouget and Launois 
\cite{pouget76} is shown in Fig.\ \ref{fig:phasdg}. 
This phase diagram differs slightly from that given by other authors,  
which fact has been attributed to differences in sample preparation 
and stoichiometry \cite{marezio72,pouget74}. Lateron, a similar phase 
diagram was reported for $ {\rm V_{1-x}Al_xO_2} $ by Ghedira {\em et al.}\ 
\cite{ghedira77a,ghedira77b}. According to Fig.\ \ref{fig:phasdg}, Cr-doped 
$ {\rm VO_2} $ enters, on cooling, the monoclinic $ {\rm M_2} $ phase in 
a first order metal-insulator transition \cite{pouget74}. On further 
lowering of the temperature, transition to the triclinic T phase occurs, 
which, for chromium concentrations smaller than 0.015, is followed by a 
first order transition to the monoclinic $ {\rm M_1} $ phase \cite{marezio72}. 
According to X-ray measurements by Marezio {\em et al.}, each of the 
three low-temperature phases shows a particular distortion pattern of 
the atomic arrangement especially of the characteristic vanadium chains 
\cite{marezio72}. Whereas in the $ {\rm M_1} $ phase both the metal-metal 
pairing and the zigzag-type lateral displacement are observed on each 
chain, in the $ {\rm M_2} $ phase only half of the chains dimerize and 
the zigzag-type deviations are reserved to the other half of the chains. 
Finally, the T phase is intermediate, as those chains, which in the 
$ {\rm M_2} $ phase dimerize, gradually start to tilt, whereas the zigzag 
chains start to dimerize until distortions in both chains are identical 
and eventually the $ {\rm M_1} $ phase is reached. Hence, the T and 
$ {\rm M_1} $ phases can be regarded as superpositions of two 
$ {\rm M_2} $-type displacements with unequal and equal weights, respectively. 
The role of the latter displacement pattern as the ``fundamental'' one may 
be understood from the symmetry considerations presented by Pouget 
{\em et al.}\ as well as by Paquet and Leroux-Hugon 
\cite{pouget74,paquet80,rice94}. According to these ideas 
the pairing on one chain is intimately connected with 
the zigzag-type displacement on the neighbouring chains: If the vanadium 
atoms pair on one chain, the atoms on the neighbouring chains, which from 
the outset are vertically displaced by half the rutile $ c $ axis, move off 
the chain axis towards the apical oxygen atom whose V neighbours of the first 
sublattice have pulled apart. For this reason, the zigzag-like displacement 
may be viewed as being coupled to the metal-metal pairing along the rutile 
$ c $ axis. 

$ {\rm V^{51}} $-NMR measurements by Pouget {\em et al.}\ confirmed the 
X-ray experiments performed by Marezio {\em et al.}. They revealed a 
positive Knight shift for the $ {\rm M_1} $ phase of $ {\rm V_{1-x}Cr_xO_2} $, 
which is similar to that of pure $ {\rm VO_2} $. In contrast, both a positive 
and a negative Knight shift were observed in the $ {\rm M_2} $ phase 
indicating two different types of vanadium atoms \cite{pouget74}. Whereas 
the positive 
Knight shift again points to paired V sites the negative Knight shift gives 
evidence for $ {\rm V^{4+}} $ atoms with localized $ 3d $ electrons. 
Finally, as D'Haenens {\em et al.}\ found from ESR data, the chromium atoms 
enter as substitutional $ {\rm Cr^{3+}} $ ions on the zigzag-type chains 
\cite{dhaenens74,dhaenens75}. 

Magnetic susceptibility curves taken by Pouget {\em et al.}\ for 
$ {\rm V_{1-x}Cr_xO_2} $ showed a striking similarity to corresponding 
data for Fe-doped $ {\rm VO_2} $ as measured by Kosuge \cite{kosuge67}. 
They could be decomposed into a van Vleck contribution for the paired 
vanadium chains, a Curie-Weiss term representing the $ {\rm Cr^{3+}} $ 
spins and an almost constant residual term. Using a model by Bonner and 
Fisher \cite{bonner64}, Pouget {\em et al.}\ were able to attribute the 
latter contribution to noninteracting spin-$ \frac{1}{2} $ linear 
Heisenberg chains, formed by the equispaced vanadium atoms, with an 
exchange constant of the order of room temperature \cite{pouget74}. 
The $ {\rm M_2} $ to T transition was thus viewed as a bonding or 
dimerization transition of a linear Heisenberg chain. This interpretation 
was confirmed by the observed decrease of the magnetic susceptibility at 
the transition, which is accompanied by an increase of the electrical 
conductivity by a factor of two \cite{ghedira77b}. From the existence 
of well localized $ 3d $ electrons at the $ {\rm V^{4+}} $ sites, Pouget 
{\em et al.}\ concluded that the results should be interpreted in terms 
of a Mott-Hubbard-type picture. In doing so, they attributed the 
insulating gap to electronic correlations \cite{pouget74}. 

Lateron, Pouget {\em et al.}\ demonstrated that both the $ {\rm M_2} $ and T 
phase could be likewise stabilized by applying uniaxial stress along either 
the $ [110] $ or the $ [1\bar{1}0] $ direction in stoichiometric $ {\rm VO_2} $ 
samples \cite{pouget75}. Such uniaxial stress can be viewed as suppressing 
the zigzag-type displacements on those chains, which have their tilting 
along the direction of the applied stress, while the respective other chains 
are not affected. However, from the above symmetry considerations it follows 
that, as a consequence of the reduced tilt on one half of the chains, the 
pairing on the other chains will be also reduced. Additional NMR and EPR data 
revealed that the local symmetries as well as the magnetic properties of the 
two vanadium sites are identical in pure $ {\rm VO_2} $ under uniaxial stress 
and in $ {\rm V_{1-x}Cr_xO_2} $ \cite{pouget75}. 

As Pouget {\em et al.}\ pointed out, the critical uniaxial stress for 
appearance of the $ {\rm M_2} $ phase is so small that the free energies 
of the $ {\rm M_1} $ and $ {\rm M_2} $ phases in pure $ {\rm VO_2} $ must 
be extremely close at temperatures just below the metal-insulator transition 
\cite{pouget75}. The same conclusion was drawn by Pouget {\em et al.}\ for 
doped $ {\rm VO_2} $. These authors pointed to the fact that, although 
the $ {\rm M_2} $ and T phases involve substantial changes in the vanadium 
atom positions, they could be stabilized by impurity concentrations as low 
as 0.2\% \cite{pouget74}. According to Pouget {\em et al.}\ the 
$ {\rm M_2} $ and the T phase thus must be interpreted as alternative 
phases of pure $ {\rm VO_2} $ whose free energies are only slightly higher 
than that of the $ {\rm M_1} $ phase of the pure material \cite{pouget74}. 
As a consequence, the $ {\rm M_2} $ phase was regarded as a metastable 
modification of the $ {\rm M_1} $ phase. In contrast, the T phase appears 
as a transitional state, which displays characteristics of both monoclinic 
phases. 

The discovery of the $ {\rm M_2} $ phase with its two different types of 
vanadium chains was difficult to reconcile with all those theoretical 
approaches, which had explained the insulating ground state as originating 
from crystal structure 
distortions with a predominant influence of either the pairing or the 
zigzag-type displacement. Furthermore, from the presence of localized 
$ 3d $ electrons on the zigzag chains and the antiferromagnetic ordering of 
the local moments within non-interacting linear Heisenberg chains many authors 
concluded that band theory were unable to correctly describe the physics of 
the $ {\rm M_2} $ and, hence, also of the $ {\rm M_1} $ phase. Instead, 
electronic correlations were regarded as essential and localized models 
called for \cite{pouget74,zylbersztejn75,rice94}. 
Yet, an explanation of the physics of $ {\rm VO_2} $ purely in terms of a 
Mott-Hubbard-type picture might be difficult in view of the complex 
structural changes at the metal-insulator transitions, which lead to 
the same crystal structures as those of the neighbouring metallic 
transition-metal dioxides. 

Although, as Rice {\em et al.}\ have pointed out again only recently 
\cite{rice94}, discussion of the monoclinic $ {\rm M_2} $ phase of 
$ {\rm VO_2} $ is essential for a comprehensive understanding of the 
physics underlying the metal-insulator transition, state-of-the-art first 
principles calculations for this phase did not seem to exist. In order to 
close this gap we included an investigation of the monoclinic $ {\rm M_2} $ 
phase in the present study \cite{habil}. Again, our aim is to provide an 
as broad as possible picture of the physics of the rutile-type transition 
metal oxides using a well defined first principles approach. 

To be specific, since the low-temperature $ {\rm M_1} $ and $ {\rm M_2} $ 
phases involve characteristic but still relative small deviations from the 
rutile phase we follow the lines of our work on $ {\rm MoO_2} $ and 
$ {\rm NbO_2} $ and start out investigating the rutile structure 
\cite{moo2pap,nbo2pap}. The results will then serve as a reference for the 
subsequent investigation of the low-temperature phases.

\section{Crystal structures and local coordinate systems}
\label{cryst}

\subsection{Rutile structure}
\label{crystrut}

\begin{figure}[htp]
\centering
\includegraphics[width=0.8\textwidth]{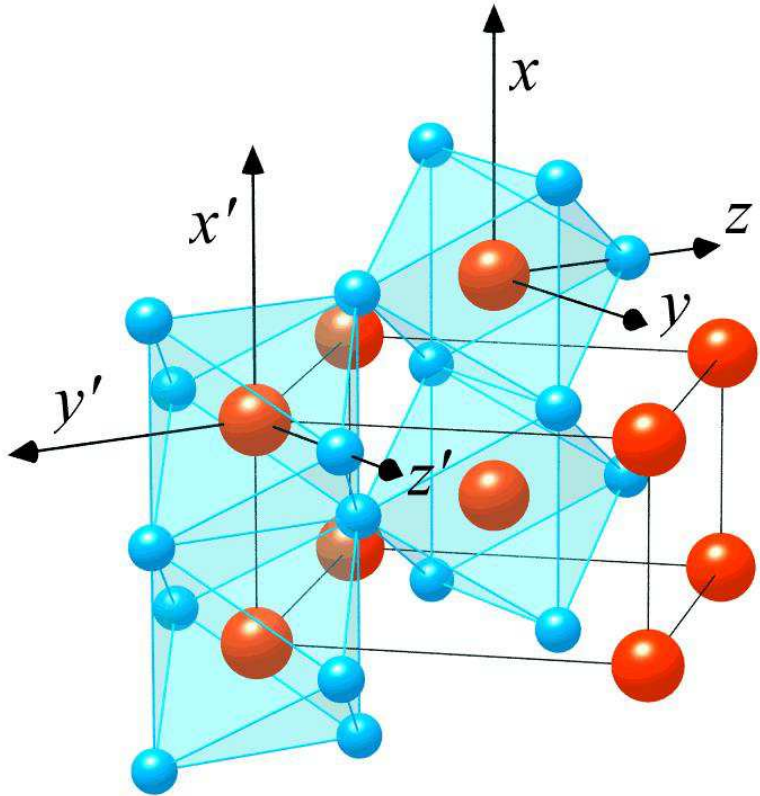}
\caption[The rutile structure.]
        {The rutile structure. Large and small spheres denote metal and 
         ligand atoms, respectively.} 
\label{fig:cryst1}   
\end{figure}

The rutile structure of metallic $ {\rm VO_2} $ is based on a simple 
tetragonal lattice with space group $ P4_2/mnm $ ($ D_{4h}^{14} $, 
No.\ 136) \cite{westman61,mcwhan74}. The metal atoms are located at the 
Wyckoff positions (2a): $ (0,0,0) $, $ (\frac{1}{2},\frac{1}{2},\frac{1}{2}) $ 
and the oxygen atoms occupy the positions (4f): 
$ \pm (u,u,0), \pm (\frac{1}{2}+u,\frac{1}{2}-u,\frac{1}{2}) $. 
The rutile structure is displayed in Fig.\ \ref{fig:cryst1}. 
According to McWhan {\em et al.}\ the lattice constants and the internal 
oxygen parameter are $ a_R $ = 4.5546\,{\AA}, $ c_R $ = 2.8514\,{\AA} 
and $ u $ = 0.3001 \cite{mcwhan74}. These numbers will be used in the 
calculations. While Westman reported slightly different lattice parameters 
for metallic $ {\rm VO_2} $ \cite{westman61}, the numbers by McWhan 
{\em et al.}\ are in very close agreement with the values given by Marezio 
{\em et al.}\ for $ {\rm V_{0.976}Cr_{0.024}O_2} $ ($ u $ = 0.3004) as 
well as with those by Ghedira {\em et al.}\ for $ {\rm VO_2} $ doped with 
1.5\% of Al ($ a_R $ = 4.5546\,{\AA}, $ c_R $ = 2.8528\,{\AA} and 
$ u $ = 0.3001) 
\cite{marezio72,ghedira77b}. Finally, we mention the parameters resulting 
from the first principles molecular dynamics calculations by Wentzcovitch 
{\em et al.}, $ a_R $ = 4.58\,{\AA}, $ c_R $ = 2.794\,{\AA} and 
$ u $ = 0.3000, which differ only slightly from the experimental values 
\cite{wentz94a}. 

As has been already outlined in our previous work on $ {\rm MoO_2} $ 
\cite{moo2pap}, the rutile structure can be alternatively visualized in terms 
of a body-centered tetragonal lattice formed by the metal atoms, where each 
metal atom is surrounded by an oxygen octahedron. Octahedra centered at the 
corners and the center of the cell are rotated by $ 90^{\circ} $ about the 
tetragonal $ c $ axis relative to each other. As a consequence, the lattice 
translational symmetry reduces to simple tetragonal. Thus, the unit cell 
contains two formula units. Octahedra neighbouring along the rutile 
$ c $ axis, share edges whereas the resulting octahedral chains are 
interlinked via corners. Each octahedron has orthorhombic symmetry. Yet, 
the deviations from tetragonal and even cubic geometry are rather small 
for most compounds and still allow for a discussion in terms of the 
latter. There exist two different metal-oxygen distances. The apical 
distance is between metal and oxygen atoms having the same $ z $ value. 
The equatorial distance is between the metal atom and the four neighbouring 
ligand atoms with $ z = z_{\rm metal} \pm 1/2 $ 
\cite{mattheiss76,bookdft,sorantin92}.

As we will realize in the following sections it is useful to discuss the 
electronic structure of rutile-type compounds in terms of local coordinate 
systems centered at each metal site. These coordinate systems have been 
already indicated in Fig.\ \ref{fig:cryst1}. Note that due to the different 
orientation of octahedra centered at the corner and in the center of the 
rutile cell, the local $ z $ axes point alternately along the $ [110] $ 
and $ [1\bar{1}0] $ direction. In contrast to the usual adjustment of the 
$ x $ and $ y $ axes parallel to the metal-ligand bonds we have rotated 
these axes by $ 45^{\circ} $ about the local $ z $ axes such that they 
are parallel and perpendicular, respectively, to the rutile $ c $ axis. 

\begin{figure}[htp]
\centering 
\subfigure[$ d_{3z^2-r^2} $]{\includegraphics[width=0.4\textwidth]{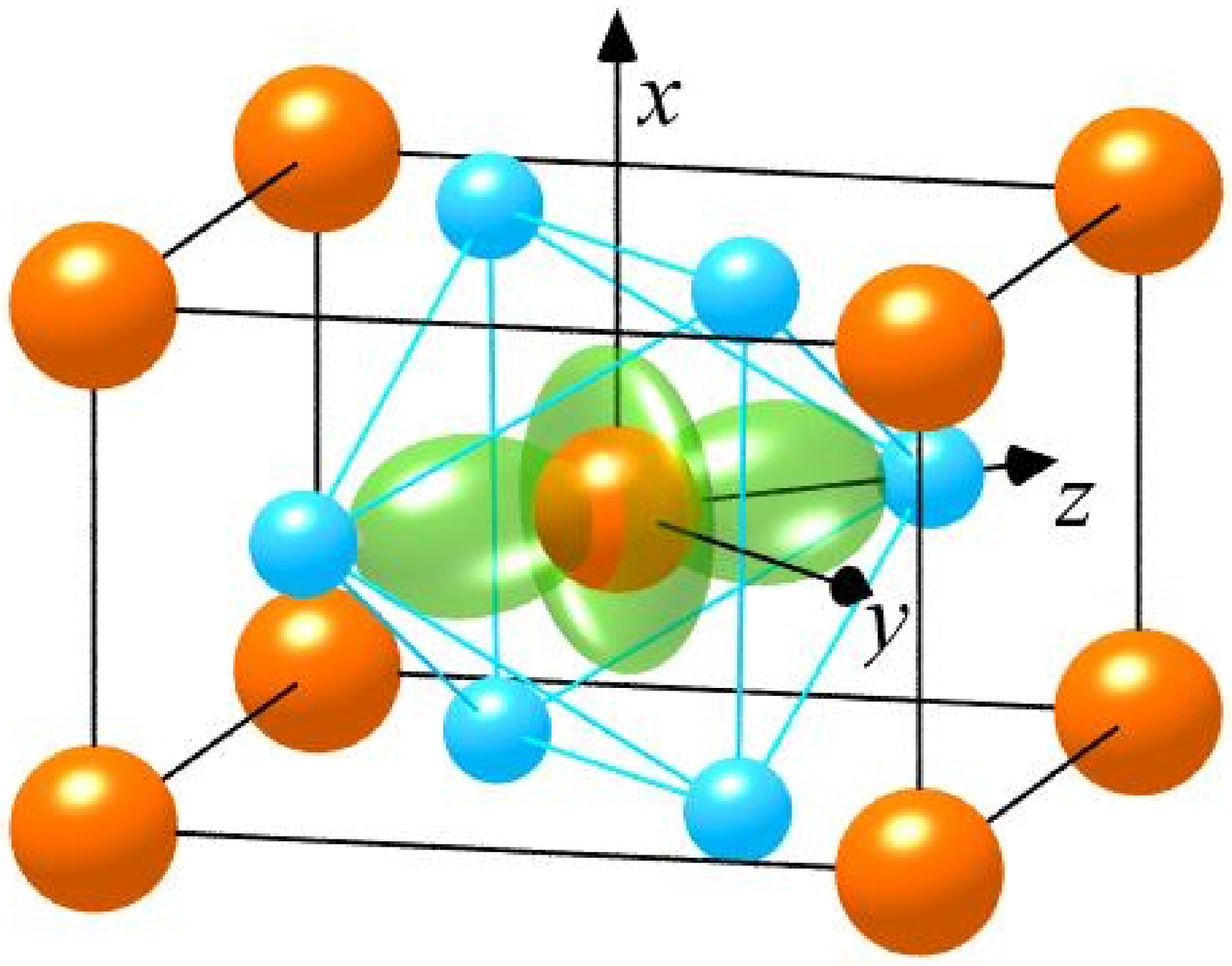}}
\hspace{0.1\textwidth}
\subfigure[$ d_{xy} $]{\includegraphics[width=0.4\textwidth]{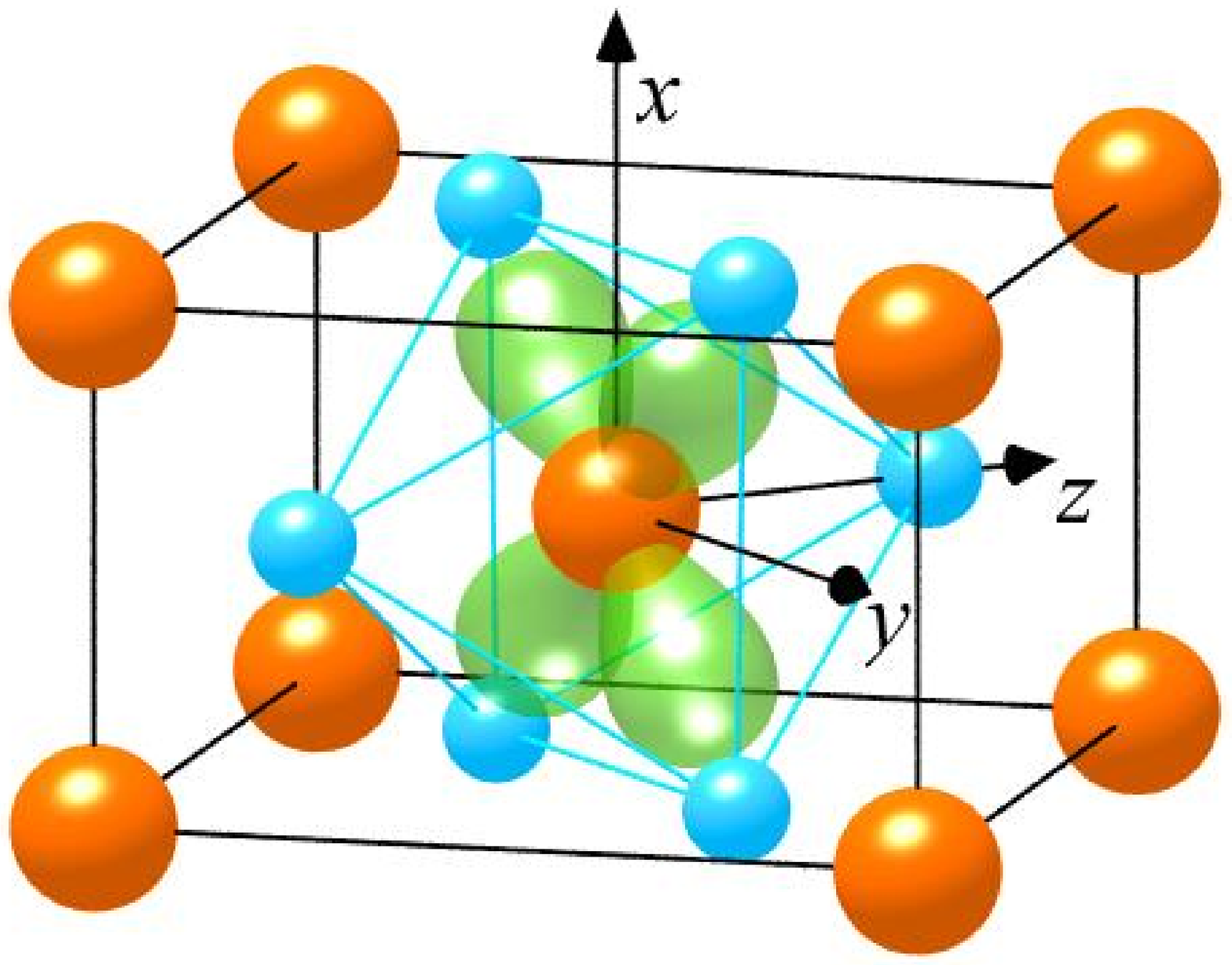}} 
\\ 
\subfigure[$ d_{x^2-y^2} $]{\includegraphics[width=0.4\textwidth]{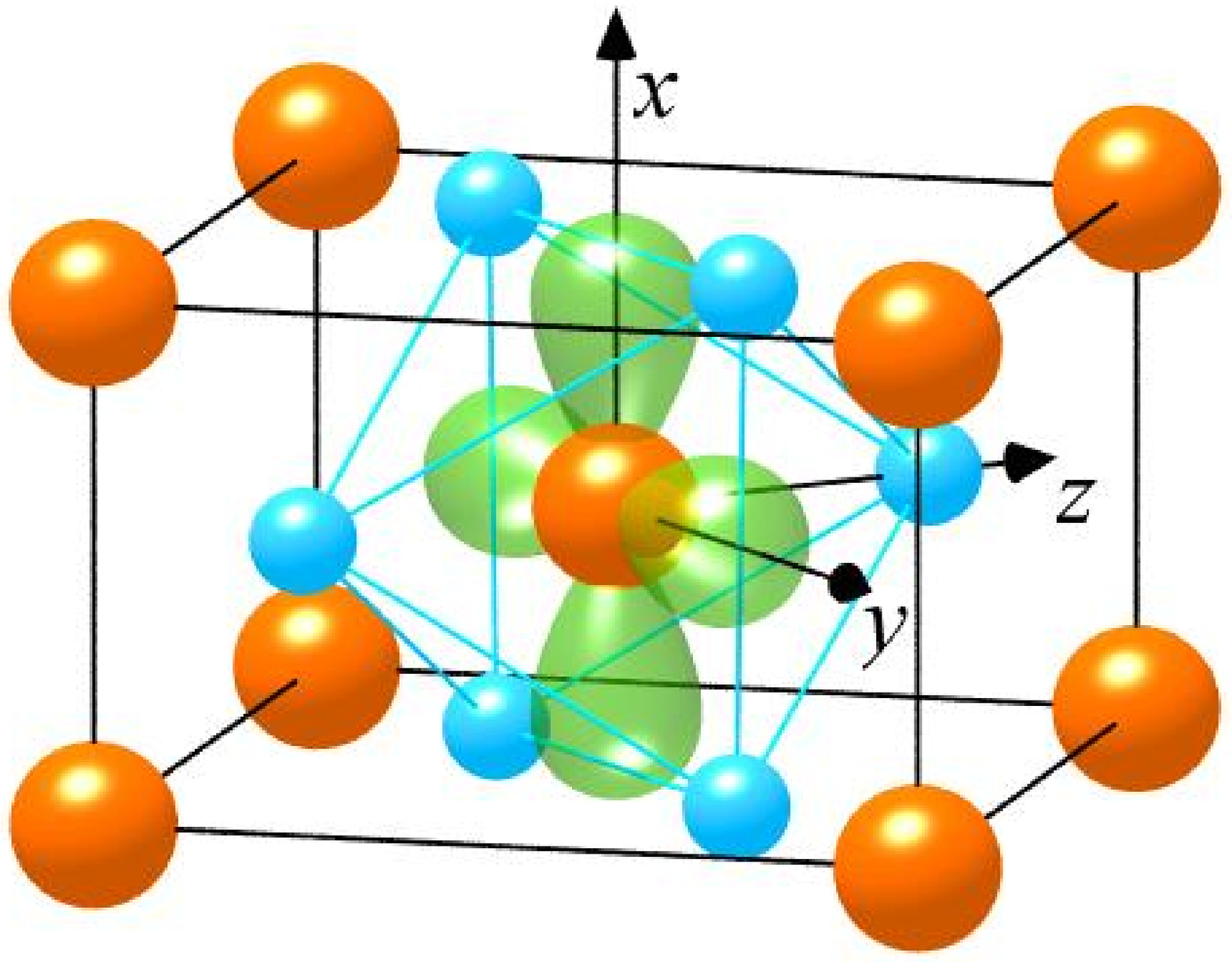}} 
\\ 
\subfigure[$ d_{xz} $]{\includegraphics[width=0.4\textwidth]{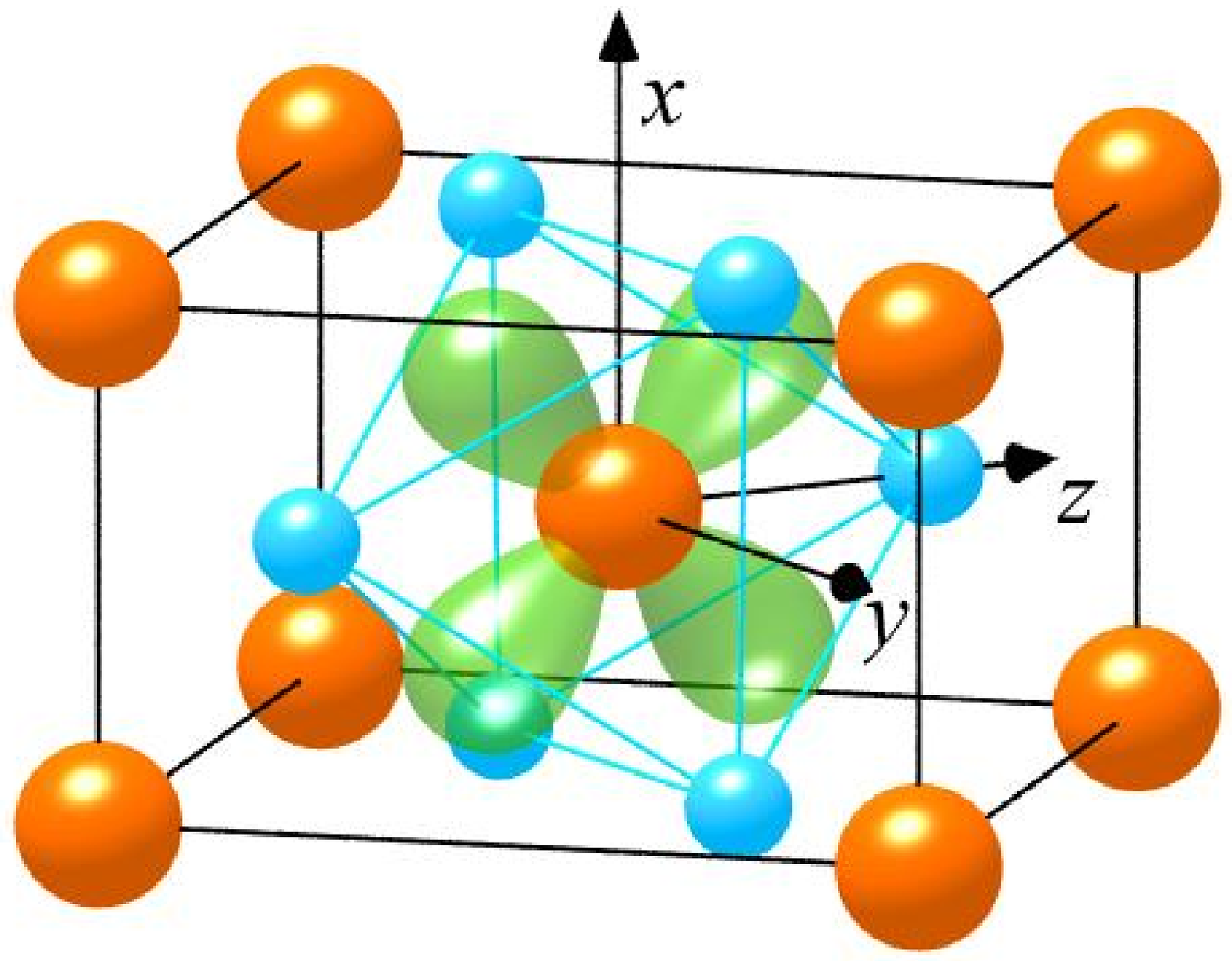}}
\hspace{0.1\textwidth}
\subfigure[$ d_{yz} $]{\includegraphics[width=0.4\textwidth]{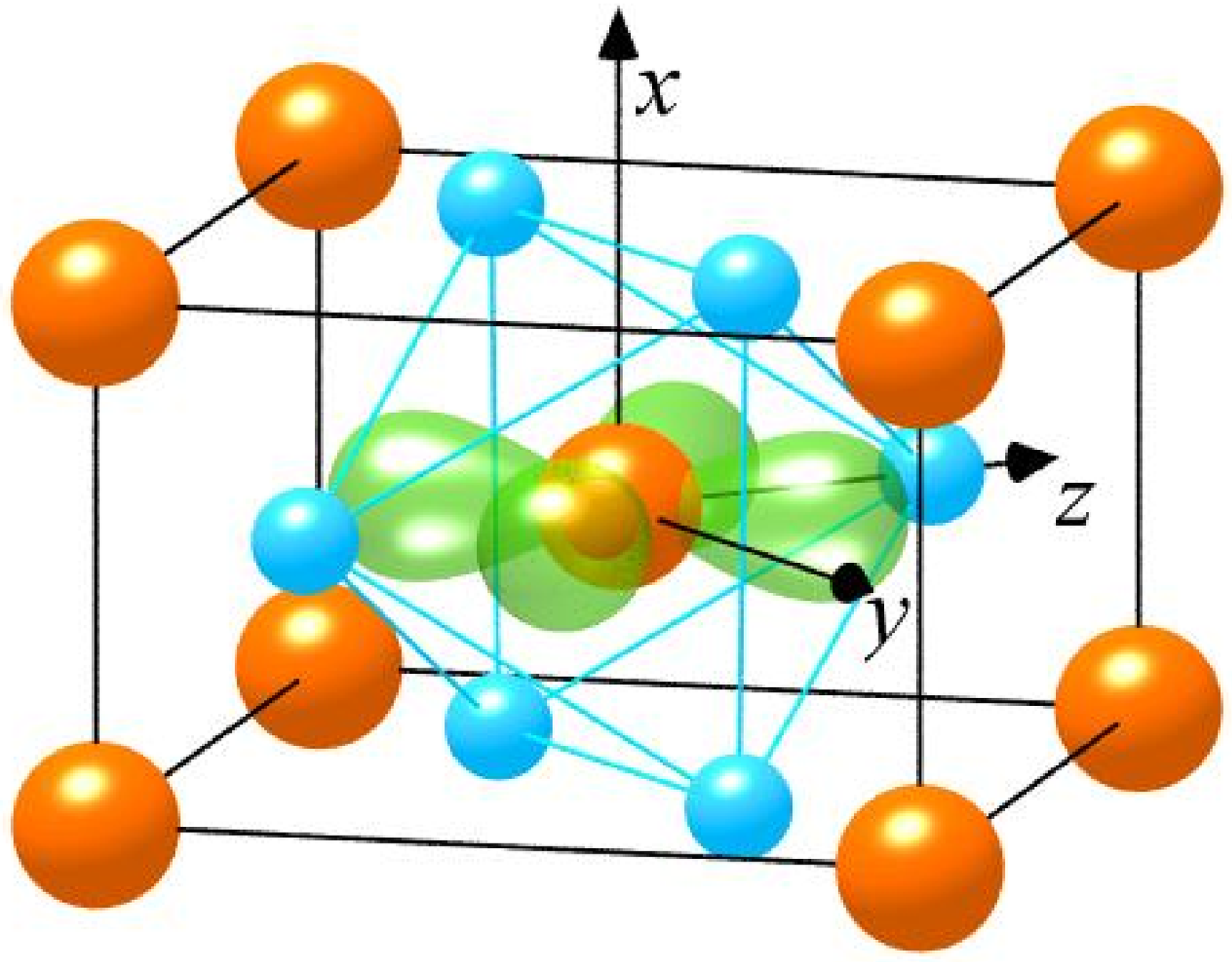}} 
\\ 
\caption{Angular parts of the $ d $ orbitals.}
\label{fig:cryst2}   
\end{figure}

In Fig.\ \ref{fig:cryst2},  
we show the angular parts of the metal $ d $ orbitals relative to the 
local reference frame of the central metal atom. With the previous choice 
of local coordinate systems the $ e_g $ states resulting from the cubic 
part of the crystal field splitting of the metal $ d $ orbitals comprise 
the $ d_{3z^2-r^2} $ and $ d_{xy} $ orbitals. In contrast, the $ t_{2g} $ 
states are built from the $ d_{x^2-y^2} $, $ d_{xz} $, and $ d_{yz} $ 
orbitals. While the $ d_{x^2-y^2} $ orbitals point along the rutile $ c $ 
and the local $ y $ axes, i.e.\ towards the edges of the basal plane of 
the octahedron, the $ d_{xz} $ and $ d_{yz} $ orbitals are directed towards 
the faces. In particular, the $ d_{yz} $ states point along the 
$ \langle100\rangle $ directions. As a consequence, the $ d_{x^2-y^2} $ 
and $ d_{xz} $ orbitals mediate $ \sigma $- and $ \pi $-type overlap, 
respectively, between metal sites within the vertical chains formed by 
the octahedra. In contrast, the $ d_{yz} $ orbitals have an albeit smaller 
$ \sigma $-type overlap with their counterparts at metal sites translated 
by the vectors $ \langle 1,0,0 \rangle $. This is due to the above 
$ 45^{\circ} $ rotation of the local coordinate system. This additional 
rotation interchanges the $ d_{x^2-y^2} $ and $ d_{xy} $ orbitals and 
adjusts the $ d_{yz} $ orbitals parallel to the axes of the rutile basal 
plane. The overlap of both the $ d_{x^2-y^2} $ and $ d_{yz} $ orbitals 
with orbitals of the same kind at neighbouring atoms thus connects atoms, 
which are separated by lattice vectors of the simple tetragonal lattice. 
In contrast, coupling between metal atoms, which are located at the 
corner and in the center of the cell, is mediated by the $ d_{xz} $ 
orbitals. These orbitals point to the voids between the metal atoms of 
the neighbouring octahedral chains where they overlap with the 
$ d_{x^2-y^2} $ orbitals of these chains.

\subsection{The $ M_1 $ structure}
\label{crystm1}

\begin{figure}[htp]
\centering
\includegraphics[width=0.8\textwidth]{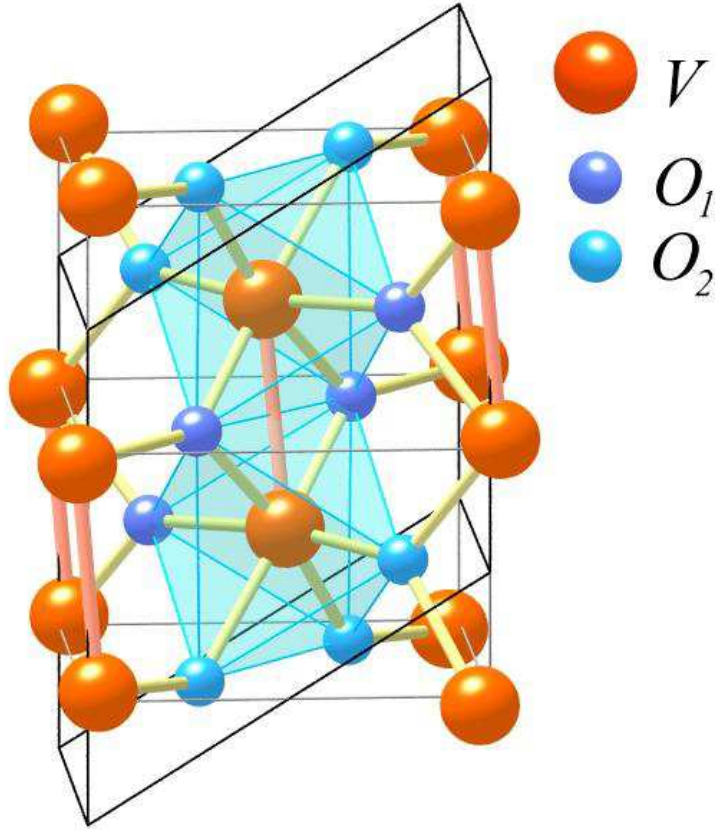}
\caption{Monoclinic $ {\rm M_1} $ structure of $ {\rm VO_2} $.}
\label{fig:cryst3}   
\end{figure}

The monoclinic $ {\rm M_1} $ structure of stoichiometric $ {\rm VO_2} $ 
is characterized by a simple monoclinic lattice with space group 
$ P2_1/c $ ($ C_{2h}^{5} $, No.\ 14) \cite{andersson56,longo70}. According 
to Andersson the lattice constants and the monoclinic angle are 
$ a_{M_1} $ = 5.743\,{\AA}, $ b_{M_1} $ = 4.517\,{\AA}, 
$ c_{M_1} $ = 5.375\,{\AA}, and $ \beta_{M_1}= 122.61^{\circ} $, respectively 
\cite{andersson56}. Lateron, Longo and Kierkegaard reported values 
$ a_{M_1} $ = 5.7517\,{\AA}, $ b_{M_1} $ = 4.5378\,{\AA}, 
$ c_{M_1} $ = 5.3825\,{\AA}, and $ \beta_{M_1}= 122.646^{\circ} $, respectively 
\cite{longo70}, which will be used in the subsequent calculations.  The 
crystal structure is displayed in Fig.\ \ref{fig:cryst3}. 
The unit cell comprises four formula units. The metal atoms as well as the 
two different types of oxygen atoms occupy the general Wyckoff position (4e): 
$ \pm (x,y,z) $, $ \pm (x,\frac{1}{2}-y,\frac{1}{2}+z) $. The latter are 
referred to the standard choice 
\begin{equation} 
{\bf a}_{M_1, 1} 
 = \left( \begin{array}{c}   0   \\ 
                             0   \\
                           - a_{M_1} 
          \end{array} 
   \right) 
\,, \; 
{\bf a}_{M_1, 2} 
 = \left( \begin{array}{c} - b_{M_1}   \\ 
                             0   \\ 
                             0     
          \end{array} 
   \right) 
\,, \; 
{\bf a}_{M_1, 3} 
 = \left( \begin{array}{c}   0   \\ 
                             c_{M_1} \sin \beta_{M_1}  \\
                           - c_{M_1} \cos \beta_{M_1} 
          \end{array} 
   \right) 
\label{eq:cryst1} 
\end{equation} 
for the primitive translations. The atomic positions given by Andersson 
as well as by Longo and Kierkegaard are listed in Tables \ref{tab:cryst1} 
\begin{table}[ht]
\begin{center}
\caption{Crystal structure parameters of $ {\rm M_1} $-$ {\rm VO_2} $ 
         as given by Andersson 
         (from Ref.\ \protect \cite{andersson56}).}
\label{tab:cryst1}   
\begin{tabular}{ccrrr} 
\\[-3mm] \hline \\[-3mm]
Atom        & Wyckoff positions & \multicolumn{3}{c}{parameters}     \\
\\[-3mm] \hline \\[-3mm]
            &                   &   $ x $   &   $ y $   &   $ z $    \\
\\[-3mm] \hline \\[-3mm]
$ \rm V   $ & (4e)              &   0.242   &   0.975   &   0.025    \\
$ \rm O_1 $ & (4e)              &   0.10    &   0.21    &   0.20     \\
$ \rm O_2 $ & (4e)              &   0.39    &   0.69    &   0.29     \\
\\[-3mm] \hline \\[-3mm]
\end{tabular} 
\end{center}
\end{table} 
and \ref{tab:cryst2}. 
\begin{table}[ht]
\begin{center}
\caption{Crystal structure parameters of $ {\rm M_1} $-$ {\rm VO_2} $ 
         as given by Longo and Kierkegaard 
         (from Ref.\ \protect \cite{longo70}).}
\label{tab:cryst2}   
\begin{tabular}{ccrrr} 
\\[-3mm] \hline \\[-3mm]
Atom        & Wyckoff positions & \multicolumn{3}{c}{parameters}     \\
\\[-3mm] \hline \\[-3mm]
            &                   &   $ x $   &   $ y $   &   $ z $    \\
\\[-3mm] \hline \\[-3mm]
$ \rm V   $ & (4e)              &  0.23947  &  0.97894  &  0.02646   \\
$ \rm O_1 $ & (4e)              &  0.10616  &  0.21185  &  0.20859   \\
$ \rm O_2 $ & (4e)              &  0.40051  &  0.70258  &  0.29884   \\
\\[-3mm] \hline \\[-3mm]
\end{tabular} 
\end{center}
\end{table} 
Note that all positions have been shifted by half of the rutile $ c $ 
axis, i.e.\ quarter of the monoclinic $ a $ axis as compared to the 
high-temperature data. 

Again we complement the measured crystal structure parameters by those 
resulting from the variable cell shape molecular dynamics calculations 
by Wentzcovitch {\em et al.}\ \cite{wentz94a,wentz94b}. These authors 
reported lattice constants $ a_{M_1} $ = 5.629\,{\AA}, 
$ b_{M_1} $ = 4.657\,{\AA}, $ c_{M_1} $ = 5.375\,{\AA}, and a monoclinic 
angle $ \beta_{M_1}= 122.56^{\circ} $, respectively \cite{wentz94a}. 
The calculated atomic parameters are listed in Table \ref{tab:cryst3}.  
\begin{table}[ht]
\begin{center}
\caption{Crystal structure parameters of $ {\rm M_1} $-$ {\rm VO_2} $ 
         as given by Wentzcovitch {\em et al.}\ 
         (from Ref.\ \protect \cite{wentz94a}).}
\label{tab:cryst3}   
\begin{tabular}{ccrrr} 
\\[-3mm] \hline \\[-3mm]
Atom        & Wyckoff positions & \multicolumn{3}{c}{parameters}     \\
\\[-3mm] \hline \\[-3mm]
            &                   &   $ x $   &   $ y $   &   $ z $    \\
\\[-3mm] \hline \\[-3mm]
$ \rm V   $ & (4e)              &   0.233   &   0.976   &   0.021    \\
$ \rm O_1 $ & (4e)              &   0.118   &   0.212   &   0.228    \\
$ \rm O_2 $ & (4e)              &   0.399   &   0.685   &   0.293    \\
\\[-3mm] \hline \\[-3mm]
\end{tabular} 
\end{center}
\end{table} 
As for the rutile structure the agreement with the measured values is 
very good. With only few small exceptions all calculated numbers are 
within the range spanned by the two sets of experimental data, which 
by themselves show some variation. 

The close relationship of the monoclinic $ {\rm M_1} $ structure to 
the high-temperature rutile structure becomes visible from a comparison 
of Fig.\ \ref{fig:cryst3} to Fig.\ \ref{fig:cryst1}. Furthermore, 
comparing the lattice parameters of the rutile and $ {\rm M_1} $ 
structures we realize that the primitive translations of the latter 
can be approximately written as 
\begin{equation} 
{\bf a}_{M_1, 1} 
 \approx \left( \begin{array}{c}   0   \\ 
                                   0   \\ 
                                 - 2 c_R   
          \end{array} 
   \right) 
\,, \;
{\bf a}_{M_1, 2} 
 \approx \left( \begin{array}{c} - a_R  \\
                                   0    \\
                                   0 
          \end{array} 
   \right) 
\,, \; 
{\bf a}_{M_1, 3} 
 \approx \left( \begin{array}{c}   0    \\ 
                                   a_R  \\
                                   c_R      
          \end{array} 
   \right)    
\,. 
\label{eq:cryst2} 
\end{equation} 
This representation emphasizes the prescription of the monoclinic cell 
in terms of a rutile cell doubled along the rutile $ c $ axis.  

Nevertheless, as for $ {\rm MoO_2} $  and $ {\rm NbO_2} $ there exist 
three distinct differences between both structures: 
i) In Fig.\ \ref{fig:cryst3} we observe the characteristic metal-metal 
pairing along the rutile $ c $ axis. It gives rise to two different 
metal-metal distances of 2.619 and 3.164\,{\AA}, which contrast the 
average value of 2.851\,{\AA} assumed in the rutile structure. 
ii) We witness the zigzag-like in-plane displacements of the vanadium 
atoms parallel to the local $ z $ axis, hence, parallel to the diagonal 
of the rutile basal plane. These shifts likewise alternate along both 
the tetragonal $ a $ and $ c $ axis. As a consequence of the zigzag-like 
distortion two different apical vanadium oxygen bond lengths of 1.77 and 
2.01\,{\AA} arise. In contrast, the metal atom pairing leads to two short 
and two long equatorial V--O bond lengths of 1.86, 1.89, 2.03, and 
2.06\,{\AA}. 
iii) There is a lattice strain which causes deviations of the ratios 
\begin{displaymath} 
\frac{     c_{M_1} \sin \beta_{M_1} }{ b_{M_1} } = 0.9988   \;, \qquad
\frac{ - 2 c_{M_1} \cos \beta_{M_1} }{ a_{M_1} } = 1.0096   
\end{displaymath} 
from unity. 

Despite the many similarities of the monoclinic structures of $ {\rm MoO_2} $ 
and $ {\rm VO_2} $ a more detailed analysis of the crystal structure data 
reveals striking differences between both compounds. They regard the 
zigzag-like in-plane displacements of the metal atoms and the distortions 
of the surrounding oxygen octahedra:  
i) In $ {\rm MoO_2} $ the in-plane shift makes an angle of $ 18^{\circ} $ 
   with the local $ z $ axis. In contrast, the vanadium atoms, in addition 
   to dimerizing along the rutile $ c $ axis, experience in-plane shifts, 
   which are parallel to the local $ z $ axis, hence, parallel to the 
   $ \langle110\rangle $ direction. 
ii) Even more striking is the different response of the oxygen sublattice 
    to the displacements of the metal atoms. In $ {\rm VO_2} $ the oxygen 
    atoms do not follow the metal atom shifts and stay almost at their original 
    positions. In contrast, the oxygen atoms in $ {\rm MoO_2} $, while 
    hardly moving parallel to the rutile $ c $ axis, follow the lateral 
    displacements of the molybdenum atoms to a large degree. Hence, in  
    $ {\rm VO_2} $ the lateral displacement of the vanadium atoms causes 
    a shift {\em relative} to the rather fixed oxygen octahedra such that 
    the $ p $--$ d $-like metal-oxygen bonding is affected. This is 
    different in $ {\rm MoO_2} $, where the molybdenum atom displacements 
    are compensated to a large degree by the simultaneous shift of the 
    surrounding oxygen atoms. In other words, the zigzag-type in-plane 
    displacement of the metal atoms does not automatically lead to an 
    (anti)ferroelectric distortion of the whole octahedron, but may still 
    be compensated by a corresponding displacement of the surrounding 
    oxygen cage. Nevertheless, according to the symmetry considerations 
    pointed out e.g.\ by Paquet and Leroux-Hugon, the zigzag-type shift 
    is still coupled to the vertical metal-metal pairing, 
The previous observation and especially the different behaviour of the 
two dioxides has strong implications for our understanding of the whole 
class of materials. As has been discussed in Sec.\ \ref{survey}, the lateral 
zigzag-like displacement of the metal atoms could lead to energetical 
upshift of the $ \pi^{\ast} $ bands. However, if compensated by 
accompanying oxygen displacements, this antiferroelectric mode is 
deactivated and the $ \pi^{\ast} $ bands stay at their original position. 
This has indeed been observed in our calculations for $ {\rm MoO_2} $ 
\cite{moo2pap}. We are thus led to the conclusion that the antiferroelectric 
mode can not explain the tendency of the transition-metal dioxides to 
form distorted variants of the rutile structure. For the same reason, this 
mode must be excluded as the driving force for the metal-insulator 
transition, which is intimately connected to the destabilization of rutile. 
These observations are difficult to reconcile with the scenarios sketched 
by Goodenough as well as Zylbersztejn and Mott 
\cite{goodenough71a,goodenough71b,zylbersztejn75}

\subsection{The $ M_2 $ structure}
\label{crystm2}

In the $ {\rm M_2} $ phase, $ {\rm VO_2} $ crystallizes in a centered 
monoclinic lattice with space group $ C2/m $ ($ C_{2h}^{3} $, No.\ 12) 
\cite{marezio72,ghedira77b}. In contrast to the monoclinic $ {\rm M_1} $ 
structure, this time the monoclinic angle distorts the basal plane of the 
original rutile cell and, hence, the setup of the new primitive translations 
is different from that of the $ {\rm M_1} $ phase. Usually, the crystal 
structure of the $ {\rm M_2} $ phase is specified in terms of the underlying 
{\em simple} monoclinic lattice as defined by the primitive translations 
\cite{marezio72} 
\begin{equation} 
{\bf a}_{M_2, 1} 
 = \left( \begin{array}{c}   a_{M_2} \\ 
                             0   \\
                             0       
          \end{array} 
   \right) 
\,, \; 
{\bf a}_{M_2, 2} 
 = \left( \begin{array}{c}   0   \\ 
                             0   \\ 
                             b_{M_2} 
          \end{array} 
   \right) 
\,, \; 
{\bf a}_{M_2, 3} 
 = \left( \begin{array}{c}   c_{M_2} \cos \beta_{M_2}  \\
                           - c_{M_2} \sin \beta_{M_2}  \\
                             0 
          \end{array} 
   \right)                  
\,. 
\label{eq:cryst3} 
\end{equation} 
The lattice constants and monoclinic angle as resulting from the single 
crystal measurements on $ {\rm V_{0.976}Cr_{0.024}O_2} $ by Marezio 
{\em et al.}\ are $ a_{M_2} $ = 9.0664\,{\AA}, $ b_{M_2} $ = 5.7970\,{\AA}, 
$ c_{M_2} $ = 4.5255\,{\AA}, and $ \beta_{M_2}= 91.88^{\circ} $, respectively 
\cite{marezio72}. There exist two different types of metal atoms and three 
types of oxygens, which occupy subsets of the general Wyckoff position (8j): 
$ \pm (x,y,z) $, $ \pm (x,-y,z) $, $ (\frac{1}{2},\frac{1}{2},0) \pm (x,y,z) $, 
and $ (\frac{1}{2},\frac{1}{2},0) \pm (x,-y,z) $. The parameters given by 
Marezio {\em et al.}, which will be used in the calculations below, are 
listed in Table \ref{tab:cryst4}.  
\begin{table}[ht]
\begin{center}
\caption{Crystal structure parameters of the $ {\rm M_2} $ phase of 
         $ {\rm V_{0.976}Cr_{0.024}O_2} $ 
         as given by Marezio {\em et al.} \protect \cite{marezio72}.}
\label{tab:cryst4}   
\begin{tabular}{ccrrr} 
\\[-3mm] \hline \\[-3mm]
Atom        & Wyckoff positions & \multicolumn{3}{c}{parameters}     \\
\\[-3mm] \hline \\[-3mm]
            &                   &   $ x $   &   $ y $   &   $ z $    \\
\\[-3mm] \hline \\[-3mm]
$ \rm V_1 $ & (4g)              &   0.0     &   0.7189  &   0.0      \\
$ \rm V_2 $ & (4i)              &   0.2314  &   0.0     &   0.5312   \\
$ \rm O_1 $ & (8j)              &   0.1482  &   0.2475  &   0.2942   \\
$ \rm O_2 $ & (4i)              &   0.3969  &   0.0     &   0.2089   \\
$ \rm O_3 $ & (4i)              &   0.1000  &   0.0     &   0.7987   \\
\\[-3mm] \hline \\[-3mm]
\end{tabular} 
\end{center}
\end{table} 
Ghedira {\em et al.}, who performed single crystal measurements on 
$ {\rm VO_2} $ doped with 1.5\% of aluminum, also found a simple 
monoclinic lattice with lattice constants $ a_{M_2} $ = 9.060\,{\AA}, 
$ b_{M_2} $ = 5.800\,{\AA}, $ c_{M_2} $ = 4.5217\,{\AA}, monoclinic angle 
$ \beta_{M_2}= 91.85^{\circ} $, and the atomic positions listed in Table 
\ref{tab:cryst5} \cite{ghedira77a,ghedira77b}. 
\begin{table}[ht]
\begin{center}
\caption{Crystal structure parameters of the $ {\rm M_2} $ phase of 
         $ {\rm V_{0.985}Al_{0.015}O_2} $ 
         as given by Ghedira {\em et al.} \protect \cite{ghedira77b}.}
\label{tab:cryst5}   
\begin{tabular}{ccrrr} 
\\[-3mm] \hline \\[-3mm]
Atom        & Wyckoff positions & \multicolumn{3}{c}{parameters}     \\
\\[-3mm] \hline \\[-3mm]
            &                   &   $ x $   &   $ y $   &   $ z $    \\
\\[-3mm] \hline \\[-3mm]
$ \rm V_1 $ & (4g)              &   0.0     &   0.7189  &   0.0      \\
$ \rm V_2 $ & (4i)              &   0.2312  &   0.0     &   0.5311   \\
$ \rm O_1 $ & (8j)              &   0.1460  &   0.2474  &   0.2865   \\
$ \rm O_2 $ & (4i)              &   0.3975  &   0.0     &   0.2284   \\
$ \rm O_3 $ & (4i)              &   0.0980  &   0.0     &   0.7862   \\
\\[-3mm] \hline \\[-3mm]
\end{tabular} 
\end{center}
\end{table} 
The striking close agreement of both data sets supports the interpretation 
of the $ {\rm M_2} $ phase as a metastable modification of the $ {\rm M_1} $ 
phase, which hardly depends on the dopant and the amount of doping. 
As already mentioned, the $ {\rm M_2} $ phase can be likewise stabilized by 
the application of uniaxial pressure along the $ [110] $ direction 
\cite{pouget75}. Unfortunetely, no crystal structure data exist for this 
situation.  

\begin{figure}[htp]
\centering
\includegraphics[width=0.8\textwidth]{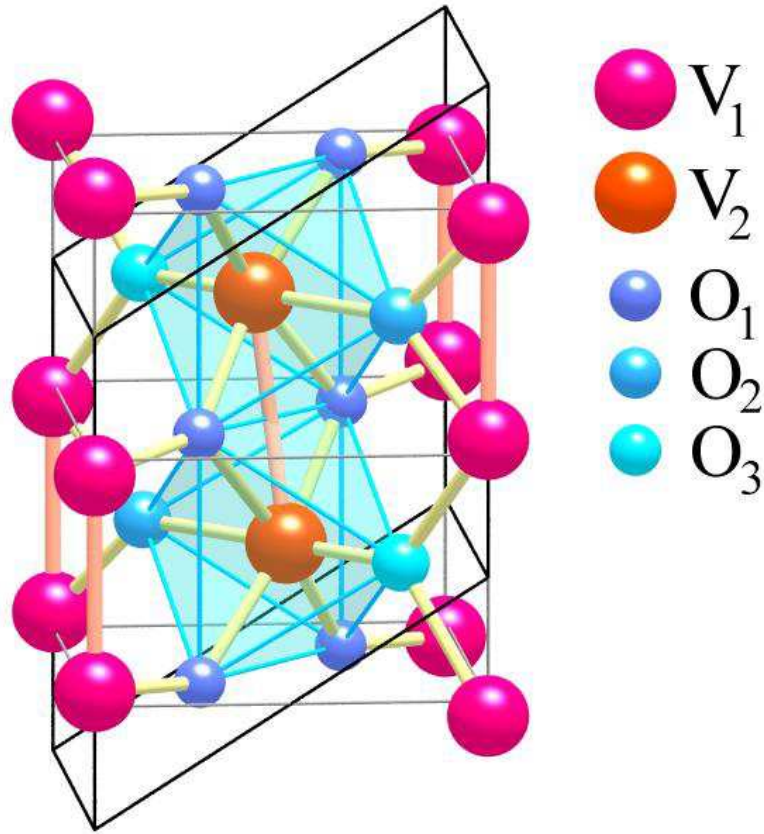}
\caption{Monoclinic $ {\rm M_2} $ structure of $ {\rm VO_2} $.}
\label{fig:cryst4}   
\end{figure}

For the purpose of the present work we prefer using the related centered 
monoclinic lattice, which comprises four instead of eight formula units. 
An alternative representation of this lattice originates from a redefinition 
of the first lattice vector as 
\begin{equation} 
{\bf a}_{M_2, 1} 
 = \frac{1}{2} 
   \left( \begin{array}{c}   a_{M_2} \\ 
                             0   \\
                             b_{M_2} 
          \end{array} 
   \right) 
\label{eq:cryst4} 
\end{equation} 
while $ {\bf a}_{M_2, 2} $ and $ {\bf a}_{M_2, 3} $ are the same as in 
Eq.\ (\ref{eq:cryst3}). Using this definition and comparing the lattice 
constants of the rutile and the $ {\rm M_2} $ phases we note the following 
approximate relation for the primitive translations \cite{marezio72} 
\begin{equation} 
{\bf a}_{M_2, 1} 
 \approx \left( \begin{array}{c}   a_R   \\ 
                                   0     \\
                                   c_R      
          \end{array} 
   \right) 
\,, \;
{\bf a}_{M_2, 2} 
 \approx \left( \begin{array}{c}   0   \\ 
                                   0   \\ 
                                   2 c_R   
          \end{array} 
   \right) 
\,, \; 
{\bf a}_{M_2, 3} 
 \approx \left( \begin{array}{c}   0  \\
                                 - a_R  \\
                                   0 
          \end{array} 
   \right)    
\,. 
\label{eq:cryst5} 
\end{equation} 
They are equivalent to the vectors given in Eq.\ (\ref{eq:cryst2}). Indeed, 
were it not for the lattice strain present in both low-temperature 
structures, which hinders writing the primitive translations as integer 
linear 
combinations of the rutile lattice vectors, the primitive translations of 
the $ {\rm M_1} $ and $ {\rm M_2} $ structure were identical (except for 
a rotation by $ 90^{\circ} $ about the Cartesian $ z $ axis). This is a 
consequence of the fact that both monoclinic phases result from the same 
R-point instability of the rutile phase. This has been pointed out by Brews 
\cite{brews70} and by Blount (as mentioned by McWhan {\em et al.}\ 
\cite{mcwhan74}). As a consequence, we are able to use the same 
setup of Brillouin zones for both monoclinic lattices. This will 
facilitate our discussion of the band structures below. The close 
relationship between the two monoclinic phases becomes also obvious 
from Fig.\ \ref{fig:cryst4},  
where we display the crystal structure of the $ {\rm M_2} $ phase using the 
unit cell spanned by the lattice vectors (\ref{eq:cryst3}) and 
(\ref{eq:cryst4}). We identify the vanadium atoms $ {\rm V_1} $ and 
$ {\rm V_2} $ at the corners and centers, respectively, of the underlying 
stacked rutile cells. Oxygen atoms of type $ {\rm O_1} $ are located at 
the apices of the octahedra centered about vanadium atoms $ {\rm V_1} $. 
They span the equatorial planes of the octahedra centered about vanadium 
atoms $ {\rm V_2} $. In contrast, oxygen atoms of types $ {\rm O_2} $ and 
$ {\rm O_3} $ are equatorial atoms for $ {\rm V_1} $ and apical atoms 
relative to $ {\rm V_2} $. Since the latter are shifted along the axis 
$ {\rm O_2} $--$ {\rm V_2} $--$ {\rm O_3} $ (see below) the identity of 
these two oxygen atoms is lost and two different apical $ {\rm V_2} $--O 
bond lengths evolve. 

As for the $ {\rm M_1} $ structure, differences between the R and 
$ {\rm M_2} $ phases arise from the distinct local deformations of the 
$ {\rm VO_6} $ octahedra. Again, basic entities are i) the metal-metal pairing 
along the rutile $ c $ axis and ii) the zigzag-like in-plane displacements of 
the vanadium atoms parallel to the local $ z $ axis, hence, parallel to the 
diagonal of the rutile basal plane (see above). As in the $ {\rm M_1} $ 
structure these shifts alternate 
along both the tetragonal $ a $ and $ c $ axis. Furthermore, 
the oxygen atoms hardly move such that the antiferroelectric mode causes a 
displacement of the vanadium atoms {\em relative} to the surrounding oxygen 
octahedron and is not suppressed as in $ {\rm MoO_2} $. 

Yet, in contrast to the $ {\rm M_1} $ phase, the above basic distortion 
patterns are not equally displayed by both chains. As already discussed 
in Sec.\ \ref{survey} and as is visible in Fig.\ \ref{fig:cryst4}, only 
half of the chains dimerize ($ {\rm V_1} $ chains) while the other half 
($ {\rm V_2} $ chains) experiences the zigzag-like displacement without 
showing any pairing. As a consequence, two different 
$ {\rm V_1} $--$ {\rm V_1} $ bond lengths of 2.538 and 3.259\,{\AA} appear 
while there is only one $ {\rm V_2} $--$ {\rm V_2} $ bond length of 
2.933\,{\AA}. Within the octahedra centered about $ {\rm V_1} $ the two 
apical $ {\rm V_1} $--$ {\rm O_1} $ distances are 1.868\,{\AA}. Due to the 
vanadium-vanadium dimerization within these chains two short 
$ {\rm V_1} $--$ {\rm O_2} $ distances of 1.852\,{\AA} and two long 
$ {\rm V_1} $--$ {\rm O_3} $ distances of 2.089\,{\AA} arise. In contrast, 
in the octahedra centered about the tilted vanadium atoms the four 
equatorial $ {\rm V_2} $--$ {\rm O_1} $ bond lengths of 1.931 and 
1.974\,{\AA} are very similar. At the same time, the distances between the 
central $ {\rm V_2} $ atom and the apical $ {\rm O_2} $ and $ {\rm O_3} $ 
disproportionate and assume the values 2.127 and 1.726\,{\AA}. 

Finally, we mention the lattice strain, which, in addition to the strains 
present in the $ {\rm M_1} $ lattice, distorts the rutile basal plane. The 
resulting parallelogram would be likewise expected to result from uniaxial 
pressure along the $ [110] $ direction. It suppresses the zigzag-like 
displacement on those chains, which have their local $ z $ axes parallel to 
the shorter diagonal of the parallelogram. Since, according to the above 
symmetry arguments, the zigzag mode is connected to the dimerization of 
metal atoms on the neighbouring chains, suppression of the in-plane 
displacement on half of the chains may well lead to a suppression of the 
metal-metal pairing on the other half.

\section{Method of calculation}
\label{method}

The calculations performed in this study are based on density functional 
theory (DFT) and the local density approximation (LDA) 
\cite{hohenberg,kohnsham}. As in our previous work on $ {\rm CrO_2} $ 
\cite{matar92,matar94}, $ {\rm MoO_2} $ \cite{moo2pap}, and $ {\rm NbO_2} $ 
\cite{nbo2pap} we employ the augmented spherical wave (ASW) method 
\cite{wkg} in its scalar-relativistic implementation (see Refs.\ 
\cite{diss,bookmat,revasw} for more recent descriptions). Since the 
ASW method uses the atomic sphere approximation (ASA) \cite{andersen75}, 
we had to insert so-called empty spheres into the open crystal structures 
of $ {\rm VO_2} $. These empty spheres are pseudo atoms without a nucleus, 
which are used to model the correct shape of the crystal potential in large 
voids. In order to minimize the sphere overlap we have recently developed 
the sphere geometry optimization (SGO) method, which solves the problem of 
finding optimal empty sphere positions as well as radii of all spheres 
automatically \cite{vpop}. The routine was applied to all three crystal 
structures under consideration. For the rutile structure, addition of 16 
empty spheres allowed to keep the linear overlap of any pair of physical 
spheres below 15\% and the overlap of any pair of physical and empty 
spheres below 20\%. The positions of the empty spheres are listed in Table 
\ref{tab:method1}. 
\begin{table}[ht]
\begin{center}
\caption{Empty sphere positions for the rutile structure.}
\label{tab:method1}   
\begin{tabular}{ccrrr} 
\\[-3mm] \hline \\[-3mm]
Atom        & Wyckoff positions & \multicolumn{3}{c}{parameters}    \\
\\[-3mm] \hline \\[-3mm]
            &                   &   $ x $   &   $ y $   &   $ z $   \\
\\[-3mm] \hline \\[-3mm]
$ \rm E_1 $ & (4c)              &           &           &           \\
$ \rm E_2 $ & (4g)              &   0.3238  &  -0.3238  &           \\
$ \rm E_3 $ & (16k)             &   0.0109  &   0.2973  &   0.2358  \\
\\[-3mm] \hline \\[-3mm]
\end{tabular} 
\end{center}
\end{table} 
In addition to the empty sphere positions the SGO algorithm proposed atomic 
sphere radii for all spheres, which are listed in Table \ref{tab:method2} 
\begin{table}[ht]
\begin{center}
\caption{Atomic sphere radii and basis set orbitals used for the rutile 
         structure.}
\label{tab:method2}   
\begin{tabular}{ccccccc} 
\\[-3mm] \hline \\[-3mm]
Atom         & Radius/$ a_B $  & \multicolumn{4}{c}{Orbitals}     \\
\\[-3mm] \hline \\[-3mm]
$ \rm V   $  & 2.230  &  $ 4s $  &  $ 4p $  &  $ 3d $  & ($ 4f $) \\
$ \rm O   $  & 1.831  &  $ 2s $  &  $ 2p $  & ($ 3d $) &          \\
$ \rm E_1 $  & 1.572  &  $ 1s $  & ($ 2p $) &          &          \\
$ \rm E_2 $  & 1.692  &  $ 1s $  &  $ 2p $  & ($ 3d $) &          \\
$ \rm E_3 $  & 0.949  &  $ 1s $  & ($ 2p $) &          &          \\
\\[-3mm] \hline \\[-3mm]
\end{tabular} 
\end{center}
\end{table} 
together with the orbitals used as the basis set for the present 
calculations. States given in parentheses were included as tails of 
the other orbitals (see Refs.\ \cite{wkg,diss,revasw} for more 
details on the ASW method). 

Applying the SGO algorithm to the monoclinic $ {\rm M_1} $ structure we 
were able, by inserting 48 empty spheres, to keep the linear overlap of 
any pair of physical spheres below 16\%, and the overlap of any pair of 
physical and empty spheres below 21\%. The positions of the empty spheres, 
the atomic sphere radii for all spheres as proposed by the SGO algorithm 
as well as the basis set orbitals are listed in Tables \ref{tab:method3} 
\begin{table}[htp]
\begin{center}
\caption{Empty sphere positions for the $ {\rm M_1} $ structure.}
\label{tab:method3}   
\begin{tabular}{ccrrr} 
\\[-3mm] \hline \\[-3mm]
Atom        & Wyckoff positions & \multicolumn{3}{c}{parameters}     \\
\\[-3mm] \hline \\[-3mm]
            &                   &   $ x $   &   $ y $   &   $ z $    \\
\\[-3mm] \hline \\[-3mm]
$ \rm E_1    $ & (2d)              &   0.5     &   0.5     &   0.0      \\
$ \rm E_2    $ & (4e)              &   0.1346  &  -0.0037  &   0.4998   \\
$ \rm E_3    $ & (4e)              &   0.4026  &   0.1483  &  -0.2102   \\
$ \rm E_4    $ & (4e)              &   0.0833  &  -0.2432  &   0.1602   \\
$ \rm E_5    $ & (4e)              &   0.3274  &   0.2118  &   0.4983   \\
$ \rm E_6    $ & (4e)              &   0.2726  &  -0.1788  &   0.4289   \\
$ \rm E_7    $ & (4e)              &   0.0141  &  -0.0326  &   0.2357   \\
$ \rm E_8    $ & (4e)              &   0.2317  &  -0.0336  &   0.2812   \\
$ \rm E_9    $ & (4e)              &   0.1289  &  -0.0016  &  -0.2788   \\
$ \rm E_{10} $ & (2b)              &   0.5     &   0.0     &   0.0      \\
$ \rm E_{11} $ & (4e)              &   0.4375  &   0.1054  &   0.2706   \\
$ \rm E_{12} $ & (4e)              &   0.1704  &   0.2552  &   0.0292   \\
$ \rm E_{13} $ & (4e)              &   0.3768  &  -0.0322  &  -0.3779   \\
\\[-3mm] \hline \\[-3mm]
\end{tabular} 
\end{center}
\end{table} 
and \ref{tab:method4}.  
\begin{table}[htp]
\begin{center}
\caption{Atomic sphere radii and basis set orbitals for the 
         $ {\rm M_1} $ structure.}
\label{tab:method4}   
\begin{tabular}{cccccc} 
\\[-3mm] \hline \\[-3mm]
Atom         & Radius/$ a_B $  & \multicolumn{4}{c}{Orbitals}     \\
\\[-3mm] \hline \\[-3mm]
$ \rm V      $  & 2.042  &  $ 4s $  &  $ 4p $  &  $ 3d $  & ($ 4f $) \\
$ \rm O_1    $  & 1.841  &  $ 2s $  &  $ 2p $  & ($ 3d $) &          \\
$ \rm O_2    $  & 1.739  &  $ 2s $  &  $ 2p $  & ($ 3d $) &          \\
$ \rm E_1    $  & 1.832  &  $ 1s $  &  $ 2p $  & ($ 3d $) &          \\
$ \rm E_2    $  & 1.753  &  $ 1s $  &  $ 2p $  & ($ 3d $) &          \\
$ \rm E_3    $  & 1.772  &  $ 1s $  &  $ 2p $  & ($ 3d $) &          \\
$ \rm E_4    $  & 1.734  &  $ 1s $  &  $ 2p $  & ($ 3d $) &          \\
$ \rm E_5    $  & 1.236  &  $ 1s $  & ($ 2p $) &          &          \\
$ \rm E_6    $  & 1.213  &  $ 1s $  & ($ 2p $) &          &          \\
$ \rm E_7    $  & 0.978  &  $ 1s $  & ($ 2p $) &          &          \\
$ \rm E_8    $  & 0.974  &  $ 1s $  & ($ 2p $) &          &          \\
$ \rm E_9    $  & 0.992  &  $ 1s $  & ($ 2p $) &          &          \\
$ \rm E_{10} $  & 0.841  &  $ 1s $  & ($ 2p $) &          &          \\
$ \rm E_{11} $  & 0.794  &  $ 1s $  & ($ 2p $) &          &          \\
$ \rm E_{12} $  & 0.791  &  $ 1s $  & ($ 2p $) &          &          \\
$ \rm E_{13} $  & 0.841  &  $ 1s $  & ($ 2p $) &          &          \\
\\[-3mm] \hline \\[-3mm]
\end{tabular} 
\end{center}
\end{table} 

Finally, for the monoclinic $ {\rm M_2} $ structure, we were able, by 
inserting 29 empty spheres, to keep the linear overlap of any pair of 
physical spheres below 17\%, and the overlap of any pair of physical 
and empty spheres below 22\%. The positions of the empty spheres are 
listed in Table \ref{tab:method5} 
\begin{table}[ht]
\begin{center}
\caption{Empty sphere positions for the $ {\rm M_2} $ structure.}
\label{tab:method5}   
\begin{tabular}{ccrrr} 
\\[-3mm] \hline \\[-3mm]
Atom        & Wyckoff positions & \multicolumn{3}{c}{parameters}     \\
\\[-3mm] \hline \\[-3mm]
            &                   &   $ x $   &   $ y $   &   $ z $    \\
\\[-3mm] \hline \\[-3mm]
$ {\rm E_1}    $ & (8j)             &   0.2491  &   0.1251  &  -0.0019   \\
$ {\rm E_2}    $ & (2c)             &   0.0     &   0.0     &   0.5      \\
$ {\rm E_3}    $ & (4h)             &   0.0     &   0.3149  &   0.5      \\
$ {\rm E_4}    $ & (4i)             &   0.0765  &   0.0     &   0.1518   \\
$ {\rm E_5}    $ & (8j)             &   0.1597  &   0.2525  &  -0.3270   \\
$ {\rm E_6}    $ & (4i)             &   0.3736  &   0.0     &  -0.1461   \\
$ {\rm E_7}    $ & (8j)             &   0.0075  &   0.1525  &  -0.2867   \\
$ {\rm E_8}    $ & (4i)             &   0.3932  &   0.0     &  -0.4322   \\ 
$ {\rm E_9}    $ & (4i)             &   0.2241  &   0.0     &   0.2109   \\ 
$ {\rm E_{10}} $ & (4i)             &   0.4763  &   0.0     &  -0.3109   \\ 
$ {\rm E_{11}} $ & (8j)             &   0.1604  &   0.2567  &  -0.0648   \\ 
\\[-3mm] \hline \\[-3mm]
\end{tabular} 
\end{center}
\end{table} 
and the corresponding atomic sphere radii as well as the basis set orbitals 
are given in Table \ref{tab:method6}. 
\begin{table}[ht]
\begin{center}
\caption{Atomic sphere radii and basis set orbitals for the 
         $ {\rm M_2} $ structure.}
\label{tab:method6}   
\begin{tabular}{ccccccc} 
\\[-3mm] \hline \\[-3mm]
Atom         & Radius/$ a_B $  & \multicolumn{5}{c}{Orbitals}     \\
\\[-3mm] \hline \\[-3mm]
$ {\rm V_1}    $  & 2.213  &  $ 4s $  &  $ 4p $  &  $ 3d $  & ($ 4f $) &  \\
$ {\rm V_2}    $  & 2.085  &  $ 4s $  &  $ 4p $  &  $ 3d $  & ($ 4f $) &  \\
$ {\rm O_1}    $  & 1.848  &  $ 2s $  &  $ 2p $  & ($ 3d $) &          &  \\
$ {\rm O_2}    $  & 1.817  &  $ 2s $  &  $ 2p $  & ($ 3d $) &          &  \\
$ {\rm O_3}    $  & 1.705  &  $ 2s $  &  $ 2p $  & ($ 3d $) &          &  \\
$ {\rm E_1}    $  & 1.642  &  $ 1s $  &  $ 2p $  & ($ 3d $) &          &  \\
$ {\rm E_2}    $  & 1.892  &  $ 1s $  &  $ 2p $  & ($ 3d $) &          &  \\
$ {\rm E_3}    $  & 1.978  &  $ 1s $  &  $ 2p $  & ($ 3d $) &          &  \\
$ {\rm E_4}    $  & 1.929  &  $ 1s $  &  $ 2p $  &  $ 3d $  & ($ 4f $) &  \\
$ {\rm E_5}    $  & 1.795  &  $ 1s $  &  $ 2p $  & ($ 3d $) &          &  \\
$ {\rm E_6}    $  & 1.781  &  $ 1s $  &  $ 2p $  & ($ 3d $) &          &  \\
$ {\rm E_7}    $  & 1.104  &  $ 1s $  & ($ 2p $) &          &          &  \\
$ {\rm E_8}    $  & 1.195  &  $ 1s $  & ($ 2p $) &          &          &  \\
$ {\rm E_9}    $  & 1.143  &  $ 1s $  & ($ 2p $) &          &          &  \\
$ {\rm E_{10}} $  & 0.892  &  $ 1s $  & ($ 2p $) &          &          &  \\
$ {\rm E_{11}} $  & 0.868  &  $ 1s $  & ($ 2p $) &          &          &  \\
\\[-3mm] \hline \\[-3mm]
\end{tabular} 
\end{center}
\end{table} 

Self-consistency was achieved by an efficient algorithm for convergence
acceleration \cite{mixpap}. The Brillouin zone sampling was done using an
increased number of $ {\bf k} $-points ranging from 18 to 1800 points, 
54 to 6750 points, and 63 to 1088 points, respectively, within the 
irreducible wedges of the three Brillouin zones. This way we were able 
to ensure convergence of our results with respect to the fineness of the 
$ {\bf k} $ space grid.

In addition to analyzing the band structure and the (partial) densities 
of states we will also address chemical bonding in terms of the crystal 
orbital overlap population (COOP) as based on the notions introduced by 
Hoffmann \cite{hoffmann88} as well as related concepts. Among the latter 
are the crystal orbital Hamiltonian population (COHP) as proposed by 
Dronskowski and Bl\"ochl \cite{dronskowski93} and the covalence energy 
by B\"ornsen {\em et al.}\ \cite{boernsen99,boernsen00}. All these 
quantities have been implemented in the ASW method \cite{coop} (see also 
Refs.\ \cite{bookdft,csmp}) and have been validated for a large number 
of compounds \cite{bookdft}.

\section{Results for metallic $ {\rm \bf VO_2}$}
\label{resrut}

\subsection{Molecular orbital picture}
\label{mopict}

\begin{figure}[htp]
\begin{center}
\unitlength0.6mm
\begin{picture}(150,90)
\thicklines
\put( 40.0,  0.0){\textcolor{blue}{\framebox(20.0,20.0){ }}}
\put( 40.0, 20.0){\textcolor{blue}{\framebox(20.0,15.0){ }}}
\put( 40.0, 50.0){\textcolor{red}{\framebox(20.0,10.0){ }}}
\put( 40.0, 65.0){\textcolor{green}{\framebox(20.0,20.0){ }}}
\dashline{2}( 35.0, 53.0)(65.0, 53.0)
\put(  0.0,  0.0){\makebox(30.0,35.0){\large O $ 2p $}}
\put(  0.0, 50.0){\makebox(30.0,10.0){\large V $ 3d $ $ t_{2g} $}}
\put(  0.0, 65.0){\makebox(30.0,20.0){\large V $ 3d $ $ e_g^{\textcolor{green}{\sigma}} $}}
\put( 65.0, 45.0){\makebox(16.0,16.0){\large $ {\rm E_F} $}}
\put( 85.0,  0.0){\makebox(10.0,20.0){\textcolor{blue}{\large $ \sigma $}}}
\put( 85.0, 20.0){\makebox(10.0,15.0){\textcolor{blue}{\large $ \pi $}}}
\put( 85.0, 50.0){\makebox(10.0,10.0){\textcolor{red}{\large $ \pi^{\ast} $}}}
\put( 85.0, 65.0){\makebox(10.0,20.0){\textcolor{green}{\large $ \sigma^{\ast} $}}}
\put( 95.0, 55.0){\line( 1, 2){5.5}}
\put( 95.0, 55.0){\line( 1,-2){5}}
\put(100.0, 45.0){\line( 1, 0){5}}
\put( 99.5, 64.0){\line( 1, 0){5.5}}
\put(100.5, 66.0){\line( 1, 0){4.5}}
\put(105.0, 40.0){\makebox(45.0,10.0){\textcolor{magenta}{\large
                  $ b_{1g} = "d_{\parallel}" $}}}
\put(105.0, 60.0){\makebox(45.0,10.0){\textcolor{red}{\large
                  $ e_g^{\pi} = "\pi^{\ast}" $}}}
\end{picture} 
\unitlength1pt
\caption{Molecular orbital scheme for rutile $ \rm VO_2 $.}
\label{fig:mopict1}   
\end{center}
\end{figure}
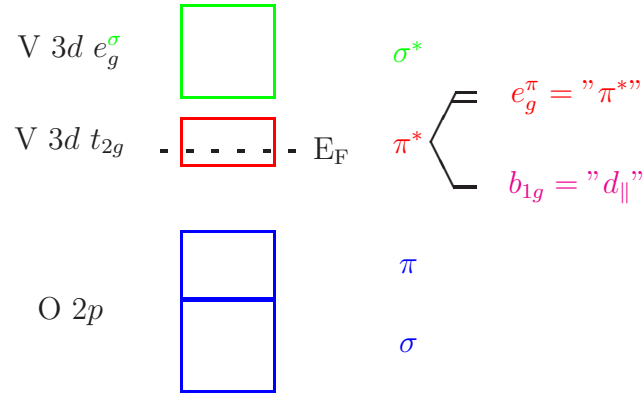

Our expectations on the electronic structure of $ {\rm VO_2} $ are easily 
stated within the molecular orbital picture proposed by Goodenough 
(see Sec.\ \ref{theory}) \cite{goodenough60,goodenough71a,goodenough71b}. 
The general situation is visualized in Fig.\ \ref{fig:mopict1}. 
First, hybridization between the oxygen $ 2p $ and vanadium $ 3d $ 
orbitals will lead to $ \sigma $- and $ \pi $-type overlap. This 
gives rise to states of $ \sigma $ and $ \sigma^{\ast} $ as well 
as $ \pi $ and $ \pi^{\ast} $ character. Since the $ p $--$ d $ 
overlap is stronger for the former these states experience the 
larger bonding-antibonding splitting. While the $ \sigma $ and 
$ \pi $ states will be filled and primarily of O $ 2p $ character, 
the corresponding antibonding bands will be dominated by the V $ 3d $ 
orbitals. Due to the electron count, the latter will be found at and above 
the Fermi energy. In the nearly perfect cubic octahedral surrounding of 
the vanadium atoms, the $ \sigma^{\ast} $ and $ \pi^{\ast} $ states 
are of $ e_g^{\sigma} $ and $ t_{2g} $ symmetry, respectively. We 
identify these orbitals in Fig.\ \ref{fig:cryst2} as the 
$ d_{3z^2-r^2} $/$ d_{xy} $ and the $ d_{x^2-y^2} $/$ d_{xz} $/$ d_{yz} $ 
orbitals, respectively. Note that the albeit small orthorhombic 
distortions of the oxygen octahedra lift all the degeneracies among 
the $ d $ orbitals. As already pointed out by Goodenough, the exact 
position and width of the $ d $ bands is subject not only to the 
$ p $--$ d $ hybridization but also strongly influenced by direct 
metal-metal interactions \cite{goodenough60,goodenough71a}. In 
particular, in the rutile structure such interactions involve the 
$ d_{x^2-y^2} $ orbitals, which experience strong overlap parallel 
to the rutile $ c $ axis. These orbitals are of $ b_{1g} $ symmetry 
but are usually designated as the $ d_{\parallel} $ bands. The 
remaining $ t_{2g} $ orbitals are of $ e_g^{\pi} $ symmetry and usually 
subsummed under the name $ \pi^{\ast} $ (not to be mixed with the same 
notation used above for the $ t_{2g} $ states as a whole). While the 
$ d_{xz} $ orbitals experience hardly any overlap with orbitals of 
neighbouring metal atoms the $ d_{yz} $ states mediate metal-metal 
bonding parallel to the in-plane axes of the tetragonal cell. The 
general situation was already sketched in Fig.\ \ref{fig:intro2}. 

The validity of the just outlined general band scheme for the rutile-type 
transition-metal dioxides has been already confirmed in our previous 
calculations for hypothetical rutile $ {\rm MoO_2} $ as well as metallic 
$ {\rm NbO_2} $ \cite{moo2pap,nbo2pap}. Furthermore, from the close 
relationship of the crystal structures of all three compounds we expect a 
very similar electronic structure also for $ {\rm VO_2} $. Yet, as compared 
to the $ 4d $ systems the vanadium $ d $ bands will display smaller 
bandwidths.

\subsection{Band structure and density of states}
\label{resrut1}
  
\begin{figure}[htp]
\centering
\includegraphics[width=0.8\textwidth]{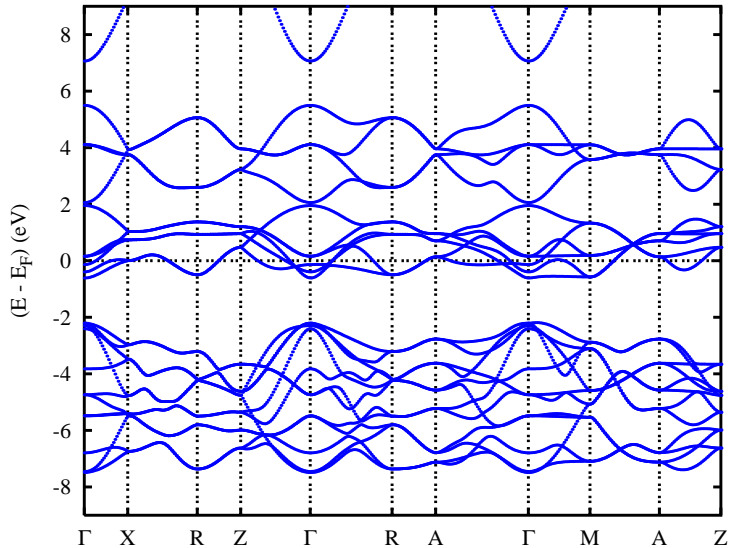}
\caption[Electronic bands of rutile $ \rm VO_2 $.]
        {Electronic bands of rutile $ \rm VO_2 $ along selected symmetry 
         lines within the first Brillouin zone of the simple tetragonal 
         lattice, Fig.\ \protect\ref{fig:bzones}(a).}
\label{fig:resrut1}   
\end{figure}

\begin{figure}[htp]
\centering
\subfigure[]{\includegraphics[width=0.3\textwidth]{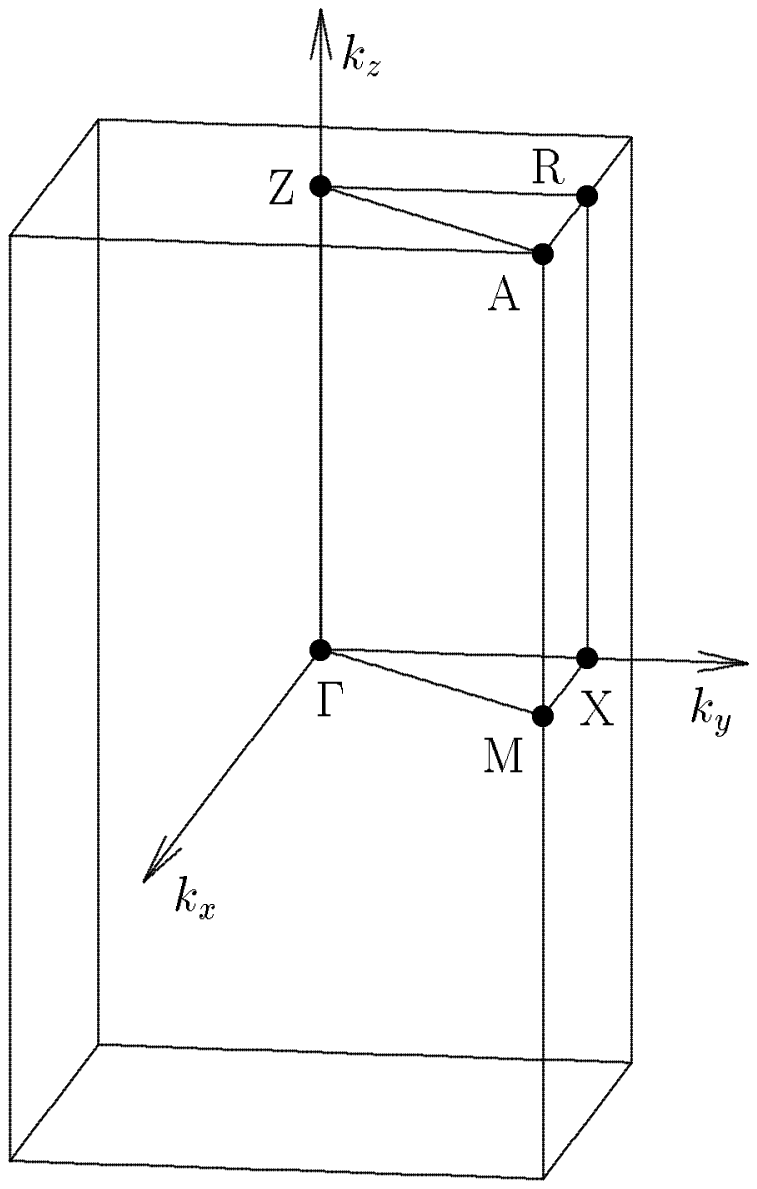}}
\hspace{0.1\textwidth}
\subfigure[]{\includegraphics[width=0.3\textwidth]{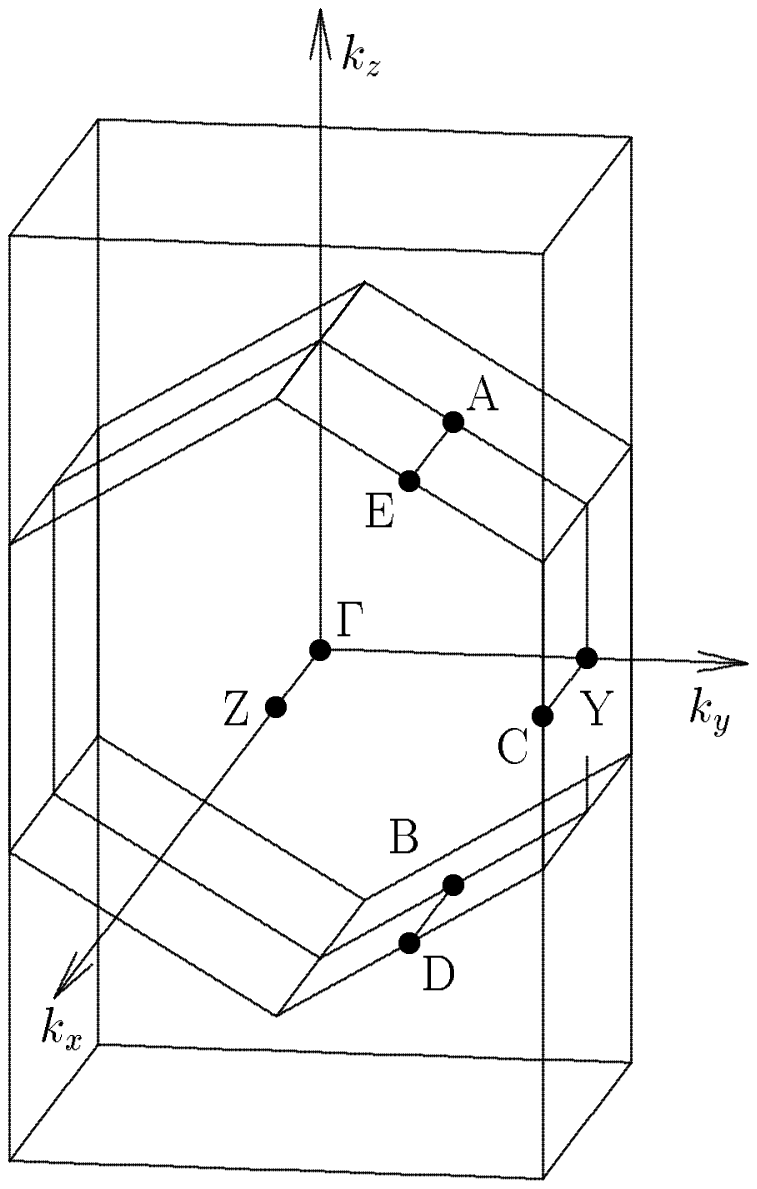}} 
\\ 
\caption[First Brillouin zones of the simple tetragonal and simple monoclinic
         lattices.]
        {First Brillouin zones of the (a) simple tetragonal and (b) simple
         monoclinic lattices. Backfolding of the tetragonal Brillouin zone
         implies the following transformation of high symmetry points:
         $ {\rm X_T \to Y_M, Z_M} $; $ {\rm M_T \to C_M} $;
         $ {\rm Z_T \to Y_M} $; $ {\rm R_T \to \Gamma} $;
         $ {\rm A_T \to Z_M} $.}
\label{fig:bzones}
\end{figure}

\begin{figure}[htp]
\centering
\includegraphics[width=0.8\textwidth]{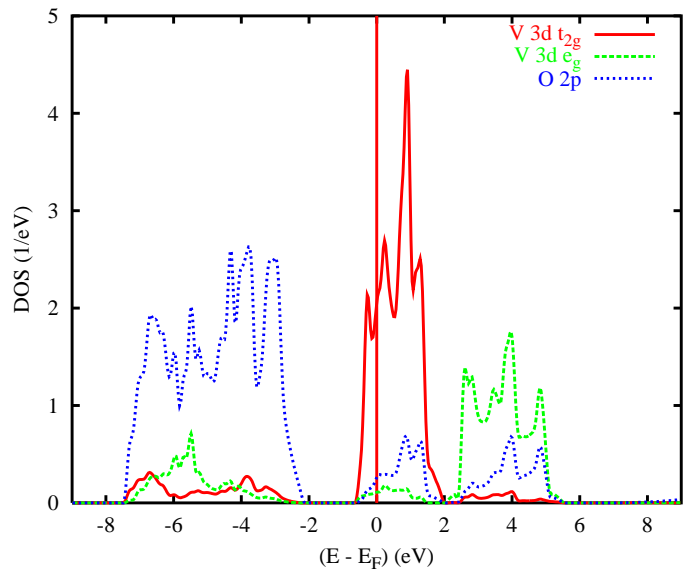}
\caption[Partial densities of states (DOS) of rutile $ {\rm VO_2} $.]
        {Partial densities of states (DOS) of rutile $ {\rm VO_2} $ 
	 per formula unit.} 
\label{fig:resrut2}   
\end{figure}

We display in Fig.\ \ref{fig:resrut1} 
the electronic states along selected high symmetry lines within the first 
Brillouin zone of the simple tetragonal lattice, Fig.\ \ref{fig:bzones}(a).
The corresponding partial densities of states (DOS) are given in Fig.\ 
\ref{fig:resrut2}.  
The total density of states at the Fermi energy, $ {\rm N ( E_F )} $, 
is 2.63 states/f.u./eV. Not shown are low lying oxygen $ 2s $ states. 

In Figs.\ \ref{fig:resrut1} and \ref{fig:resrut2} we identify four groups 
of bands. In the energy range from -7.6 to -2.2\,eV we observe 12 bands, 
which trace back mainly to O $ 2p $ states but have a non-negligible 
contribution due to the V $ 3d $ states. Bands are most easily counted 
along the direction X-R where they are twofold degenerate. The next two 
groups, which extend from -0.6 to 2.0\,eV and from 2.0 to 5.5\,eV, contain 
six and four bands, respectively. They originate mainly from V $ 3d $ 
states. Yet, $ p $--$ d $ hybridization causes additional O $ 2p $ 
contributions in this energy range. From the above molecular-orbital 
point of view we interprete the bands between -7.6 and -2.2\,eV as the 
O $ 2p $-dominated lower $ \sigma $ and higher $ \pi $ states. The 
respective $ \pi^{\ast} $ and $ \sigma^{\ast} $ states are found in the 
energy intervals from -0.6 to 2.0\,eV and 2.0 to 5.5\,eV. Finally, we 
observe V $ 4s $ states starting at 7.0\,eV. 

\begin{figure}[htp]
\centering
\includegraphics[width=0.8\textwidth]{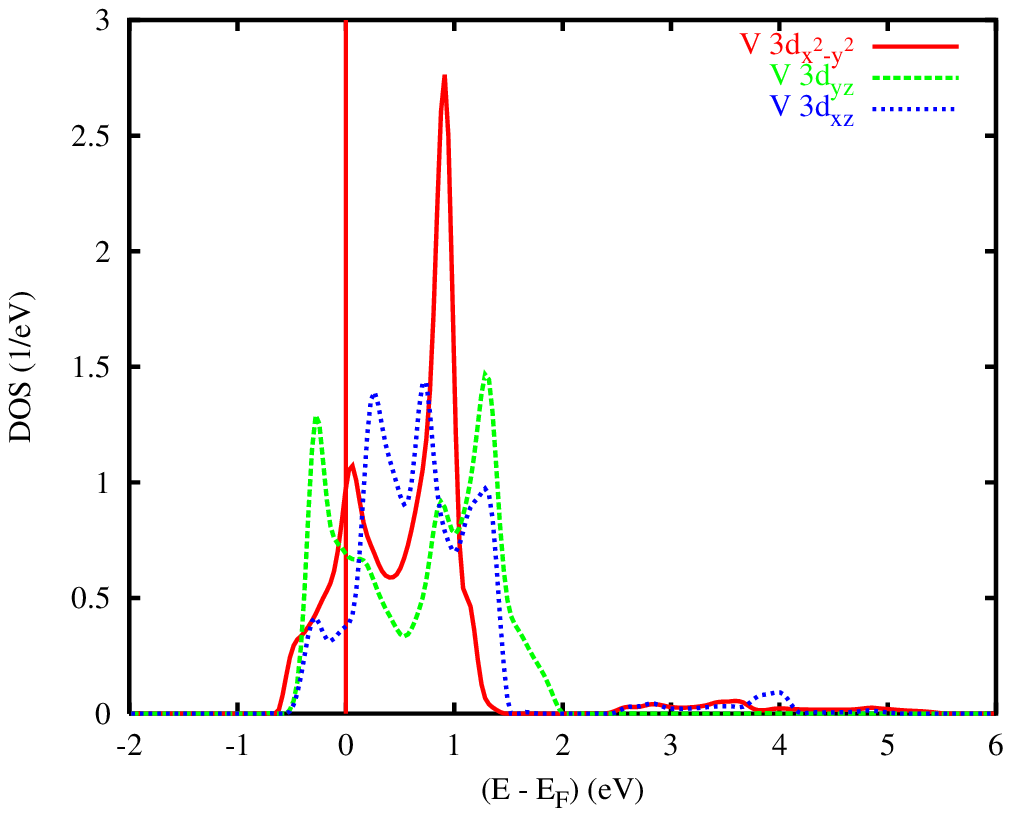}
\includegraphics[width=0.8\textwidth]{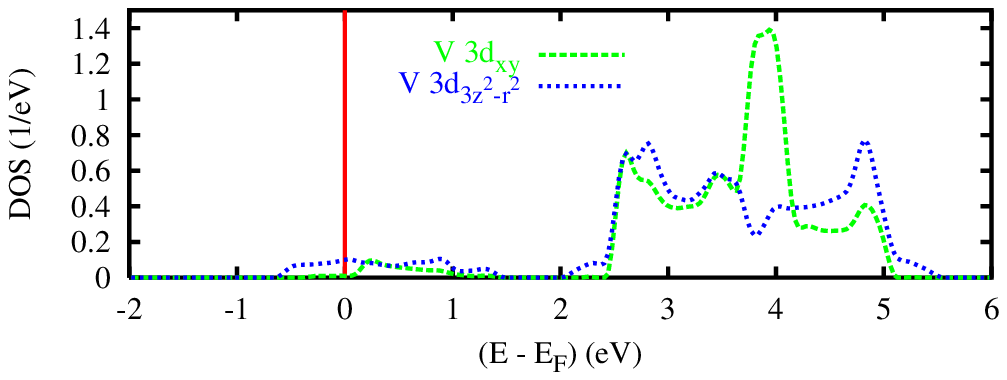}
\caption[Partial V $ 3d $ $ t_{2g} $ and $ e_g $ densities of states (DOS) 
         of rutile $ {\rm VO_2} $.]
        {Partial V $ 3d $ $ t_{2g} $ and $ e_g $ densities of states (DOS) 
         of rutile $ {\rm VO_2} $. Selection of orbitals is relative to the 
         local rotated reference frame.} 
\label{fig:resrut3}   
\end{figure}

Crystal field splitting expected from the fact that the metal atoms are 
located at the centers of slightly distorted $ {\rm VO_6} $ octahedra is 
observed in the partial V $ 3d $ DOS shown in Fig.\ \ref{fig:resrut3}. 
There we have included only the single vanadium atom at the corner of the 
rutile cell and used the local rotated reference frame introduced in Sec.\ 
\ref{crystrut}.  Fig.\ \ref{fig:resrut3} clearly reveals the almost perfect 
energetical separation of the $ 3d $ $ t_{2g} $ and $ e_g $ groups of bands. 
The former states appear almost exclusively in the energy range from -0.6 to 
2.0\,eV. The $ e_g $ states dominate the bands between 2.0 and 5.5\,eV. 
The small but finite $ t_{2g} $--$ e_g $ configuration mixing is a 
measure of octahedral distortions. Contributions of the V $ 3d $ states to 
the oxygen bands are slightly larger for the $ e_g $ states which, forming 
$ \sigma $ bonds, experience a larger overlap with the O $ 2p $ states. For 
the same reason, the bonding-antibonding splitting is larger for the 
$ e_g $ states as compared to the $ t_{2g} $ states, which give rise to 
$ \pi $ bonds. 

In Fig.\ \ref{fig:resrut3} the differences between the single symmetry 
components of the $ t_{2g} $ and $ e_g $ orbitals, respectively, reflect 
the orthorhombic site symmetry. Worth mentioning is the pronounced double 
peak structure of the $ d_{yz} $ partial DOS. Since the respective 
orbitals point along the two in-plane $ a $ axes of the tetragonal cell 
these double maxima result from metal-metal bonding. Finally, the 
$ d_{x^2-y^2} $ partial DOS displays an albeit smaller separation into 
two peaks, which is due to the overlap of these  orbitals parallel to 
the rutile $ c $ axis. 

Our findings are in perfect accordance with the molecular orbital 
picture proposed by Goodenough \cite{goodenough71a,goodenough71b}, 
which was sketched in Sec.\ \ref{mopict}. There the $ d_{x^2-y^2} $ 
states were designated as the $ d_{\parallel} $ states and the 
notation $ \pi^{\ast} $ was used for the $ d_{xz} $ and $ d_{yz} $ 
states. Deviations from the band diagram of Goodenough show up with 
respect to the widths of the different bands. Goodenough and many 
authors after him assumed the $ d_{\parallel} $ band to be 
considerably more narrow than the $ \pi^{\ast} $ bands. Yet, our 
calculation reveals similar widths of all three $ t_{2g} $ bands. In 
particular the $ d_{\parallel} $ band has a width of $ \approx 1.5 $\,eV. 
Of course, this has strong implications for model approaches building 
on strong correlations within this band. 

While the results for rutile $ {\rm VO_2} $ agree very well with those 
for hypothetical rutile $ {\rm MoO_2} $ \cite{moo2pap} differences 
appear in a more detailed analysis of the single symmetry components 
of the $ t_{2g} $ orbitals as given in Fig.\ \ref{fig:resrut3}. In 
$ {\rm MoO_2} $, the Mo $ 4d_{x^2-y^2} $ band shows a strong 
tendency towards bonding-antibonding splitting while the remaining $ t_{2g} $ 
bands give rise to very similar densities of states. In contrast, in 
$ {\rm VO_2} $ bonding-antibonding splitting of the V $ 3d_{x^2-y^2} $ band 
is much weaker. At the same time, the $ 3d_{xz} $ and $ 3d_{yz} $ partial 
DOS show distinct deviations. In particular, the latter reveals some 
splitting into two peaks at about -0.3 and 1.3\,eV. These features may be 
traced back to the much larger $ c $ axis of $ {\rm VO_2} $ (2.8514\,{\AA} 
vs.\ 2.805\,{\AA} in $ {\rm MoO_2} $), which leads to a larger separation of 
the metal atoms along this axis. In contrast, the rutile $ a $ axis is reduced 
(4.5546\,{\AA} vs.\ 4.856\,{\AA} in $ {\rm MoO_2} $). This causes a larger 
overlap of metal $ d $ orbitals within the planes and, hence, a visible 
bonding-antibonding splitting of the local $ d_{yz} $ orbitals. 

Whereas our calculations are in good  agreement with the calculations 
by Caruthers {\em et al.}\ as well as by Gupta {\em et al.}\ 
\cite{caruthers73a,gupta77}, a closer look reveals small but distinct 
deviations concerning the exact band positions and, consequently, the 
Fermi surface. We attribute these differences to the approximations inherent 
in the old calculations. Good agreement, however, is found in a comparison with 
the existing state-of-the-art calculations by Wentzcovitch {\em et al.}\ 
\cite{wentz94a} as well as Kurmaev {\em et al.}\ \cite{kurmaev98}. 
Minor differences in the band structures may result from the 
use of the crystal structure data of Ref.\ \cite{westman61} or from 
Brillouin zone samplings with different numbers of $ {\bf k} $-points. Finally, 
we observe distinct deviations in the recent calculations by Nikolaev 
{\em et al.}\ with respect to the band positions \cite{nikolaev92}. 
While our calculations give an occupied V $ 3d $ bandwidth of 0.6\,eV,  
in close agreement with Wentzcovitch {\em et al.}, Nikolaev {\em et al.}\ 
report a value of 0.77\,eV. At the same time, the latter authors locate the 
upper edge of the O $ 2p $ bands 1.39\,eV below the Fermi level. In our 
calculation the band maximum is found at -2.4\,eV in almost perfect 
agreement with both the optical measurements by Verleur {\em et al.}\ and 
the photoemission data by Powell {\em et al.}, Shin {\em et al.}, Bermudez 
{\em et al.}\ as well as Goering {\em et al.}. These authors find the 
O $ 2p $ states 2.4-2.5\,eV below the Fermi energy 
\cite{verleur68,powell69,shin90,bermudez92,goering96,goering97b}. The total 
width of the O $ 2p $ band of 5.37\,eV reported by Nikolaev {\em et al.} 
is again in agreement with our value of $ \approx 5.6 $\,eV. 
Finally, we point to the almost equal densities of states at the Fermi 
energy of all three $ t_{2g} $ orbitals, which confirm the reported 
isotropic conductivity \cite{bongers65,kosuge67}.

\subsection{Chemical bonding}
\label{resrut2}

\begin{figure}[htp]
\centering
\includegraphics[width=0.8\textwidth]{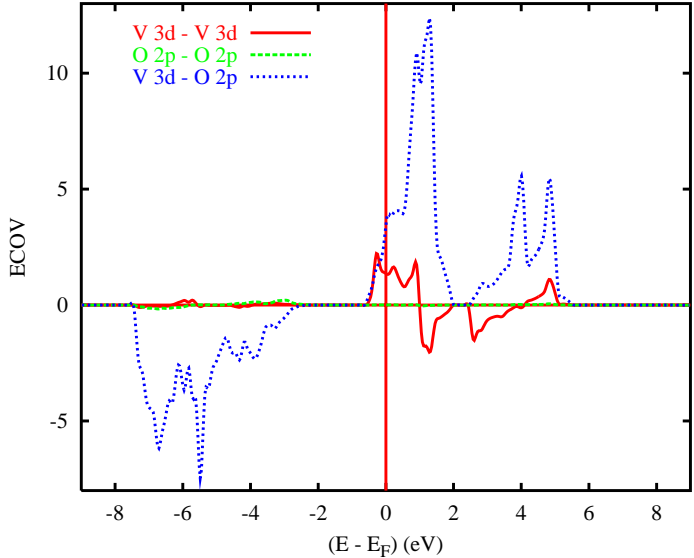}
\caption{Partial covalence energies ($ {\rm E_{cov}} $) of rutile 
         $ {\rm VO_2} $.}
\label{fig:resrut4}
\end{figure}

Chemical bonding is addressed via the covalence energy shown in Fig.\ 
\ref{fig:resrut4}.
The curves display the ``canonical'' behaviour being negative (bonding) 
and positive (antibonding) in the low and high energy regions. While 
the oxygen-oxygen overlap is almost negligible we observe considerable 
metal-metal bonding in the V $ 3d $ dominated groups of bands. Obviously,  
the dominating contribution to the total bonding results from the 
V $ 3d $--O $ 2p $ bonding. Below $ {\rm E_F} $, the corresponding 
$ E_{cov} $ curve is negative below -2.2\,eV and positive only in the 
small energy range of the occupied V $ 3d $ derived bands.

\subsection{Comparison to experiment}
\label{resrut3}

\begin{figure}[htp]
\centering
\includegraphics[width=0.8\textwidth]{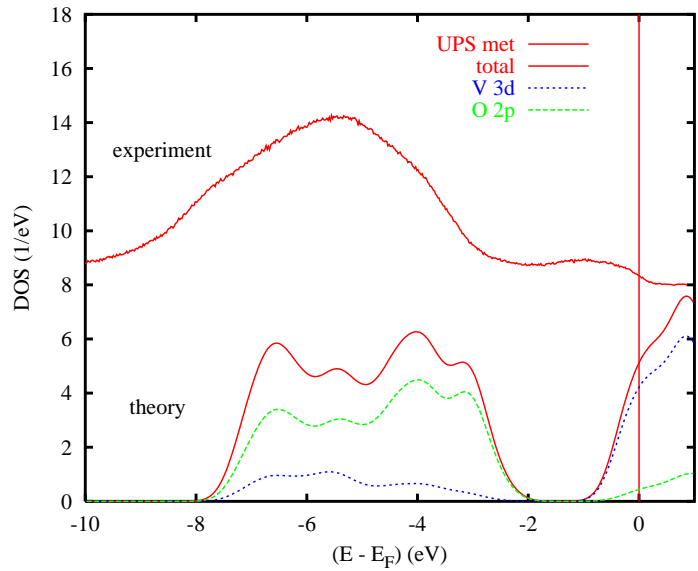}
\caption[Total and partial densities of states (DOS) of rutile $ {\rm VO_2} $ 
         folded with a 0.25\,eV wide Gaussian and UPS spectra.]
        {Total and partial densities of states (DOS) of rutile $ {\rm VO_2} $ 
         folded with a 0.25\,eV wide Gaussian (lower set of curves) and 
         UPS spectra (upper curve; note the offset introduced in order 
         to distinguish experimental and theoretical results; from Ref.\ 
         \protect \cite{goering96,goering97b}).} 
\label{fig:resrut5}   
\end{figure}

\begin{figure}[htp]
\centering
\includegraphics[width=0.8\textwidth]{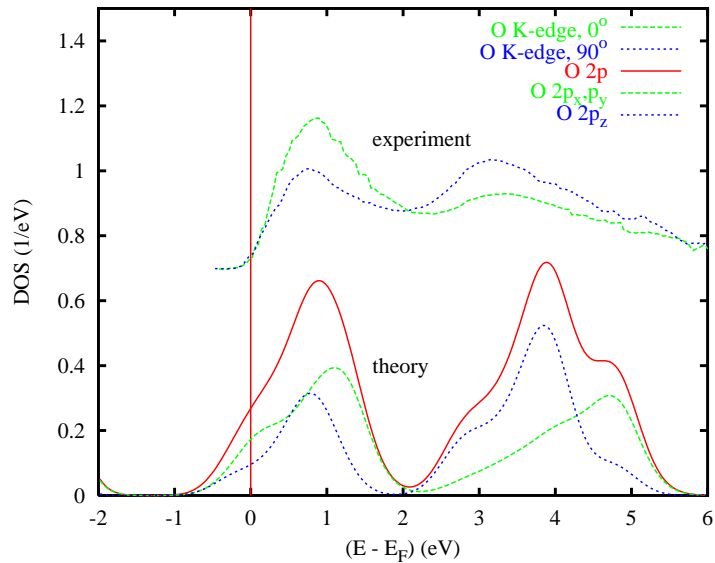}
\caption[Partial O $ 2p $ densities of states (DOS) of rutile $ {\rm VO_2} $ 
         folded with a 0.5\,eV wide Gaussian and XAS O K edge spectra.]
        {Partial O $ 2p $ densities of states (DOS) of rutile $ {\rm VO_2} $ 
         folded with a 0.5\,eV wide Gaussian (lower set of curves) and 
         XAS O K edge spectra (upper set of curves; note the offset 
         introduced in order to distinguish experimental and theoretical 
         results; from Ref.\ \protect \cite{mueller96,mueller97}; data 
         shifted by 529.5\,eV).}
\label{fig:resrut6}   
\end{figure}

We display in Figs.\ \ref{fig:resrut5}  
and \ref{fig:resrut6} 
total and partial V $ 3d $ and O $ 2p $ densities of states folded with 
a Gaussian of 0.25 and 0.5\,eV width, respectively, for the occupied and 
unoccupied part of the spectrum. In Fig.\ \ref{fig:resrut5} we have 
added UPS spectra as measured by Goering {\em et al.}\ 
\cite{goering96,goering97b}. Good agreement of the calculated and measured 
curves is found. In addition, the calculated total DOS compares very well 
with the XPS and UPS spectra by Blaauw {\em et al.}, Shin {\em et al.}\ and 
Bermudez {\em et al.}\ \cite{blaauw75,shin90,bermudez92}. The occupied 
bandwidth of $ \approx 8.5 $\,eV deduced from the experiments is close to the 
calculated 8.0\,eV. According to the XPS and UPS spectra the valence band is 
split into a low and high binding part of about 1.5 and 6\,eV width, 
respectively \cite{blaauw75,shin90,goering97b}. Whereas the low binding region 
results from a single peak centered at $ \approx -1.0 $\,eV, which is attributed 
to the V $ 3d $ states, the high binding region, resulting mainly from the 
O $ 2p $ states, is dominated by a large peak at $ \approx -5.5 $\,eV and shows 
two slight shoulders at -4.0 and -7.5\,eV \cite{goering96,goering97b}. In 
contrast, as compared to the UPS data, the XPS spectra display a considerably 
higher intensity at energies below -5.5\,eV and a strong decrease in intensity 
around the top of the O $ 2p $ band \cite{blaauw75,shin90}. Since the 
V $ 3d $ states have a larger cross section in XPS \cite{yeh85} this shift of 
intensity reveals a considerable V $ 3d $ contribution especially to the bands 
below -5.5\,eV and, hence, a strong $ p $--$ d $ hybridization in this energy 
region. This is well reflected by the calculations. 

In Fig.\ \ref{fig:resrut6} we have complemented the theoretical results by 
soft-X-ray absorption spectra as measured by M\"uller {\em et al.}\ 
\cite{mueller96,mueller97}. The experimental data are shifted by 529.5\,eV. 
In order to probe the angular dependence, in these experiments the polarization 
vector $ {\rm \bf E} $ was oriented either parallel ($ \phi = 90^{\circ} $) or 
else perpendicular ($ \phi = 0^{\circ} $) to the rutile $ c $ axis. In the 
former case, dipole selection rules allow for transitions from O $ 1s $ to the 
O $ 2p_z $ state. In contrast, for $ {\rm \bf E} $ perpendicular to the $ c $ 
axis transitions to the O $ 2p_x $ and $ 2p_y $ states may occur. The curves 
given by M\"uller conform with the spectra by Abbate {\em et al.}\ who found, 
for metallic $ {\rm VO_2} $, two peaks at 529.9 and 532.5\,eV photon energy for 
$ {\rm \bf E} \parallel {\bf c} $ \cite{abbate91}. In the spectra given by 
M\"uller {\em et al.}\ these peaks are located at 530.3 and 532.7\,eV, 
respectively \cite{mueller97}. However, as M\"uller has pointed out previously, 
the curves actually can be fitted to four peaks at 530.1, 530.7, 532.7, and 
534.7\,eV \cite{mueller96}. In the energy-shifted spectra of Fig.\ 
\ref{fig:resrut6} these peaks are located at 0.6, 1.2, 3.2, and 5.2\,eV.  

In Fig.\ \ref{fig:resrut6}, the partial DOS correspond to the final 
states, which are occupied in the XAS experiments. Note that the curve 
marked O $ 2p_x, p_y $ comprises only the mean average of these orbitals 
since the component perpendicular to $ {\rm \bf E} $ is not seen in 
experiment. For the same reason, the curve marked O $ 2p $ actually 
contains the full contribution from the $ p_z $ orbital but only half 
of the contributions from the $ p_x $ and $ p_y $ orbitals. As for the 
occupied part of the spectrum we observe overall good agreement between 
experiment and theory. This holds for the positions of the peaks, their 
relative intensity and their angular dependence. Still, agreement seems 
to be less satisfactory in the high energy region between 4 and 5\,eV. 
However, the discrepancy is resolved by the aforementioned four peak 
analysis of the experimental spectra. Finally, differences with respect 
to intensities may be attributed to matrix element and core-hole effects, 
which might be important but are not accounted for in our present 
analysis \cite{degroot89,abbate91}.

\subsection{Fermi surface}
\label{resrut5}

Despite the good agreement of our results with previous experimental and 
calculated data we are still seeking for more conclusive arguments for the 
destabilization of the rutile structure of $ {\rm VO_2} $ at low temperatures. 
Since the metal-insulator transition is accompanied by a structural 
transformation one might be tempted to deduce more information about the 
origin of the transition from investigating the Fermi surface. This 
has been already done by Gupta {\em et al.}\ in the course of their 
non-selfconsistent calculations \cite{gupta77}. Furthermore, investigation of 
the Fermi surface of hypothetical rutile $ {\rm MoO_2} $ in our previous 
work revealed strong nesting on flat portions at height 
$ q_z = \pm \frac{\pi}{2c} $. This indicated a possible splitting of the 
$ d_{x^2-y^2} $, which was indeed observed in monoclinic $ {\rm MoO_2} $ 
\cite{moo2pap}. 

\begin{figure}[htp]
\centering
\includegraphics[height=0.4\textheight]{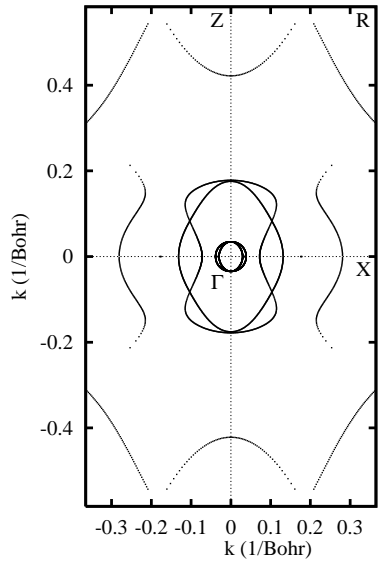}
\caption{$ \{100\} $ cut through the Fermi surface of rutile $ {\rm VO_2} $.} 
\label{fig:resrut7}   
\end{figure}

\begin{figure}[htp]
\centering
\includegraphics[height=0.4\textheight]{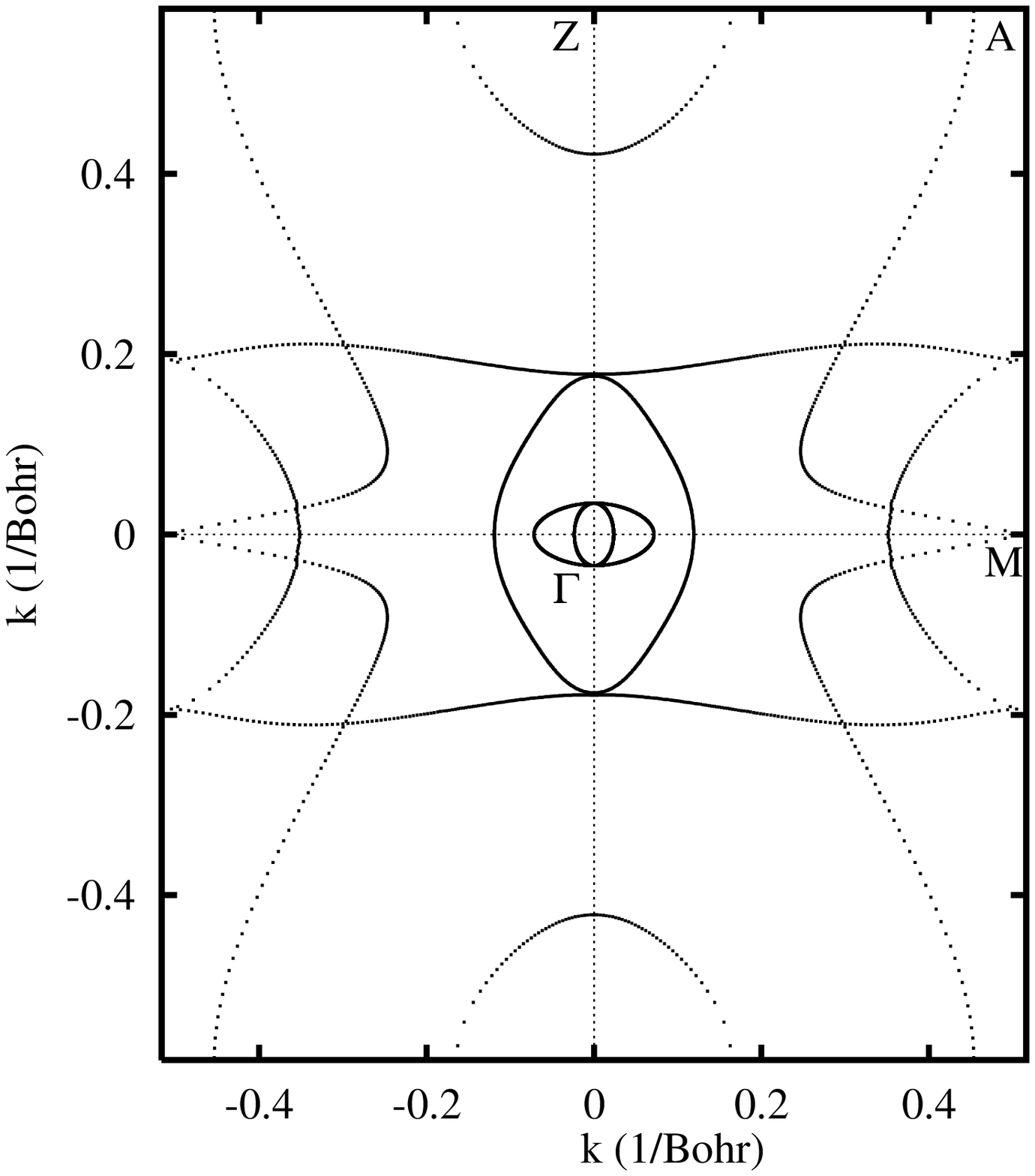}
\caption{$ \{110\} $ cut through the Fermi surface of rutile $ {\rm VO_2} $.} 
\label{fig:resrut8}   
\end{figure}

\begin{figure}[htp]
\centering
\includegraphics[height=0.4\textheight]{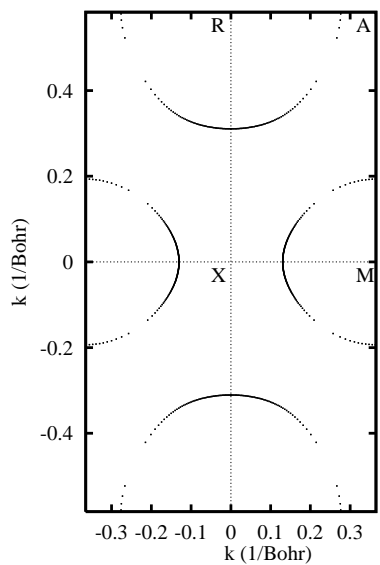}
\caption{$ \{100\} $ cut through the Fermi surface of rutile $ {\rm VO_2} $.} 
\label{fig:resrut9}   
\end{figure}

Cuts through the Fermi surface parallel to the $ \{100\} $ and $ \{110\} $ 
planes as resulting from the present calculations are shown in Figs.\ 
\ref{fig:resrut7},  
\ref{fig:resrut8},  
and \ref{fig:resrut9}.  
They contain the $ {\rm \Gamma} $-X-Z-R, $ {\rm \Gamma} $-M-Z-A, and X-M-R-A 
planes, which are the vertical borders of the irreducible wedge of the first 
Brillouin zone, Fig.\ \ref{fig:bzones}(a). As before, interpretation of the 
figures is greatly facilitated by combining them with the band structure 
shown in Fig.\ \ref{fig:resrut1}. We recognize electron- and hole-like 
regions near the $ {\rm \Gamma} $- and Z-point, respectively. Nevertheless, 
in contrast to the non-selfconsistent calculations by Gupta {\em et al.}, 
our calculations do not allow to identify any Fermi surface nesting in Figs.\ 
\ref{fig:resrut7} to \ref{fig:resrut9}, which might be indicative of 
a destabilization of the rutile structure. This is in contrast to the 
aforementioned findings for hypothetical rutile $ {\rm MoO_2} $. 
However, since $ {\rm VO_2} $ has one $ d $ electron less, the differences 
between both compounds were to be expected. We thus conclude, that 
investigation of the Fermi surface does not allow for an explanation of 
the instability of the rutile structure. In so far, our result resembles the 
above mentioned argument of Goodenough against a Jahn-Teller-type distortion 
as the major source for the crystal structure deformations.

\subsection{Symmetry analysis of the band structure}
\label{resrut6}

\begin{figure}[htp]
\centering
\includegraphics[width=0.75\textwidth]{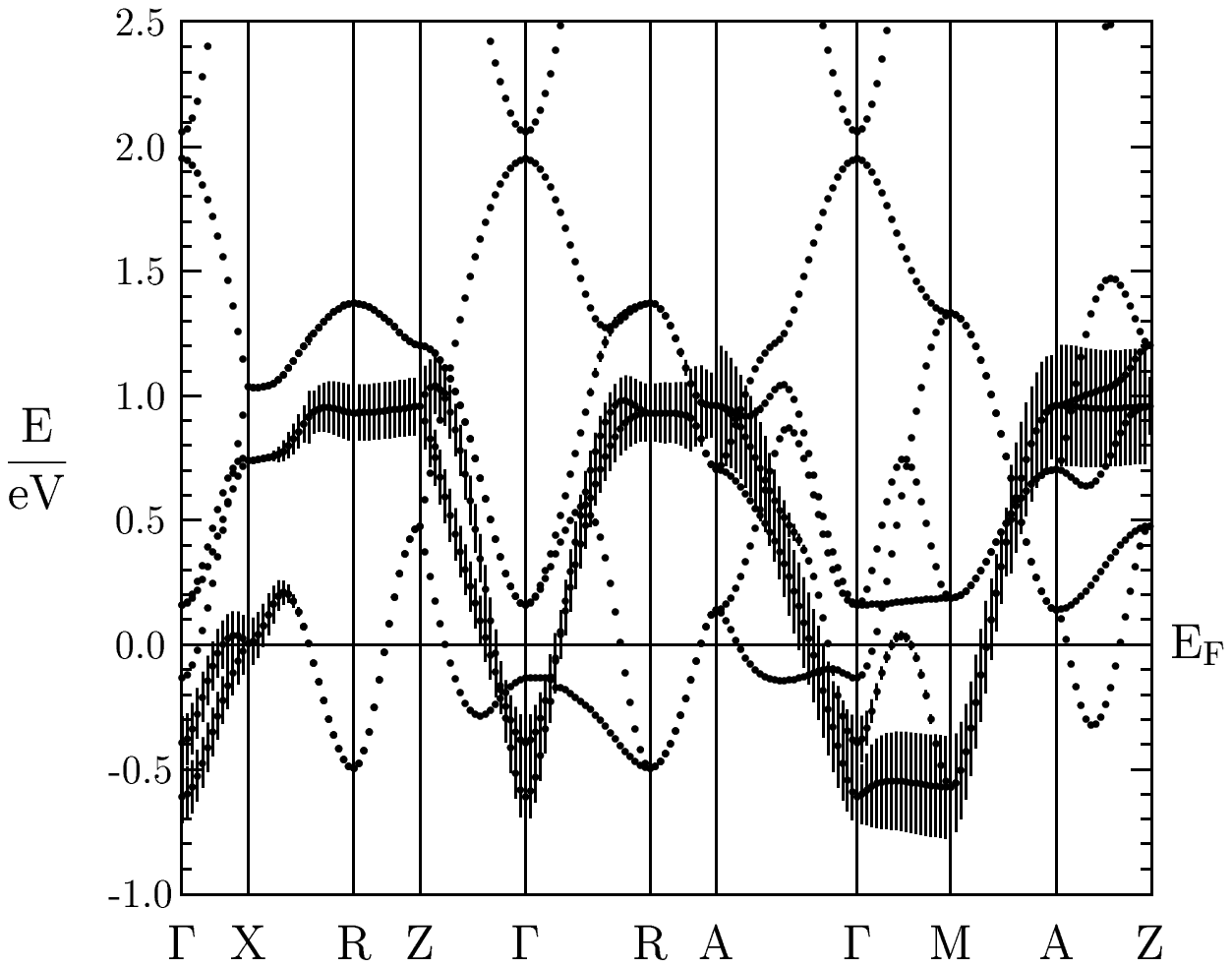}
\caption[Weighted electronic bands of rutile $ {\rm VO_2} $.]
        {Weighted electronic bands of rutile $ {\rm VO_2} $. The width of 
         the bars given for each band indicates the contribution due to the 
         $ 3d_{x^2-y^2} $ orbital of the V atom at (0,0,0) relative to the 
         local rotated reference frame.} 
\label{fig:resrut10}   
\end{figure}

\begin{figure}[htp]
\centering
\includegraphics[width=0.75\textwidth]{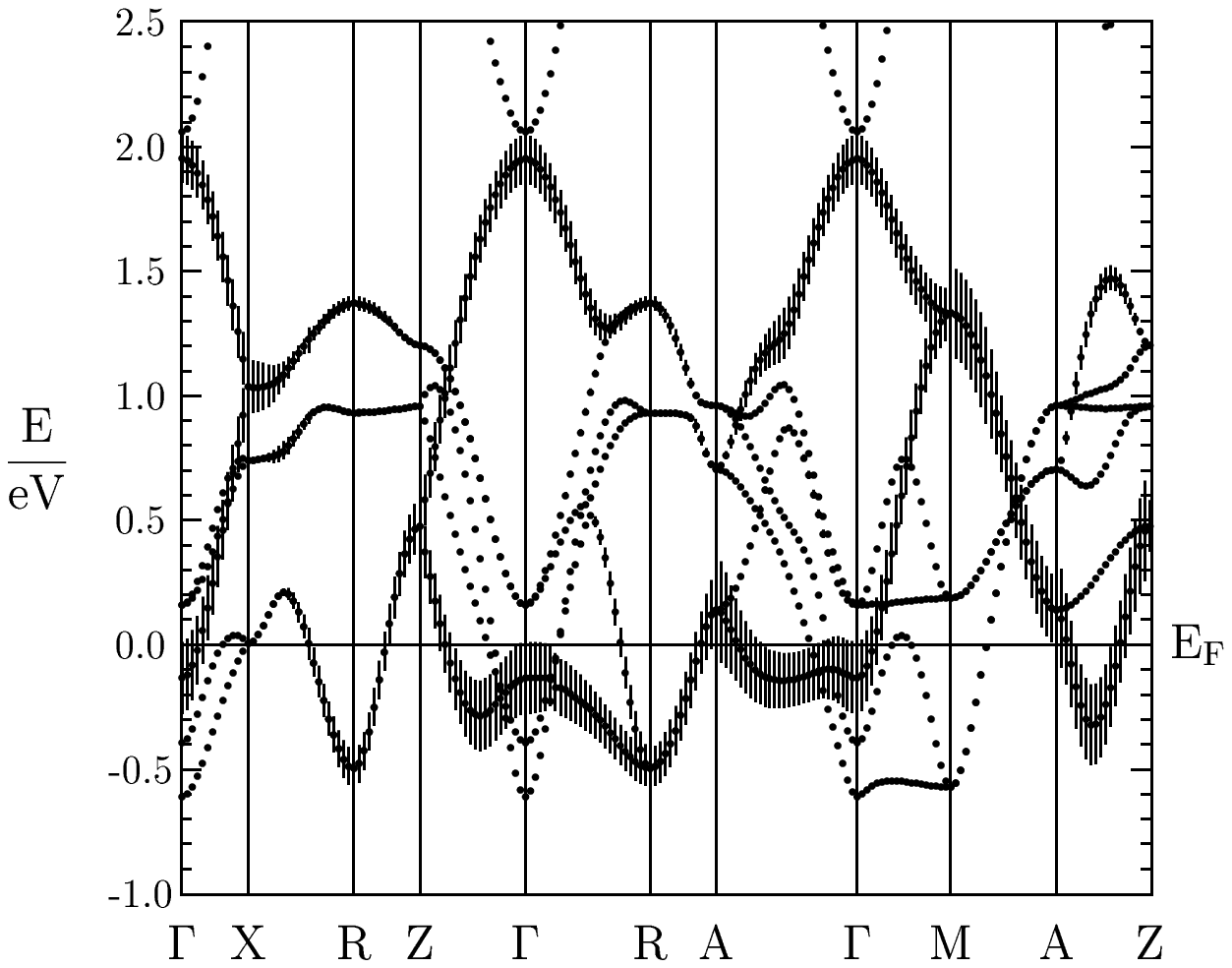}
\caption[Weighted electronic bands of rutile $ {\rm VO_2} $.]
        {Weighted electronic bands of rutile $ {\rm VO_2} $. The width of 
         the bars given for each band indicates the contribution due to the 
         $ 3d_{yz} $ orbital of the V atom at (0,0,0) relative to the 
         local rotated reference frame.} 
\label{fig:resrut11}   
\end{figure}

\begin{figure}[htp]
\centering
\includegraphics[width=0.75\textwidth]{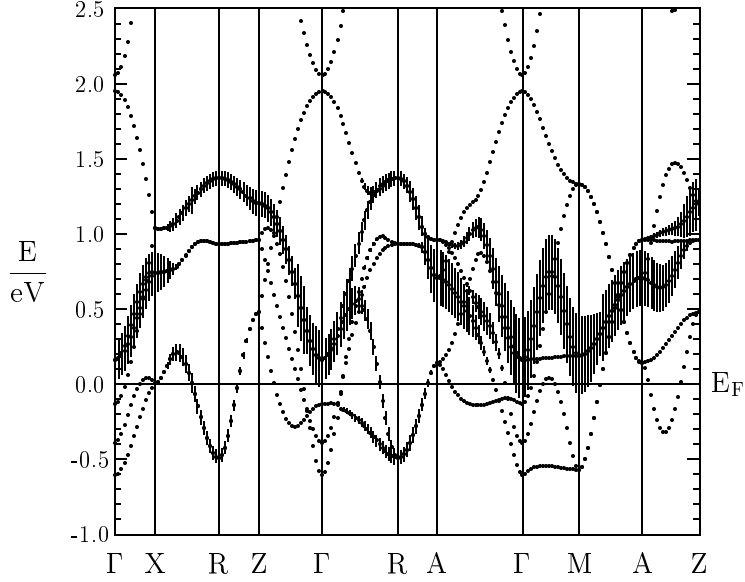}
\caption[Weighted electronic bands of rutile $ {\rm VO_2} $.]
        {Weighted electronic bands of rutile $ {\rm VO_2} $. The width of 
         the bars given for each band indicates the contribution due to the 
         $ 3d_{xz} $ orbital of the V atom at (0,0,0) relative to the 
         local rotated reference frame.} 
\label{fig:resrut12}   
\end{figure}

In order to find a common origin of the instability of the rutile structure 
in the $ d^1 $, $ d^2 $, and $ d^3 $ transition-metal dioxides we proceed 
along the same lines as in our work on $ {\rm MoO_2} $ and $ {\rm NbO_2} $ 
and perform a detailed analysis of the electronic states. To this end we 
display in Figs.\ \ref{fig:resrut10},  
\ref{fig:resrut11},  
and \ref{fig:resrut12} 
the electronic bands in the narrow energy range from -1.0 to 2.5\,eV about 
the Fermi energy in the same special representation as in our work on 
$ {\rm MoO_2} $ \cite{moo2pap}: In all three figures, each band at each 
$ {\bf k} $-point is given a bar, which length is a measure for the 
contribution from a specified orbital. Since we again refer to the 
local frame of reference, Figs.\ \ref{fig:resrut10} to \ref{fig:resrut12} 
correspond to the partial V $ 3d $ $ t_{2g} $ DOS shown in Fig.\ 
\ref{fig:resrut3}, which likewise made use of the local coordinate system. 

As in $ {\rm MoO_2} $ and $ {\rm NbO_2} $ we observe in Fig.\ 
\ref{fig:resrut10} a strong dispersion of the metal $ 3d_{x^2-y^2} $ 
derived bands along all lines parallel to $ {\rm \Gamma} $-Z. In contrast,  
the dispersion perpendicular to this line is almost negligible. Again, 
we attribute this dispersion to the overlap of $ d_{x^2-y^2} $ orbitals 
at metal sites neighbouring in $ z $ direction. As compared to 
$ {\rm MoO_2} $ the width of these bands relative to the total $ t_{2g} $ 
band width is reduced. This is due to the increased value of the rutile 
$ c $ axis and, hence, the larger metal-metal distance in this direction. 
In contrast, the smaller value of the $ a $ axis in $ {\rm VO_2} $ causes 
a larger interchain interaction. This leads to an increased splitting of 
the $ d_{x^2-y^2} $ bands along the lines $ {\rm \Gamma} $-M and A-Z as 
well as along the lines where the strong dispersion is observed. 

The V $ 3d_{yz} $ derived bands are highlighted in Fig.\ \ref{fig:resrut11}. 
They show a large dispersion in all directions and, as in $ {\rm MoO_2} $, 
a large splitting at the $ {\rm \Gamma} $-point. The latter is a consequence 
of the underlying body-centered tetragonal lattice formed by the metal atoms 
and results from backfolding the bands from the Brillouin zone of this 
lattice to the smaller one of the simple tetragonal lattice. This effect can 
be read off from the bands along the lines connecting the $ {\rm \Gamma} $-  
point to the X-, M-, A-, and Z-point, respectively. 

Finally, the V $ 3d_{xz} $ derived bands display the smallest dispersion. 
This is due to the fact that these orbitals induce no substantial 
$ \sigma $-type overlap with neighbouring metal atoms. Instead, they 
mediate the small coupling between the octahedral chains. Still, the 
larger width of these bands as compared to $ {\rm MoO_2} $ results from 
the increased interchain coupling coming with the smaller value of the 
$ a $ lattice constant. 

Of course, the previous analysis of the $ t_{2g} $ bands with its nearly 
unique assignment of a single character to each band as well as the subsequent 
interpretation in terms of the geometric coordination of the atoms and 
their orbitals benefitted from the very small mixing among these states. 
Exceptions are the hybridization of the $ d_{x^2-y^2} $ bands with the 
$ d_{yz} $- and $ d_{xz} $-derived bands along the symmetry line X-R and, 
to a lesser degree, along the line $ {\rm \Gamma} $-A. Nevertheless, we 
point out that the three $ t_{2g} $ bands disperse almost completely 
independently and are coupled mainly via charge conservation. For this 
reason the $ d_{\parallel} $ band may be regarded as a one-dimensional 
band in a three-dimensional embedding background of $ \pi^{\ast} $ bands.
The situation is thus identical to that found in rutile $ {\rm NbO_2} $ 
as well as in hypothetical rutile $ {\rm MoO_2} $ \cite{moo2pap,nbo2pap}

Eventually, the finding of an only weak hybridization of the different 
types of bands has a strong impact on our further understanding since 
it will allow to establish a relation between particular atomic 
displacements and the response of the electronic states. This way, we 
may find a route to an understanding of the mechanisms leading to the 
low-temperature monoclinic structure.

\section{Results for insulating $ {\rm \bf M_1} $-$ {\rm \bf VO_2}$}
\label{resm1}

\subsection{Band structure and density of states}
\label{resm11}
  
\begin{figure}[htp]
\centering
\includegraphics[width=0.8\textwidth]{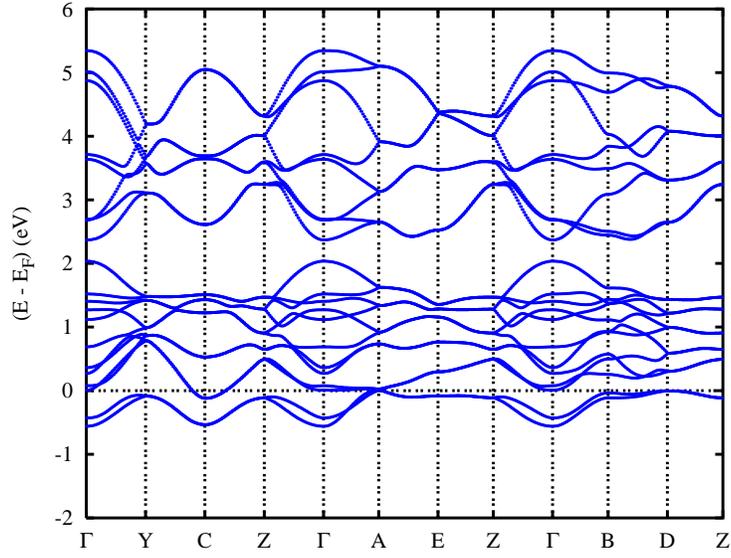}
\caption[Electronic bands of monoclinic $ {\rm VO_2} $.]
        {Electronic bands of monoclinic $ {\rm VO_2} $ along selected 
         symmetry lines within the simple monoclinic Brillouin zone, 
         Fig.\ \protect\ref{fig:bzones}(b).}
\label{fig:resm11}   
\end{figure}

\begin{figure}[htp]
\centering
\includegraphics[width=0.8\textwidth]{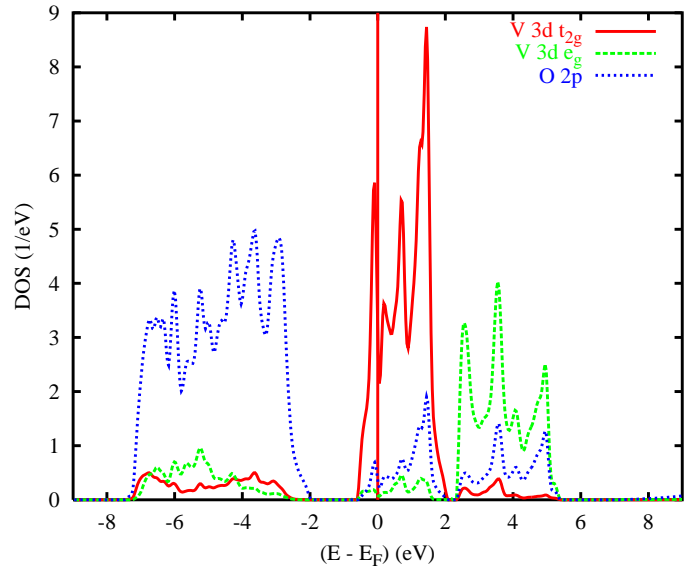}
\caption[Partial densities of states (DOS) of monoclinic $ {\rm VO_2} $.] 
        {Partial densities of states (DOS) of monoclinic $ {\rm VO_2} $ 
	 per unit cell.} 
\label{fig:resm12}   
\end{figure}

The electronic bands of monoclinic $ {\rm VO_2} $ in the energy range of 
the V $ 3d $ states are displayed in Fig.\ \ref{fig:resm11} 
along selected high symmetry lines within the simple monoclinic Brillouin 
zone, Fig.\ \ref{fig:bzones}(b). The dominant partial densities of states 
(DOS) are given in Fig.\ \ref{fig:resm12}.  
As before low lying oxygen $ 2s $ states are not included. 

In Figs.\ \ref{fig:resm11} and \ref{fig:resm12} we identify the same groups 
of bands as for rutile $ {\rm VO_2} $. They comprise 24 oxygen $ 2p $ 
dominated bands well below the Fermi level and two groups of 12 and 8 bands, 
respectively, at and above $ {\rm E_F} $. These latter two groups trace back 
mainly to V $ 3d $ states. Finally, 
$ p $--$ d $ hybridization leads to V $ 3d $ and O $ 2p $ contributions of 
the order of about 10\% in the energy regions where the respective other 
partner dominates. The energetical separation between the O $ 2p $ and 
V $ 3d $ derived bands turns out to be slightly smaller than in the rutile 
phase. This is in agreement with the XPS results by Blaauw {\em et al.}\ 
\cite{blaauw75}. Again, these general findings conform well with the 
molecular orbital picture drawn in Sec.\ \ref{mopict}. 

Nevertheless, on going into more detail we find distinct differences between  
the results for the rutile and the monoclinic structure. In particular, we 
witness the evolution of a sharp peaks in the V $ 3d $ $ t_{2g} $ partial 
DOS just below $ {\rm E_F} $ and at $ \approx 1.3 $\,eV. In general, spectral 
weight is shifted from the central region of this group of bands to the 
edges. As a consequence, the density of states at the Fermi 
energy is lowered as compared to the rutile value and amounts to 2.08 
states/f.u./eV. 

\begin{figure}[htp]
\centering
\includegraphics[width=0.8\textwidth]{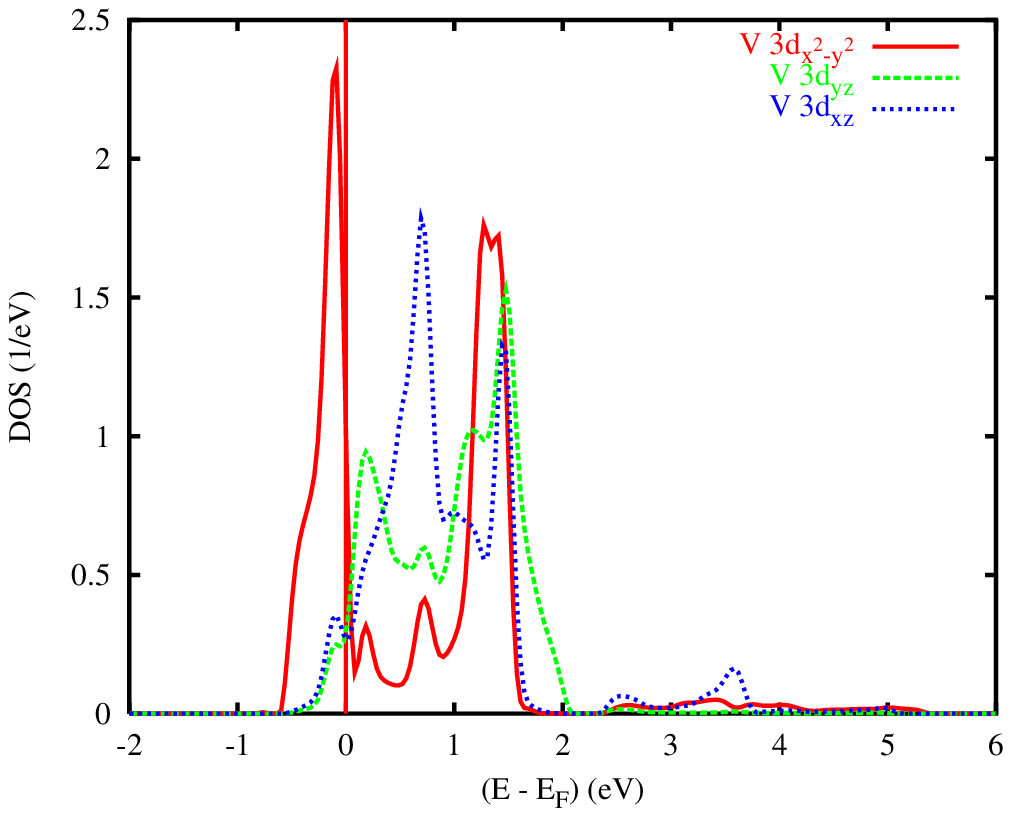}
\includegraphics[width=0.8\textwidth]{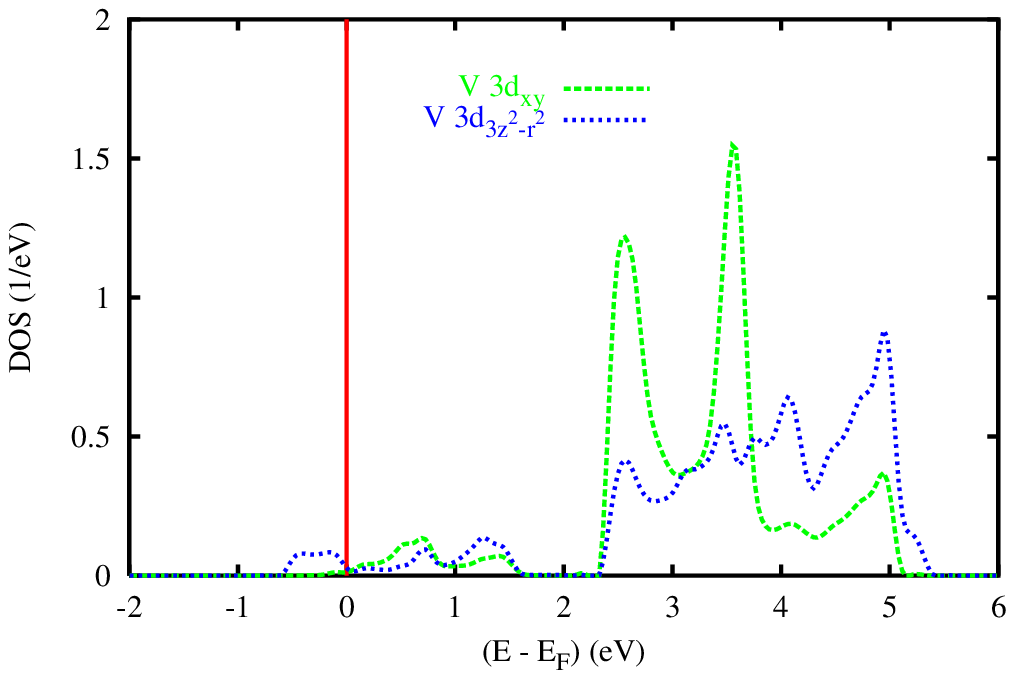}
\caption[Partial V $ 3d $ $ t_{2g} $ and $ e_g $ densities of states (DOS) 
         of monoclinic $ {\rm VO_2} $.]
        {Partial V $ 3d $ $ t_{2g} $ and $ e_g $ densities of states (DOS) 
         of monoclinic $ {\rm VO_2} $. Selection of orbitals is relative 
         to the local rotated reference frame.} 
\label{fig:resm13}   
\end{figure}

In order to understand these differences we proceed displaying in Fig.\ 
\ref{fig:resm13} 
the single symmetry components of the V $ 3d $ partial DOS. In doing so we 
include only the single vanadium atom near the corner of the underlying 
rutile cell and use the local rotated reference frame as defined in Sec.\ 
\ref{crystrut}. As in Fig.\ \ref{fig:resm12} we recognize the almost perfect 
energetical separation of the $ 3d $ $ t_{2g} $ and $ e_g $ groups of bands, 
which results from the octahedral crystal field splitting. The small 
$ t_{2g} $--$ e_g $ configuration mixing again points to the slight 
distortions of the octahedra. 

Concentrating on the $ t_{2g} $ partial DOS, we find close similarities 
of the $ d_{xz} $ and $ d_{yz} $ partial DOS to those calculated for the 
rutile structure given in Fig.\ \ref{fig:resrut3}. Yet, both bands have 
experienced a considerable energetical upshift and are left almost 
unoccupied. 

While the $ d_{xz} $ and $ d_{yz} $ partial DOS still resemble each other 
the $ d_{x^2-y^2} $ partial DOS display striking deviations from both these 
two curves and the results obtained for the rutile structure. Comparing 
Fig.\ \ref{fig:resrut3} to Fig.\ \ref{fig:resm13} we recognize strong 
splitting of the $ d_{x^2-y^2} $ partial DOS into two peaks, which are 
located just below $ {\rm E_F} $ and at $ \approx 1.3 $\,eV. In addition, 
the partial DOS in between is considerably reduced. This leads to the 
decrease of the density of states at the Fermi energy. 

Our findings for the $ t_{2g} $ states of monoclinic $ {\rm VO_2} $ are 
in good agreement with the band scheme proposed by Goodenough
\cite{goodenough71a}, which we sketched in Fig.\ \ref{fig:intro2}. We 
thus attribute the striking behaviour of the $ d_{x^2-y^2} $ (or 
$ d_{\parallel} $) bands to the pairing of the vanadium atoms along the 
rutile $ c $ axis and the resulting bonding-antibonding splitting. In 
contrast, the energetical upshift of the $ d_{xz} $ and $ d_{yz} $ (or 
$ \pi^{\ast} $) bands results from the lateral antiferroelectric 
displacement of the vanadium atoms, which increases the $ p $--$ d $ 
bonding. 

\begin{figure}[htp]
\centering
\includegraphics[width=0.8\textwidth]{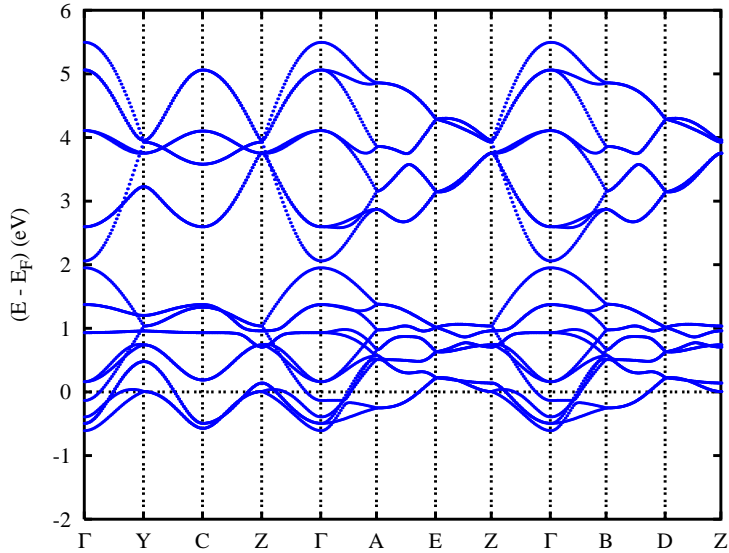}
\caption[Electronic bands of rutile $ {\rm VO_2} $.]
        {Electronic bands of rutile $ {\rm VO_2} $ along selected symmetry 
         lines within the simple monoclinic Brillouin zone, 
         Fig.\ \protect\ref{fig:bzones}(b).}
\label{fig:resm14}   
\end{figure}

The situation becomes clearer from a comparison to the band structure 
of rutile $ {\rm VO_2} $ shown in Fig.\ \ref{fig:resm14}.  
Here we have used the first Brillouin zone of the simple monoclinic lattice, 
Fig.\ \ref{fig:bzones}(b) and folded back all bands from the larger Brillouin 
zone of the tetragonal lattice, Fig.\ \ref{fig:bzones}(a). Just in the 
same manner as for $ {\rm MoO_2} $, those two bands in Fig.\ \ref{fig:resm14}, 
which give rise to the lowest states within the $ t_{2g} $-derived group of 
bands at the $ {\rm \Gamma} $-point, bend upwards along the line 
$ {\rm \Gamma} $-A and cross the higher lying bands. In monoclinic 
$ {\rm VO_2} $ these two bands are separated from the other states and form 
a split-off doublet, which, however, is still touching the higher lying bands 
at the A-point. For this reason, we end up with a finite density of states 
at the Fermi energy, which spoils the experimental finding of an insulating 
ground state with an optical band gap of about 0.6\,eV. Instead, from the 
touching bands at the A-point and the occupied band width of the 
conduction band at the C-point of about 0.1\,eV we would rather regard 
monoclinic $ {\rm VO_2} $ as a semimetal, in agreement with the results by 
Wentzcovitch {\em et al.}\ \cite{wentz94a}. Like these authors we attribute 
this failure of the calculations to the shortcomings of the local density 
approximation, which usually underestimates the band gap by about 50\% and 
in some small gap semiconductors like Ge misses the gap at all. This seems 
to be also the case for monoclinic $ {\rm VO_2} $. Like in the 
zincblende-type semiconductors the optical band gap originates to a large 
part from bonding-antibonding splitting of hybridized bands, which, 
obviously, is not well enough accounted for by the LDA.

\subsection{Chemical bonding}
\label{resm12}

\begin{figure}[htp]
\centering
\includegraphics[width=0.8\textwidth]{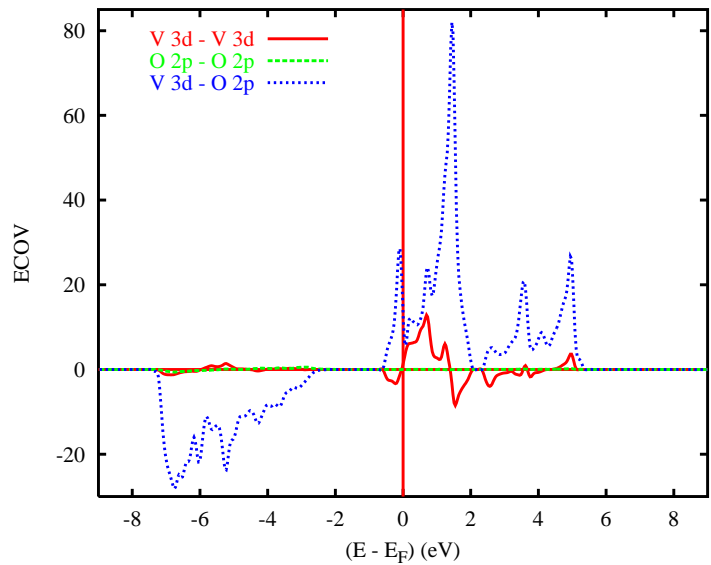}
\caption{Partial covalence energies ($ {\rm E_{cov}} $) of monoclinic 
         $ {\rm VO_2} $.}
\label{fig:resm15}
\end{figure}

Again we investigate the chemical bonding by determining the covalence 
energy. The result is shown in Fig.\ \ref{fig:resm15}.
The $ E_{cov} $ curves for the O $ 2p $--O $ 2p $ and the V $ 3d $--O $ 2p $ 
bonding agree fairly well with the respective curves for rutile $ {\rm VO_2} $ 
as shown in Fig.\ \ref{fig:resrut4}. Qualitative differences show up in the 
V $ 3d $--V $ 3d $ bonding. In particular, 
we observe a change in sign in the energy region of the bonding 
V $ 3d_{x^2-y^2} $ bands. While the respective $ E_{cov} $ curve in the 
lower half of the V $ 3d $ $ t_{2g} $ bands was positive (antibonding) 
for rutile $ {\rm VO_2} $ it has become negative (bonding) just below 
the Fermi energy in the $ {\rm M_1} $ structure. As for monoclinic 
$ {\rm MoO_2} $ this signals the increased metal-metal bonding due to 
the pairing of the metal atoms. Furthermore, the formation of bonding 
and antibonding V $ 3d_{x^2-y^2} $ states below and above the Fermi 
energy, respectively, decreases the value of the total integrated 
$ E_{cov} $ at the Fermi level.

\subsection{Comparison to experiment}
\label{resm13}

\begin{figure}[htp]
\centering
\includegraphics[width=0.8\textwidth]{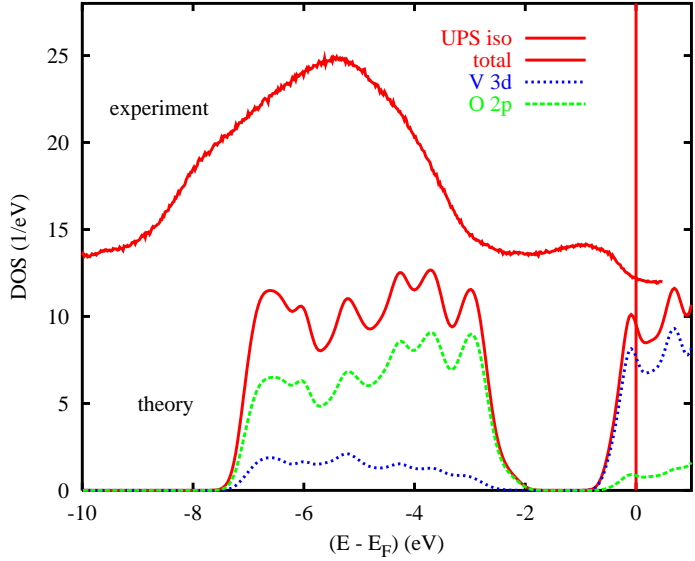}
\caption[Total and partial densities of states (DOS) of monoclinic 
         $ {\rm VO_2} $ folded with a 0.25\,eV wide Gaussian and UPS 
         spectra.]
        {Total and partial densities of states (DOS) of monoclinic 
         $ {\rm VO_2} $ folded with a 0.25\,eV wide Gaussian (lower set 
         of curves) and UPS spectra (upper curve; note the offset 
         introduced in order to distinguish experimental and theoretical 
         results; from Ref.\ \protect \cite{goering96,goering97b}).} 
\label{fig:resm16}   
\end{figure}

\begin{figure}[htp]
\centering
\includegraphics[width=0.8\textwidth]{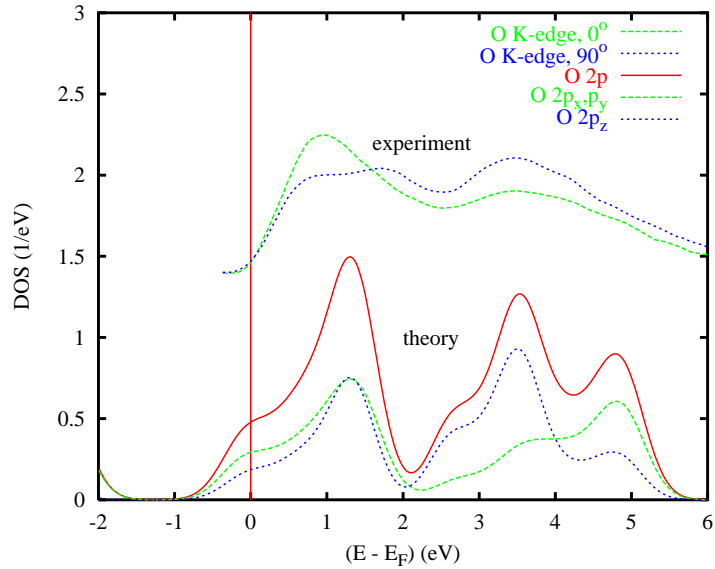}
\caption[Partial O $ 2p $ densities of states (DOS) of monoclinic 
         $ {\rm VO_2} $ folded with a 0.5\,eV wide Gaussian and 
         XAS O K edge spectra.] 
        {Partial O $ 2p $ densities of states (DOS) of monoclinic 
         $ {\rm VO_2} $ folded with a 0.5\,eV wide Gaussian (lower set 
         of curves) and XAS O K edge spectra (upper set of curves; 
         note the offset introduced in order to distinguish experimental 
         and theoretical results; from Ref.\ \protect 
         \cite{mueller96,mueller97}, data shifted by 529.5\,eV).}
\label{fig:resm17}   
\end{figure}

As for metallic $ {\rm VO_2} $ we display in Figs.\ \ref{fig:resm16}  
and \ref{fig:resm17} 
total and partial V $ 3d $ and O $ 2p $ densities of states. The curves 
are folded with a Gaussian of 0.25 and 0.5\,eV width, respectively, for 
the occupied and unoccupied part of the spectrum. Good agreement with 
the XPS and UPS spectra by Blaauw {\em et al.}, Shin {\em et al.}, 
Bermudez {\em et al.}, and Goering {\em et al.}\ is found 
\cite{blaauw75,shin90,bermudez92,goering96,goering97b}. Data by the latter 
group were added in Fig.\ \ref{fig:resm16}. 
According to these experiments the UPS spectra taken in the metallic and the 
insulating phase are very similar with respect to the oxygen dominated bands 
but show distinct differences in the low-binding region. They arise from a 
sharpening and slight downshift of the V $ 3d $-derived peak 
\cite{blaauw75,shin90,goering96,goering97b}. As a comparison of Fig.\ 
\ref{fig:resm16} with Fig.\ \ref{fig:resrut5} reveals, this experimental 
result is well accounted for by our calculations except for the fact that 
the V $ 3d $ downshift is underestimated.

More striking differences between metallic and insulating $ {\rm VO_2} $ 
are observed in the soft-X-ray absorption spectra (XAS) as measured by 
M\"uller {\em et al.}\ \cite{mueller96,mueller97}, which we display in 
Fig.\ \ref{fig:resm17}. The experimental data were shifted by 529.5\,eV. 
As already described in Sec.\ \ref{resrut3}, the XAS spectra allowed to 
probe the angular dependence by varying the orientation of the 
polarization vector $ {\rm \bf E} $ from parallel ($ \phi = 90^{\circ} $) 
to perpendicular 
($ \phi = 0^{\circ} $) to the rutile $ c $ axis. Dipole selection rules 
then allow for transitions from O $ 1s $ to either the O $ 2p_z $ or the 
O $ 2p_x $ and $ 2p_y $ states. This is reflected by the calculated partial 
DOS of the corresponding final states. However, note that the curve 
marked O $ 2p_x, p_y $ actually comprises only the mean average of these 
orbitals since the component perpendicular to $ {\rm \bf E} $ is not seen in 
experiment. For the same reason, the curve marked O $ 2p $ actually contains 
the full contribution from the $ p_z $ orbital but only half of the 
contributions from the $ p_x $ and $ p_y $ orbitals. 

According to the experiments by M\"uller {\em et al.}\ as well as by Abbate 
{\em et al.}\ major differences between the XAS spectra of metallic and 
insulating $ {\rm VO_2} $ originate from the appearance of an additional 
peak in the low-temperature phase \cite{mueller96,mueller97,abbate91}. In 
the spectra by Abbate {\em et al.}\ this new peak is located at 530.9\,eV 
and seems to have split off from the peak at 529.9\,eV, which, like the one 
at 532.5\,eV, exists also in the metallic phase \cite{abbate91}. Abbate 
{\em et al.}\ interpreted the additional peak as arising from an upshift 
of the antibonding $ d_{\parallel} $ band by about 1\,eV. The other two 
$ t_{2g} $ bands, namely the $ \pi^{\ast} $ bands, stay at their original 
position. The XAS curves taken by M\"uller {\em et al.}\ are almost identical 
to the results of Abbate {\em et al.}. However, in a more detailed fit 
M\"uller assigned the spectra to five different peaks at 529.9, 530.4, 
531.2, 532.7, and 533.9\,eV \cite{mueller96}. 

Again, there is an overall good agreement between experiment and theory 
with respect to positions, intensities, and angular dependencies of the 
peaks. As a comparison with Fig.\ \ref{fig:resrut6} reveals, the 
calculated curves reflect the upshift of the O $ 2p_z $ peak from 0.7\,eV 
to 1.3\,eV, which is the position of the upper $ d_{x^2-y^2} $ peak in 
Fig.\ \ref{fig:resm13}. At the same time, all other peaks hardly move. 
Nevertheless, the O $ 2p_z $ peak has its maximum at the same energy as 
the O $ 2p_x, p_y $ curve at 1.3\,eV. This is in contrast to the XAS 
experiments, which find the maximum of the O $ 2p_z $ peak near 1.8\,eV. 
Thus, it seems that the upshift of the O $ 2p_z $ states and, hence, of 
the upper $ d_{x^2-y^2} $ states, is underestimated by the calculation 
in the same manner as in our previous work on $ {\rm MoO_2} $ \cite{moo2pap}. 
From a comparison of the calculated DOS with both angle-resolved UPS and 
XAS data for the molybdenum dioxide, we obtained perfect agreement with 
experiment except for the bonding and antibonding Mo $ 4d_{x^2-y^2} $ 
bands. Their splitting was underestimated by about 1\,eV. 
For $ {\rm VO_2} $ the calculations yield a separation of $ \approx 1.7 $ 
eV between the bonding and antibonding $ d_{x^2-y^2} $ bands. A value of 
2.0 to 2.5\,eV has been extracted from UPS and XAS measurements 
\cite{shin90,abbate91}.

Finally, as for 
rutile $ {\rm VO_2} $ the calculated high-energy peak at $ \approx 4.8 $ 
eV gives rise to only a very slight shoulder in the experimental curve. 
Yet, it can be fully resolved by the aforementioned five peak analysis 
by M\"uller \cite{mueller96}.

\subsection{Embedded Peierls instability}
\label{resm14}

\begin{figure}[htp]
\centering
\includegraphics[width=0.75\textwidth]{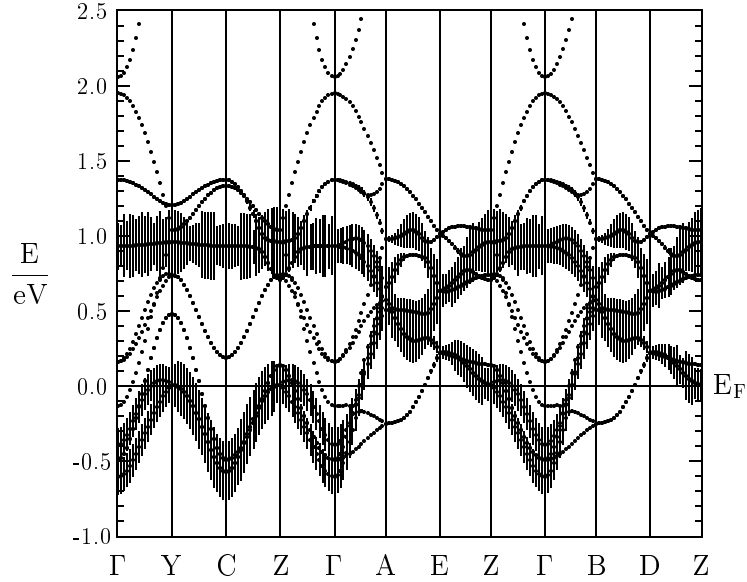}
\caption[Weighted electronic bands of rutile $ {\rm VO_2} $.]
        {Weighted electronic bands of rutile $ {\rm VO_2} $. The width of 
         the bars given for each band indicates the contribution due to the 
         $ 3d_{x^2-y^2} $ orbital of the V atom at the corner of the rutile 
         cell relative to the local rotated reference frame.} 
\label{fig:resm18}   
\end{figure}

The upshift of the V $ 3d_{xz} $ and $ d_{yz} $ bands as well as the 
bonding-antibonding splitting of the $ d_{x^2-y^2} $ band as observed 
in the partial DOS already signaled the general validity of the mechanism 
proposed by Goodenough and, more recently, by Wentzcovitch {\em et al.} 
\cite{goodenough71a,goodenough71b,wentz94a}. Nevertheless, we still aim 
at obtaining more direct support for these ideas from the electronic 
structure. In order to prepare for the discussion we display in Fig.\ 
\ref{fig:resm18} 
the weighted band structure of rutile $ {\rm VO_2} $ corresponding to the 
$ d_{x^2-y^2} $ states. It results from that shown in Fig.\ 
\ref{fig:resrut10} by backfolding to the monoclinic Brillouin zone, Fig.\ 
\ref{fig:bzones}(b). As before the length of a bar given with a band at a 
$ {\bf k} $-point is a measure of the contribution from the $ d_{x^2-y^2} $ 
orbital. Note that again we refer to the local reference frame. 

In Fig.\ \ref{fig:resm18} we observe an arrangement of the $ d_{x^2-y^2} $ 
bands in two parts below and above 0.4\,eV, respectively. They are connected 
by dispersing bands especially along the lines $ {\rm \Gamma} $-A and 
$ {\rm \Gamma} $-B. Both lines have components perpendicular to the rutile 
basal plane and correspond to half the line $ {\rm \Gamma} $-R of the 
tetragonal Brillouin zone. In contrast, the dispersion parallel to the 
rutile basal plane is suppressed to a large degree. The respective bands 
stay at energies either well below or above 0.4\,eV, even if there is some 
dispersion within the lower group. Finally, $ d_{x^2-y^2} $ bands along 
the lines A-E and B-D, while likewise displaying a reduced dispersion, 
stay at an intermediate energy in the middle between the two subgroups. 
Note that these lines lie completely within horizontal planes at height 
$ \frac{\pi}{2c} $. The general characteristics of these bands are thus 
very similar to those of the $ 4d_{x^2-y^2} $ bands of hypothetical 
rutile $ {\rm MoO_2} $. In that system the dispersionless bands along 
A-E and B-D were responsible for the flat portions of the Fermi surface. 
They were thus interpreted as driving the Peierls-type instability of 
this structure. 

The V $ 3d_{xz} $ and $ d_{yz} $ bands can be identified in Fig.\ 
\ref{fig:resm18} as the states without a bar. This interpretation is 
based on the fact that the O $ 2p $ and the V $ 3d $ $ e_g $ states 
play only a minor role in the energy range shown. In contrast to the 
$ d_{x^2-y^2} $ bands these $ \pi^{\ast} $ bands do not fall into 
energetically lower and upper bands. Instead they rather show 
isotropic dispersion. Finally, we note again the much reduced 
hybridization between both types of bands, which we already mentioned 
at the end of Sec.\ \ref{resrut6}. As a consequence, the 
$ d_{\parallel} $ and $ \pi^{\ast} $ bands can, as before, be regarded 
as nearly independent. 

\begin{figure}[htp]
\centering
\includegraphics[width=0.75\textwidth]{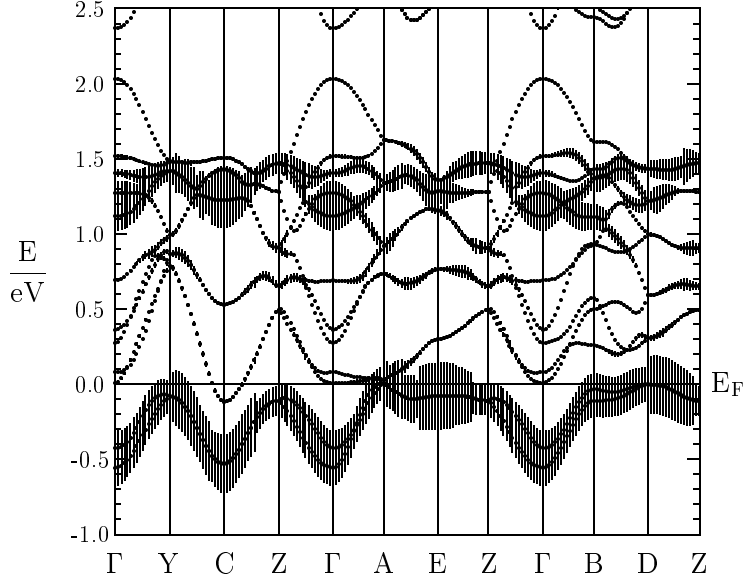}
\caption[Weighted electronic bands of monoclinic $ {\rm VO_2} $.]
        {Weighted electronic bands of monoclinic $ {\rm VO_2} $. The width of 
         the bars given for each band indicates the contribution due to the 
         $ 3d_{x^2-y^2} $ orbital of the V atom at the corner of the rutile 
         cell relative to the local rotated reference frame.} 
\label{fig:resm111}
\end{figure}

The corresponding weighted band structure for monoclinic $ {\rm VO_2} $ 
is shown in Fig.\ \ref{fig:resm111}. 
The differences between both figures is striking. On going from rutile 
to monoclinic $ {\rm VO_2} $ we observe a strong splitting of the 
$ d_{x^2-y^2} $ bands into two narrow subbands at the lower and upper 
edge of the $ t_{2g} $ group of bands. This splitting is a consequence of 
the metal-metal dimerization. In the partial DOS, Fig.\ \ref{fig:resm13}, 
this bonding-antibonding splitting led to the formation of the two 
distinct peaks just below $ {\rm E_F} $ and at $ \approx 1.3 $\,eV. 
In particular, the $ d_{x^2-y^2} $ bands along the lines 
$ {\rm \Gamma} $-A-E-Z and $ {\rm \Gamma} $-B-D-Z have moved downward 
and upward, respectively, by about 0.5\,eV relative to their positions 
in the rutile structure. As a consequence, the so 
far unoccupied branch of these bands between $ {\rm E_F} $ and 
$ \approx 0.4 $\,eV has become almost fully occupied and the Fermi surface 
due to the $ d_{x^2-y^2} $ states has almost vanished. Moreover, these 
changes explain the occurrence of the split-off double band at the 
lower edge of the $ t_{2g} $ group of bands, which we mentioned at the 
end of Sec.\ \ref{resm11}. 

In contrast, the V $ 3d_{xz} $ and $ d_{yz} $ bands, i.e.\ those bands 
in Fig.\ \ref{fig:resm111}, which have bars of negligible lengths, 
experience a considerable upshift by $ \approx 0.5 $\,eV due to the 
zigzag-like antiferroelectric displacement of the vanadium atoms. This 
result is in full agreement with the data by Shin {\em et al.}\ 
\cite{shin90}. Yet, with only minor exceptions, the shift of the 
$ \pi^{\ast} $ bands leaves the general topology of these bands intact 
and causes their almost complete depopulation. Our findings agree well 
with the arguments given by Goodenough, who concluded from the isotropic 
conductivity that the $ \pi^{\ast} $ states should be located at the 
lower edge of the conduction band well below the antibonding 
$ d_{\parallel} $ band \cite{goodenough71b}. Furthermore, Goodenough 
deduced from the measured low Hall mobility a narrowing of the 
$ \pi^{\ast} $ bands \cite{goodenough71b}. Indeed, our calculations 
result in a band width of about 2\,eV of the $ \pi^{\ast} $ bands, which 
is considerably less than the 2.5\,eV found in the rutile phase. 

Finally, we note that the hybridization between both types of bands is 
still very small. In other words, the just described changes of the 
$ d_{\parallel} $ and $ \pi^{\ast} $ bands occurred rather independently.  
The coupling between them is still only via the common Fermi energy. 
Just in the same manner as in $ {\rm NbO_2} $ and $ {\rm MoO_2} $ we may 
thus regard the transition to the low-temperature state as a Peierls 
instability of the $ d_{\parallel} $ bands in an embedding background 
of $ \pi^{\ast} $ electrons.

An aspect, which seemingly has not been considered before, arises from 
the fact, that metal-metal pairing along the rutile $ c $ axis, in 
addition to increasing the bonding-antibonding splitting of the 
$ d_{x^2-y^2} $ bands, shifts the vanadium atoms also {\em relative} 
to the surrounding oxygen octahedra and, hence, causes an increased 
metal-oxygen overlap. Thus, if not compensated by a corresponding 
displacement of the oxygen octahedron, the dimerization likewise brings 
in an antiferroelectric component. As a consequence, within an 
investigation of the monoclinic $ {\rm M_1} $ structure we are actually 
not able to uniquely attribute the upshift of the $ \pi^{\ast} $ bands 
to either the zigzag-type in-plane displacement or the metal-metal pairing. 
This is due to the fact that in the $ {\rm M_1} $ phase all metal chains 
experience the same atomic displacements. The issue can only be resolved 
by studying the monoclinic $ {\rm M_2} $ phase, which brings in a 
disproportionation of the vanadium chains.

\section{Results for insulating $ {\rm \bf M_2} $-$ {\rm \bf VO_2}$}
\label{resm2}

In the previous two sections we have investigated in detail both the 
metallic rutile and the insulating monoclinic ($ {\rm M_1} $) phase of 
$ {\rm VO_2} $. In addition, we proposed a possible explanation for 
the origin of the metal-insulator transition in terms of an embedded 
Peierls-like instability. Our conclusions confirmed the previous 
findings of Wentzcovitch {\em et al.}, who, using a molecular dynamics 
type approach, arrived at very similar results \cite{wentz94a,wentz94b}. 
Still, the discussion by these authors was criticized by Rice 
{\em et al.}, who focused attention on the monoclinic $ {\rm M_2} $ 
phase of $ {\rm VO_2} $ \cite{rice94}. 

The arguments given by Rice {\em et al.}\ aimed at attempts to explain 
the metal-insulator transition of stoichiometric $ {\rm VO_2} $ purely 
in terms of the dimerization in the monoclinic $ {\rm M_1} $ phase. If 
this were the only or the dominating driving force there would be no 
reason why $ {\rm VO_2} $ in the $ {\rm M_2} $ phase would be insulating, 
since only half of the chains dimerize. Yet, as has been already pointed out 
by Goodenough, metal-metal pairing is not the only source of an efficient 
splitting of the $ d_{\parallel} $ band but the same effect could be 
achieved by exchange splitting coming with the antiferromagnetic order 
\cite{goodenough71a,goodenough71b}. The latter was indeed observed on the 
zigzag chains of the $ {\rm M_2} $ structure. Nevertheless, from the 
antiferromagnetic ordering several authors concluded that the respective 
$ 3d $ electrons would be rather localized and, while being accessible 
to strong electron-electron correlations, resist a proper description 
within band theory. Due to the close relation between the low-temperature 
monoclinic phases these arguments would then also hold for the 
$ {\rm M_1} $ phase. 

Since the $ {\rm M_2} $ phase has not yet been considered using 
state-of-the-art electronic structure calculations we included 
its investigation in the present study. From discussion of the 
$ {\rm M_2} $ phase we expect a deeper understanding of the 
mechanisms, which are active in the $ {\rm M_1} $ phase, since each chain 
shows only one type of octahedral deformation. Furthermore, we 
should be able to distinguish the different types of vanadium atom 
displacements with respect to their coupling to the electronic 
states and, in particular, to identify the implications of the 
two different types of antiferroelectric modes arising from the 
zigzag-type and pairing displacements. The latter can not be 
resolved in investigations of the $ {\rm M_1} $ phase. 

As has been outlined in Sec.\ \ref{m2phase}, Pouget {\em et al.}\ 
proposed to view the monoclinic $ {\rm M_2} $ phase as a metastable 
modification of the $ {\rm M_1} $ phase and to regard both the doping 
with Cr or Al and the application of uniaxial stress as only small 
perturbations of pure $ {\rm VO_2} $ under ambient pressure. These 
arguments were supported by the crystal structure data reported for 
different types of dopants. For both the rutile and the $ {\rm M_2} $ 
phase the parameters given by Marezio {\em et al.}\ for 
$ {\rm V_{0.976}Cr_{0.024}O_2} $ and by Ghedira {\em et al.}\ for 
$ {\rm V_{0.985}Al_{0.015}O_2} $ are almost identical 
\cite{marezio72,ghedira77b} and for the rutile structure they are 
identical to the numbers given for stoichiometric $ {\rm VO_2} $, see Sec.\ 
\ref{crystrut} and Tabs.\ \ref{tab:cryst4} and \ref{tab:cryst5}. It is 
thus well justified to use the crystal structure data for the doped 
material as input for the present study of the $ {\rm M_2} $ phase. 

Our investigations for the $ {\rm M_2} $ phase proceeded in two steps. 
First we performed calculations using the crystal structure data given 
by Marezio {\em at al.}\ and assuming spin-degeneracy. Spin-polarization 
was included in a second step. This way we were able to check the  
stability of the antiferromagnetic state with respect to the 
spin-degenerate case. Furthermore, our procedure allowed to separate 
the impact of the structural transformations and magnetic order 
on the electronic structure.

\subsection{Non-spinpolarized calculations}
\label{resm21}
  
\begin{figure}[htp]
\centering
\includegraphics[width=0.8\textwidth]{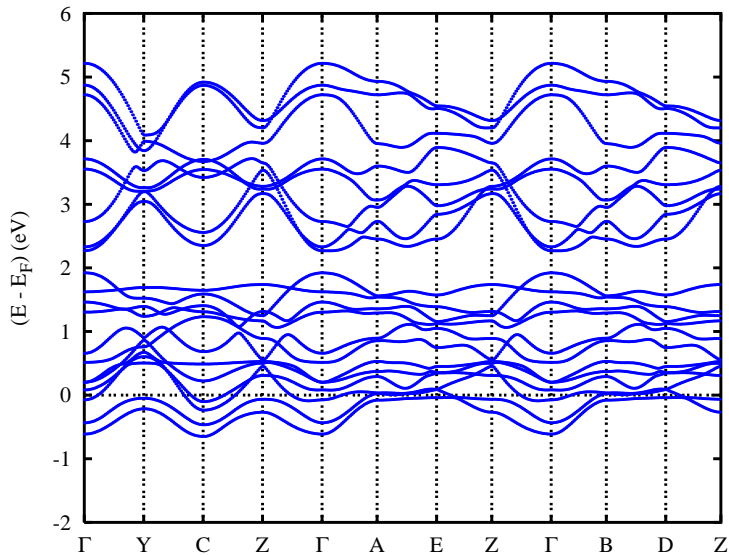}
\caption[Electronic bands of monoclinic $ {\rm M_2} $ $ {\rm VO_2} $.]
        {Electronic bands of monoclinic $ {\rm M_2} $ $ {\rm VO_2} $ along 
         selected symmetry lines within the simple monoclinic Brillouin 
         zone.}
\label{fig:resm21}   
\end{figure}

\begin{figure}[htp]
\centering
\includegraphics[width=0.8\textwidth]{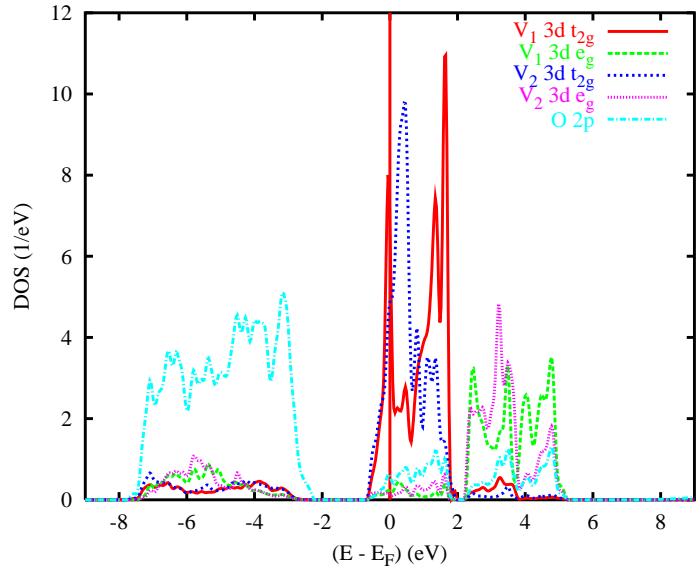}
\caption[Partial densities of states (DOS) of monoclinic $ {\rm M_2} $ 
         $ {\rm VO_2} $.]
        {Partial densities of states (DOS) of monoclinic $ {\rm M_2} $ 
	 $ {\rm VO_2} $ per unit cell.} 
\label{fig:resm22}   
\end{figure}

We display in Fig.\ \ref{fig:resm21} 
the electronic states along selected high symmetry lines within the centered  
monoclinic Brillouin zone. Except for the small lattice strain, which 
distorts the basal plane, the latter Brillouin zone is identical to the 
simple monoclinic Brillouin zone of the $ {\rm M_1} $ phase. The dominant 
partial densities of states (DOS) are given in Fig.\ \ref{fig:resm22}.  
As for the $ {\rm M_1} $ phase of $ {\rm VO_2} $ we observe many similarities 
with the rutile results as concerns the ordering, widths and separations of 
the bands. Still, we find the O $ 2p $-, V $ 3d $ $ t_{2g} $- and 
$ e_g $-dominated groups of bands below, at and above the Fermi energy, 
respectively, as well as a hybridization of these states of the order of 
10\%. 

Differences between the results for the two low-temperature phases show 
up on closer inspection of the V $ 3d $ derived partial DOS. Due to 
the presence of two different chains in the $ {\rm M_2} $ phase we have 
distinguished the two types of vanadium atoms, namely $ {\rm V_1} $ and 
$ {\rm V_2} $, which belong to the dimerizing and the zigzag chains, 
respectively. Distinct differences occur in the $ t_{2g} $ group of bands. 
The $ {\rm V_1} $ $ 3d $ $ t_{2g} $ partial DOS displays strong splitting 
into two peaks just below $ {\rm E_F} $ and at $ \approx 1.8 $\,eV similar 
to that observed for the $ {\rm M_1} $ phase, Fig.\ \ref{fig:resm12}. 
In contrast, the $ {\rm V_2} $ $ 3d $ $ t_{2g} $ partial DOS is dominated 
by a single maximum at $ \approx 0.3 $\,eV and resembles the respective 
partial DOS found for the rutile structure, Fig.\ \ref{fig:resrut2}. 
Qualitatively, the same holds for the V $ 3d $ $ e_g $ partial DOS, for 
which we note the close similarity of the partial DOS assigned to the 
$ {\rm V_1} $ and $ {\rm V_2} $ atoms to those calculated for the 
$ {\rm M_1} $ and the rutile structure. All these findings reflect our 
expectations. In a non-spinpolarized calculation the dimerizing and 
zigzag chains should display similar behaviour as that known from 
the $ {\rm M_1} $ and the rutile phase, respectively. 

\begin{figure}[htb]
\centering
\includegraphics[width=0.8\textwidth]{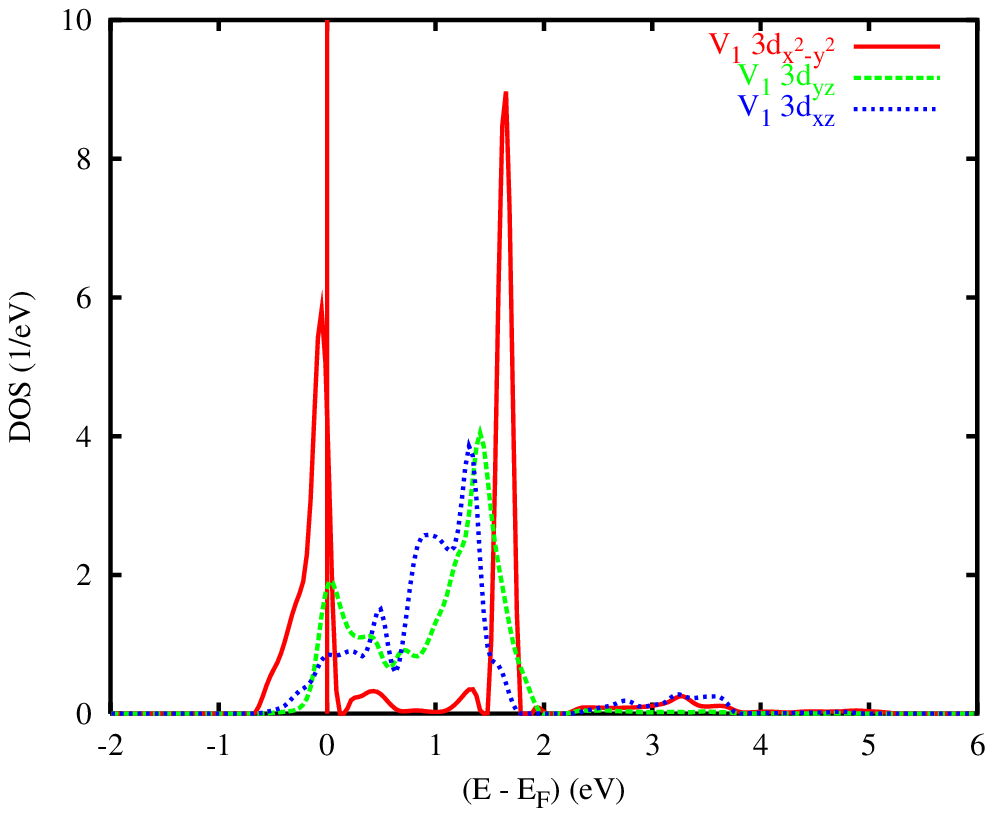}
\includegraphics[width=0.8\textwidth]{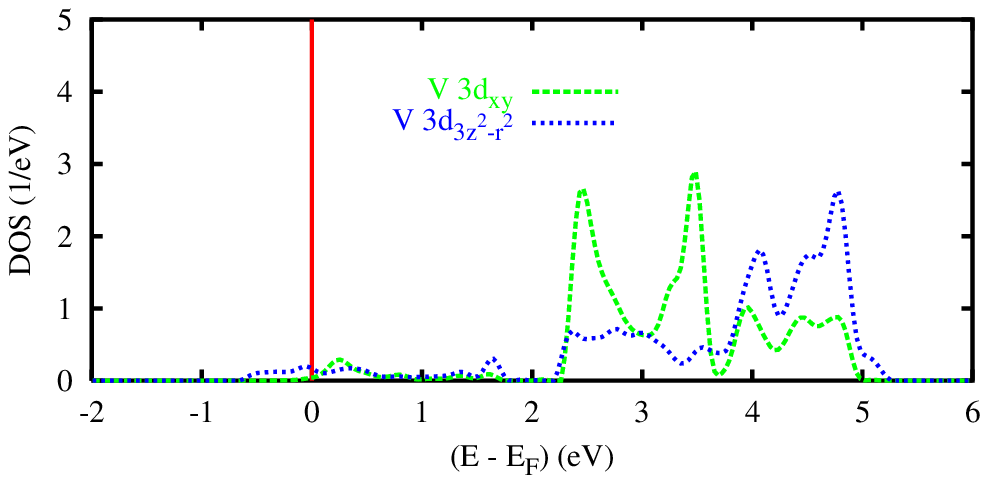}
\caption[Partial $ {\rm V_1} $ $ 3d $ $ t_{2g} $ and $ e_g $ densities of 
         states (DOS) of monoclinic $ {\rm M_2} $ $ {\rm VO_2} $.]
        {Partial $ {\rm V_1} $ $ 3d $ $ t_{2g} $ and $ e_g $ densities of 
         states (DOS) of monoclinic $ {\rm M_2} $ $ {\rm VO_2} $. 
         Selection of orbitals is relative to the local rotated reference 
         frame.} 
\label{fig:resm23}   
\end{figure}

\begin{figure}[htb]
\centering
\includegraphics[width=0.8\textwidth]{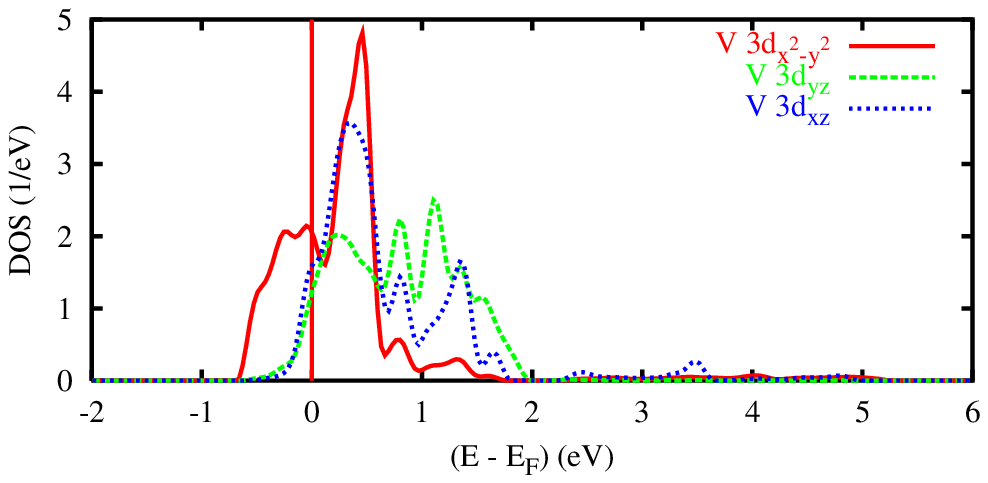}
\includegraphics[width=0.8\textwidth]{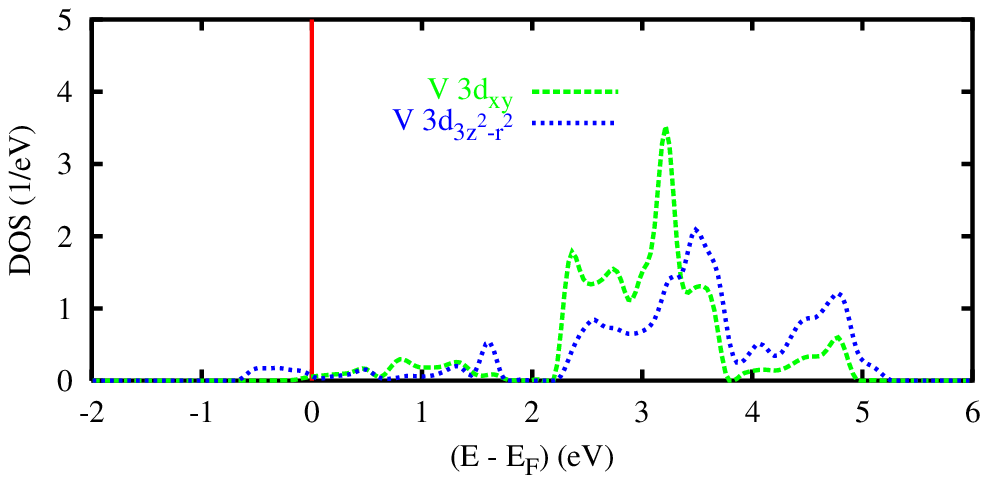}
\caption[Partial $ {\rm V_2} $ $ 3d $ $ t_{2g} $ and $ e_g $ densities of 
         states (DOS) of monoclinic $ {\rm M_2} $ $ {\rm VO_2} $.]
        {Partial $ {\rm V_2} $ $ 3d $ $ t_{2g} $ and $ e_g $ densities of 
         states (DOS) of monoclinic $ {\rm M_2} $ $ {\rm VO_2} $. 
         Selection of orbitals is relative to the local rotated reference 
         frame.} 
\label{fig:resm24}   
\end{figure}

Going into more detail we display in Figs.\ \ref{fig:resm23} 
and \ref{fig:resm24} 
the $ 3d $ partial DOS of the $ {\rm V_1} $ and $ {\rm V_2} $ atoms as 
decomposed into their symmetry components relative to the respective 
local reference frame of each chain. As for the rutile and the 
$ {\rm M_1} $ phase we observe for both types of vanadium atoms the 
energetical separation of the $ 3d $ $ t_{2g} $ and $ e_g $ groups of 
bands resulting from the crystal field splitting. However, within these  
groups of bands we also witness the aforementioned differences between 
the $ {\rm V_1} $ and $ {\rm V_2} $ atoms. 

The $ t_{2g} $ partial DOS of the paired $ {\rm V_1} $ atoms closely 
resemble the $ t_{2g} $ partial DOS of the $ {\rm M_1} $ phase as given 
by Fig.\ \ref{fig:resm13}. They both are characterized by the strong 
bonding-antibonding splitting of the $ d_{x^2-y^2} $ states, which is 
even more pronounced in the $ {\rm M_2} $ phase. This result fully 
agrees with the NMR data, which revealed the strong similarity of the 
Knight shift due to the dimerized vanadium atoms to that measured in the 
$ {\rm M_1} $ phase \cite{pouget74}. In contrast, the $ d_{x^2-y^2} $ 
partial DOS of the $ {\rm V_2} $ atoms, which belong to the zigzag 
chains, display a completely different behaviour. These states do not 
show any splitting but form a single maximum at $ \approx 0.4 $\,eV with 
a broad shoulder extending to -0.65\,eV and only minor contributions at 
energies above $ \approx 0.7 $. This shape thus bears a strong 
similarity to the corresponding DOS of the rutile phase, see Fig.\ 
\ref{fig:resrut3}, rather than to that of the monoclinic $ {\rm M_1} $ 
phase. To conclude, the $ d_{x^2-y^2} $ partial DOS can be readily 
understood from the local environment of the respective atoms as being 
more $ {\rm M_1} $- and rutile-like, respectively. 

The $ d_{xz} $ and $ d_{yz} $ curves follow this general trend in that 
the $ {\rm V_1} $ partial DOS reflect the behaviour known from the 
$ {\rm M_1} $ phase while the $ {\rm V_2} $ partial DOS are more 
rutile-like. Yet, two features are worth mentioning: 
i) The splitting of the $ 3d_{yz} $ partial DOS as observed in the 
rutile structure is almost completely suppressed on the $ {\rm V_2} $ 
chains. This is due to the zigzag-like displacement of these atoms, 
which undermines the in-plane overlap of the $ d_{yz} $ orbitals and, 
hence, reduces their bonding-antibonding splitting. 
ii) The upshift of the $ \pi^{\ast} $ states affects {\em both} the 
$ {\rm V_1} $ and the $ {\rm V_2} $ partial DOS. From the discussion 
in Sec.\ \ref{resm1}, where we attributed the upshift of these bands 
to the antiferroelectric mode of the vanadium atoms coming with the 
zigzag-like in-plane displacements, we would have expected a raise 
in energy {\em only} for the $ \pi^{\ast} $ states of the $ {\rm V_2} $ 
atoms but not for the vanadium atoms in the dimerizing chains. Yet, 
a closer look at the partial DOS in Figs.\ \ref{fig:resm23} and 
\ref{fig:resm24} reveals that the centers of gravity of the 
$ \pi^{\ast} $ bands actually is higher on the dimerizing chains. 

\begin{figure}[htp]
\centering
\includegraphics[width=0.6\textwidth]{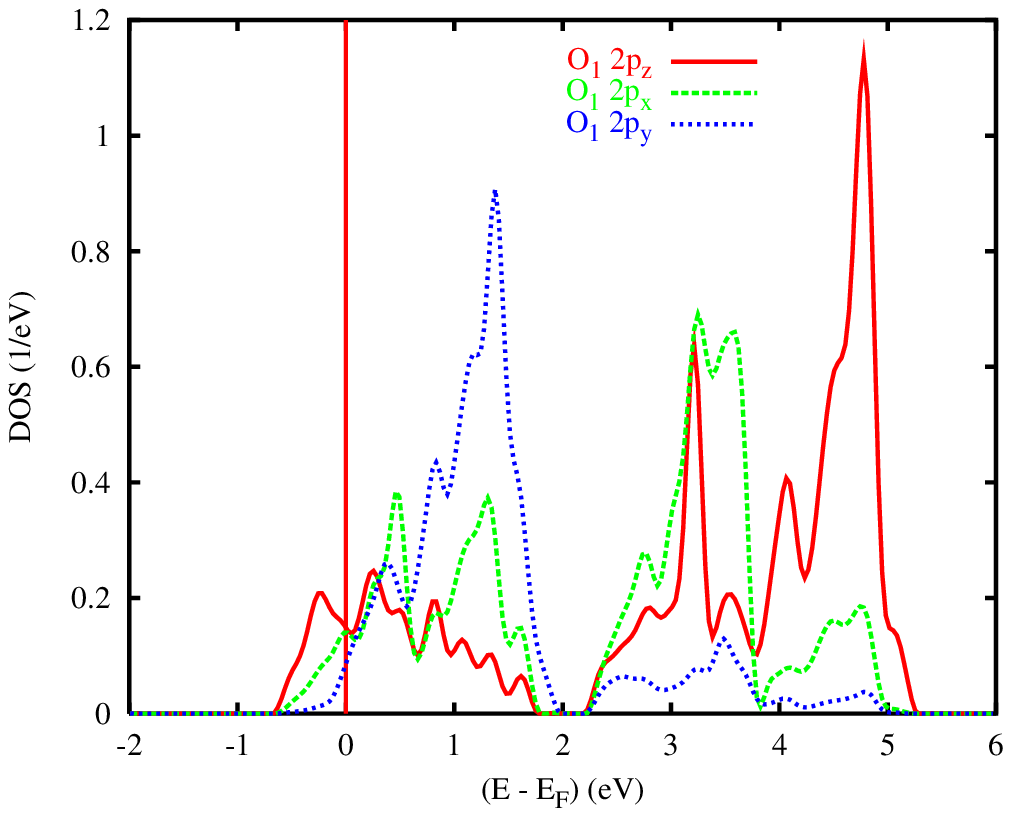}
\includegraphics[width=0.6\textwidth]{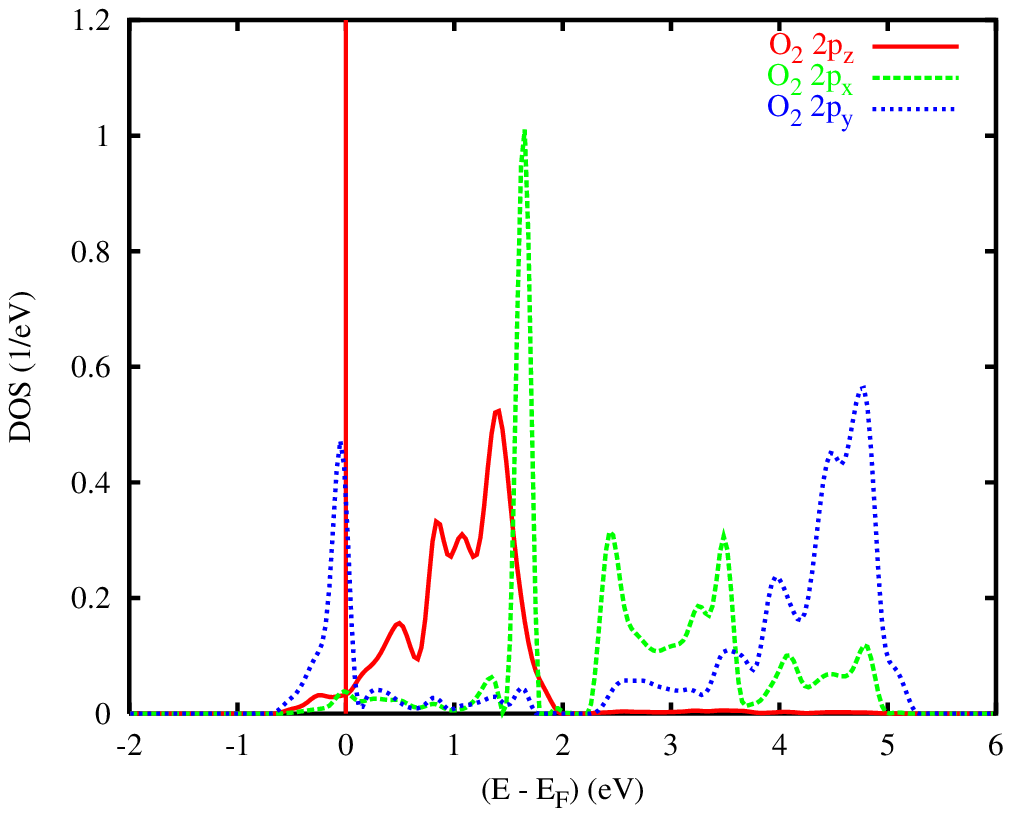}
\includegraphics[width=0.6\textwidth]{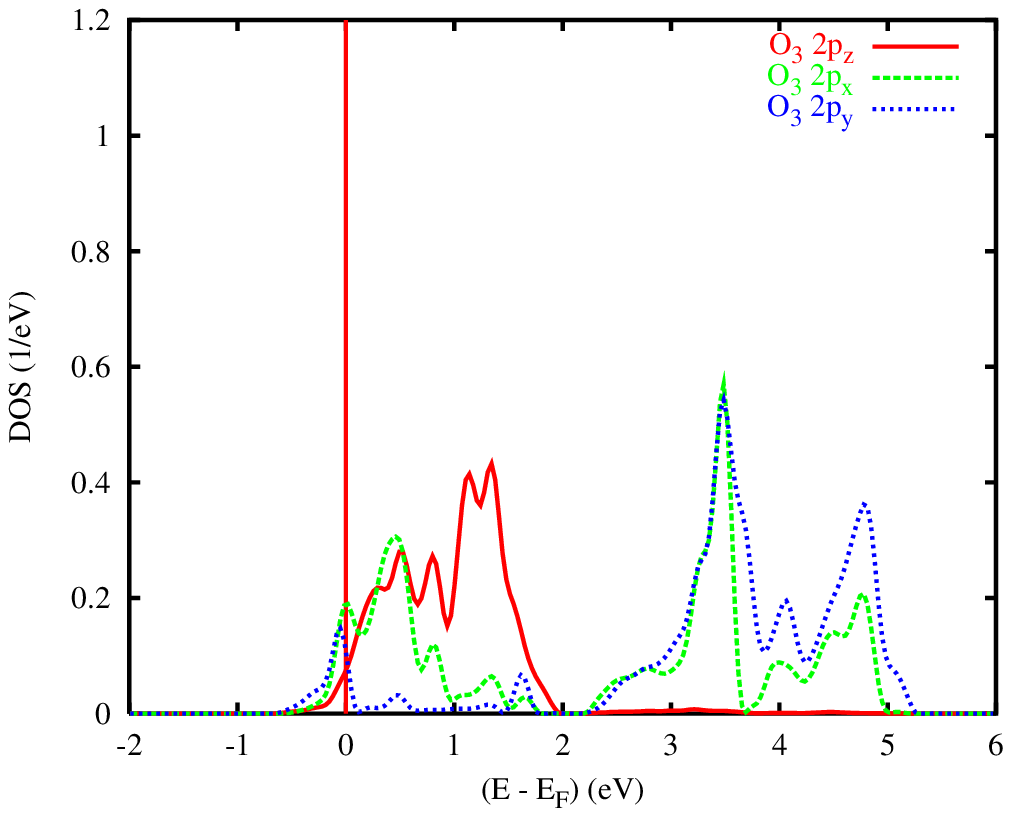}
\caption[Partial O $ 2p $ densities of states (DOS) of monoclinic 
         $ {\rm M_2} $ $ {\rm VO_2} $.]
        {Partial O $ 2p $ densities of states (DOS) of monoclinic 
         $ {\rm M_2} $ $ {\rm VO_2} $. Selection of orbitals is relative 
         to the local rotated reference frame of atom $ {\rm V_1} $.} 
\label{fig:resm25}   
\end{figure}

An explanation for this latter behaviour can be found in an analysis 
of the possible metal-oxygen overlaps using Tab.\ \ref{tab:resm21},  
\begin{table}[ht]
\begin{center}
\caption{V $ 3d $--O $ 2p $ orbital overlaps.}
\label{tab:resm21}   
\begin{tabular}{lccc} 
\\[-3mm] \hline \\[-3mm]
                 & $ {\rm O_1} $  & $ {\rm O_2} $  & $ {\rm O_3} $ \\  
\\[-3mm] \hline \\[-3mm]
$ {\rm V_1} $ 
$ 3d_{x^2-y^2} $ &                & $ 2p_x, 2p_y $ & $ 2p_x, 2p_y $    \\
$ {\rm V_1} $ 
$ 3d_{xz} $      & $ 2p_x $       & $ 2p_z $       & $ 2p_z $      \\
$ {\rm V_1} $ 
$ 3d_{yz} $      & $ 2p_y $       & $ 2p_z $       & $ 2p_z $      \\
                 &                &                &               \\
$ {\rm V_2} $ 
$ 3d_{x^2-y^2} $ & $ 2p_x, 2p_y $ &                &                   \\
$ {\rm V_2} $ 
$ 3d_{xz} $      &                & $ 2p_x $       & $ 2p_x $      \\
$ {\rm V_2} $ 
$ 3d_{yz} $      &                & $ 2p_y $       & $ 2p_y $      \\
\\[-3mm] \hline \\[-3mm]
\end{tabular} 
\end{center}
\end{table} 
where we combine those particular V $ 3d $ $ t_{2g} $ and O $ 2p $ 
orbitals, which, within a molecular-orbital picture, are expected to 
overlap. We complement the table with the partial O $ 2p $ partial 
DOS in Fig.\ \ref{fig:resm25},  
where we have again used the local rotated frame of reference for the 
selection of the orbitals. Note that we opted for the local reference 
frame corresponding to the dimerizing $ {\rm V_1} $ chains. As outlined 
in Sec.\ \ref{crystrut} interpretation in terms of the reference frame 
of the zigzag-like $ {\rm V_2} $ chains simply requires exchange of the 
$ y $ and $ z $ components. For the following discussion we recall that 
the $ {\rm O_1} $ atoms are the apical oxygen atoms of the dimerizing 
$ {\rm V_1} $ chains and occupy the equatorial positions of the 
zigzag-like $ {\rm V_2} $ chains, see Fig.\ \ref{fig:cryst4}. For 
oxygen atoms $ {\rm O_2} $ and $ {\rm O_3} $ things are reversed with 
the latter having the shorter distance to the central $ {\rm V_2} $ atom. 
In addition, due to the dimerization, $ {\rm O_2} $ have a very short 
bond with the vanadium atoms of type $ {\rm V_1} $. 

The aforementioned geometry is reflected by the partial DOS. In Fig.\ 
\ref{fig:resm25} we observe a striking similarity of the partial 
DOS corresponding to atom $ {\rm O_2} $ to the $ {\rm V_1} $ $ 3d $ 
partial DOS as shown in Fig.\ \ref{fig:resm23}. In general, all 
contributions in Figs.\ \ref{fig:resm23} to \ref{fig:resm25} can be 
well understood using the orbital overlaps listed in Tab.\ 
\ref{tab:resm21}. In particular, we observe the hybridization of the 
$ {\rm V_2} $ $ 3d_{xz} $ peak at $ \approx 0.4 $\,eV with the 
$ {\rm O_3} $ $ 2p_x $ contribution. Onset of the zigzag-like mode 
increases this overlap and shifts these states to higher energies. In 
contrast, the $ {\rm V_1} $ $ 3d_{xz} $ states, which are found mainly 
above 0.8\,eV, overlap with the $ 2p_z $ states of their equatorial 
$ {\rm O_2} $ and $ {\rm O_3} $ partners. This overlap increases with 
increasing dimerization on these chains and, hence, also causes an 
upshift in energy. We thus witness a considerable antiferroelectric 
component of the metal-metal pairing, which drives upshift of the 
$ \pi^{\ast} $ states on the corresponding chains and which we already 
proposed at the end of Sec.\ \ref{resm14}. 

\begin{figure}[htp]
\centering
\includegraphics[width=0.75\textwidth]{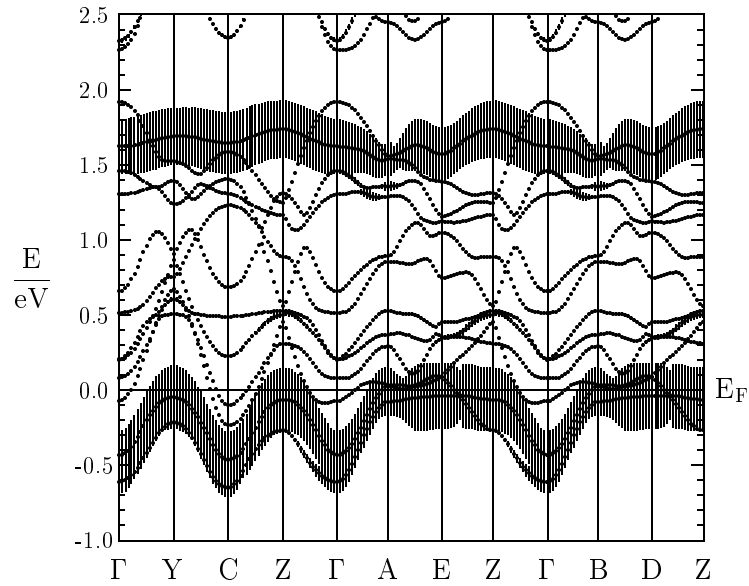}
\caption[Weighted electronic bands of monoclinic $ {\rm M_2} $ $ {\rm VO_2} $.]
        {Weighted electronic bands of monoclinic $ {\rm M_2} $ $ {\rm VO_2} $. 
         The width of the bars given for each band indicates the contribution 
         due to the $ 3d_{x^2-y^2} $ orbital of atom $ {\rm V_1} $, which 
         belongs to a dimerizing chain, relative to the local rotated reference 
         frame.} 
\label{fig:resm26}
\end{figure}

\begin{figure}[htp]
\centering
\includegraphics[width=0.75\textwidth]{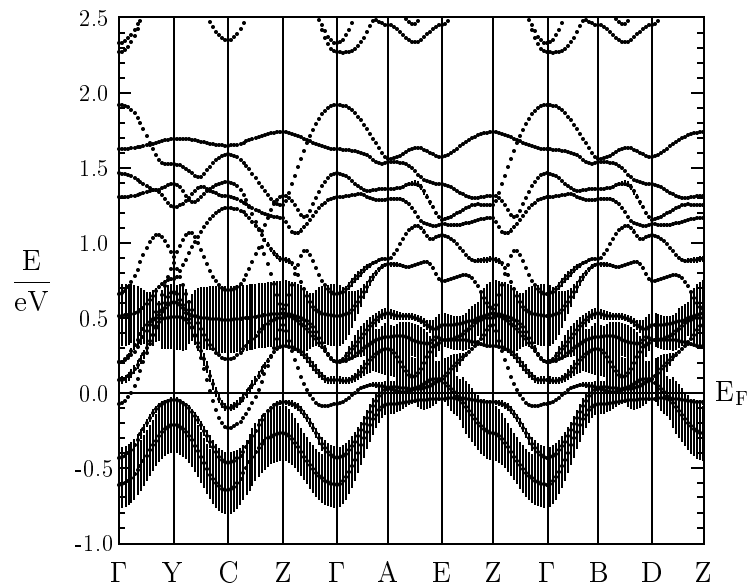}
\caption[Weighted electronic bands of monoclinic $ {\rm M_2} $ $ {\rm VO_2} $.]
        {Weighted electronic bands of monoclinic $ {\rm M_2} $ $ {\rm VO_2} $. 
         The width of the bars given for each band indicates the contribution 
         due to the $ 3d_{x^2-y^2} $ orbital of atom $ {\rm V_2} $, which 
         belongs to a zigzag chain, relative to the local rotated reference 
         frame.} 
\label{fig:resm27}
\end{figure}

The interpretation of the results found especially for the $ d_{x^2-y^2} $ 
partial DOS of the two different types of chains is confirmed by the 
weighted band structures as displayed in Figs.\ \ref{fig:resm26} 
and \ref{fig:resm27}.  
Again we note the energetical upshift of the $ d_{xz} $ and $ d_{yz} $ bands, 
which is almost independent of the type of chain, on which these orbitals 
are located. The $ d_{x^2-y^2} $ bands of the $ {\rm V_1} $ atoms show 
the strong bonding-antibonding splitting already observed in the partial 
DOS, Fig.\ \ref{fig:resm23}, in complete analogy to the weighted 
$ d_{x^2-y^2} $ bands of the $ {\rm M_1} $ phase, Fig.\ \ref{fig:resm111}. 
In both phases it results from the metal-metal pairing on the respective 
chains and leads to the semimetallic behaviour of these bands with a clear 
separation of bands of different characters. The same effect was already 
observed in the results 
for the $ {\rm M_1} $ structure. In contrast, the $ d_{x^2-y^2} $ bands of 
the $ {\rm V_2} $ atoms, which have equal metal-metal distances, resemble 
the dispersion of these bands in the rutile structure, see Fig.\ 
\ref{fig:resm18}. Their reduced band width, which is by $ \approx 0.4 $\,eV 
smaller as compared to the rutile structure, reflects the increased 
$ {\rm V_2} $--$ {\rm V_2} $ distance in the zigzag chains. Nevertheless, 
these bands still display the one-dimensional dispersion found for the 
rutile phase. As a consequence, from the spin-degenerate calculations 
we obtain half-filled $ {\rm V_2} $ $ d_{x^2-y^2} $ bands of $ \approx 1.1 $ 
eV width, which are clearly metallic. To conclude, enforcing 
spin-degeneracy we find indications for a metal-insulator transition very 
similar to those discussed in Sec.\ \ref{resm1} in the dimerizing chains 
but at the same time we obtain rutile-like metallic behaviour due to the 
bands originating from states on the zigzag chains.

\subsection{Spinpolarized calculations}
\label{resm22}
  
\begin{figure}[htp]
\centering
\includegraphics[width=0.8\textwidth]{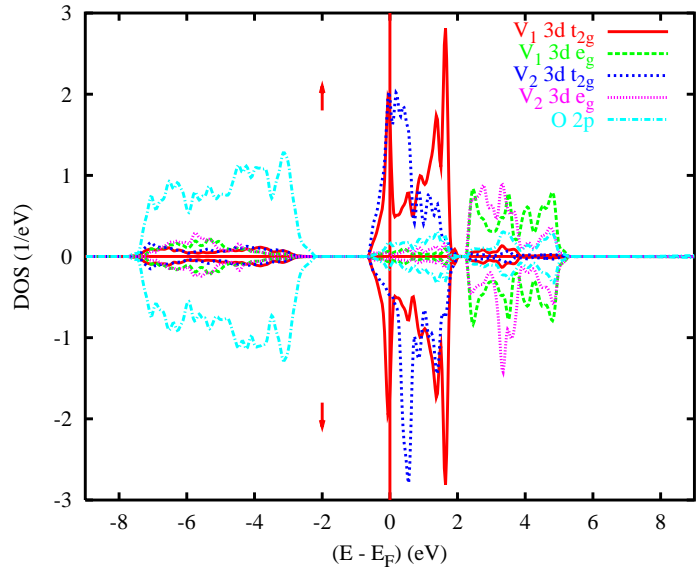}
\caption{Partial densities of states (DOS) of antiferromagnetic 
         monoclinic $ {\rm M_2} $ $ {\rm VO_2} $ per unit cell.} 
\label{fig:resm28}
\end{figure}

Next we turn to the spinpolarized calculations as suggested by the 
experimental finding of antiferromagnetic Heisenberg chains in the 
$ {\rm M_2} $ phase of $ {\rm VO_2} $. As a result we obtain a 
stable antiferromagnetic solution with a total energy, which is by 
0.7\,mRyd per unit cell lower than that growing out of the 
non-spinpolarized calculation. Magnetic moments of 0.46\,$ \mu_B $ 
are carried by the $ {\rm V_2} $ $ 3d $ states of the zigzag chains. 
The moments of all other atoms are either negligibly small or, for 
symmetry reasons, exactly zero as in case of the $ {\rm V_1} $. 
The dominant partial DOS are displayed in Fig.\ \ref{fig:resm28}. 
Note that {\em all} curves in Fig.\ \ref{fig:resm28} comprise 
contributions from one magnetic sublattice only. On comparing Fig.\ 
\ref{fig:resm28} to Fig.\ \ref{fig:resm22} for the non-spinpolarized 
case we recognize the main changes arising from spinpolarization in 
the small energy interval from -1 to 1\,eV about the Fermi energy. 
The $ {\rm V_1} $ $ 3d $ $ t_{2g} $ partial DOS are identical for 
both spin directions due to the vanishing magnetic moments at these 
sites. Spin splitting arises mainly from the $ {\rm V_2} $ $ 3d $ 
$ t_{2g} $ states. Their spin up and down DOS are dominated by a broader 
peak just above $ {\rm E_F} $ and rather sharp peak at $ \approx 0.7 $\,eV, 
respectively. 

\begin{figure}[ht]
\centering
\includegraphics[width=0.8\textwidth]{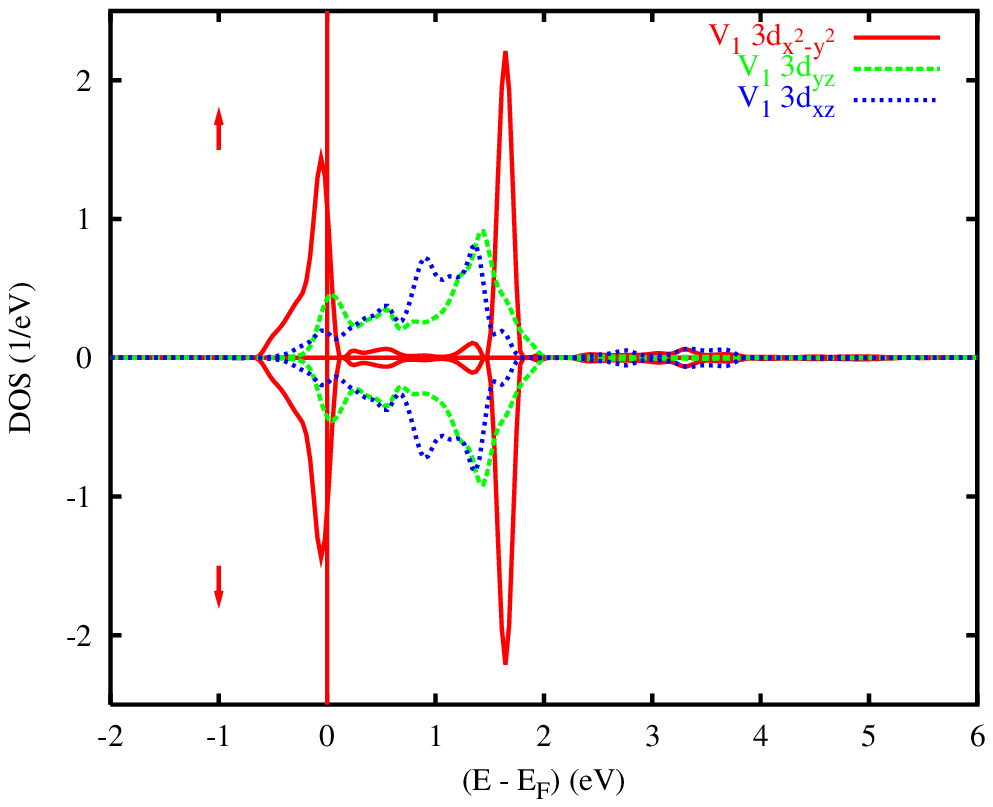}
\includegraphics[width=0.8\textwidth]{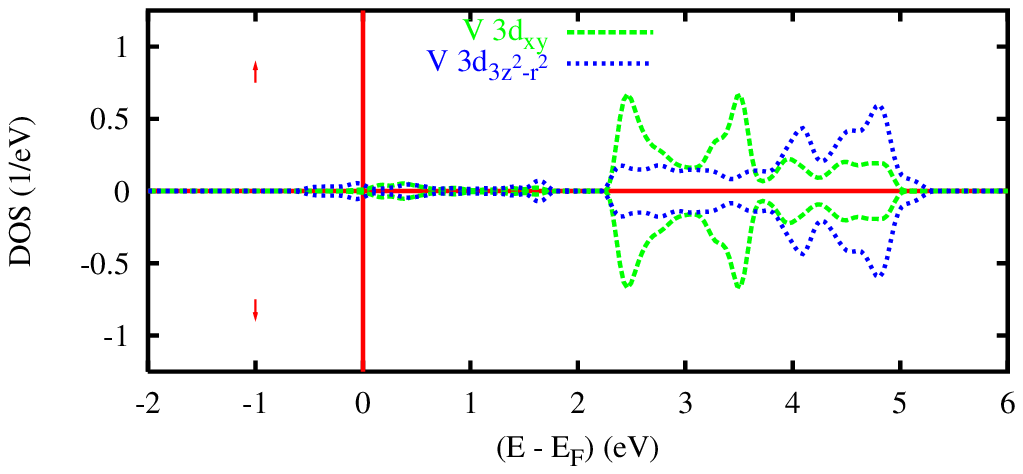}
\caption[Partial $ {\rm V_1} $ $ 3d $ $ t_{2g} $ and $ e_g $ densities 
         of states (DOS) of antiferromagnetic monoclinic $ {\rm M_2} $ 
         $ {\rm VO_2} $.]
        {Partial $ {\rm V_1} $ $ 3d $ $ t_{2g} $ and $ e_g $ densities 
         of states (DOS) of antiferromagnetic monoclinic $ {\rm M_2} $ 
         $ {\rm VO_2} $. Selection of orbitals is relative to the local 
         rotated reference frame.} 
\label{fig:resm29}
\end{figure}

\begin{figure}[ht]
\centering
\includegraphics[width=0.8\textwidth]{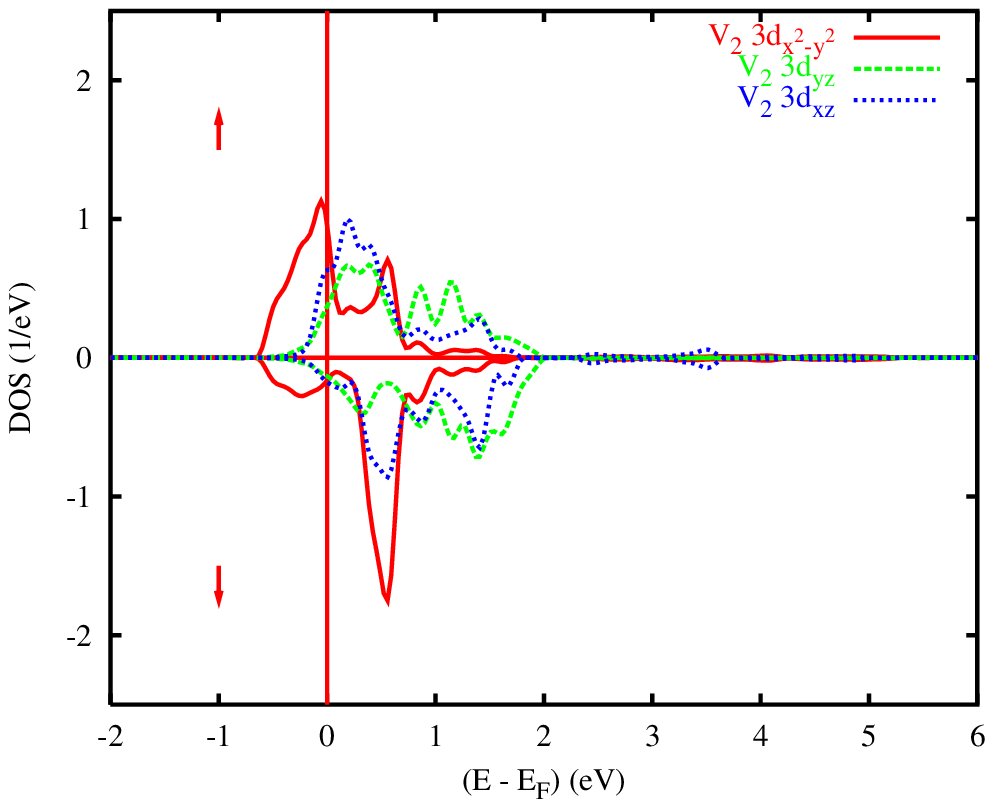}
\includegraphics[width=0.8\textwidth]{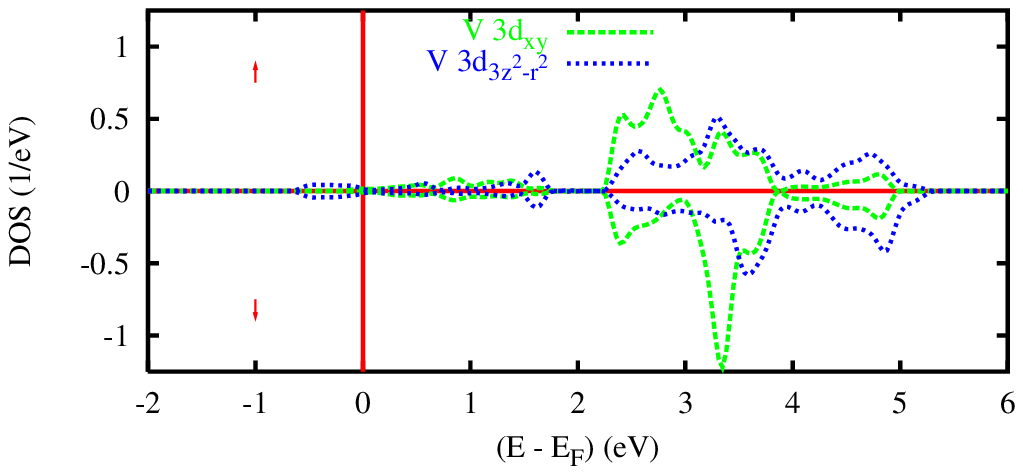}
\caption[Partial $ {\rm V_2} $ $ 3d $ $ t_{2g} $ and $ e_g $ densities 
         of states (DOS) of antiferromagnetic monoclinic $ {\rm M_2} $ 
         $ {\rm VO_2} $.]
        {Partial $ {\rm V_2} $ $ 3d $ $ t_{2g} $ and $ e_g $ densities 
         of states (DOS) of antiferromagnetic monoclinic $ {\rm M_2} $ 
         $ {\rm VO_2} $. Selection of orbitals is relative to the local 
         rotated reference frame.} 
\label{fig:resm210}   
\end{figure}

The changes originating from antiferromagnetic order are reflected by 
analysis of the V $ 3d $ partial DOS, which we display in Figs.\ 
\ref{fig:resm29} 
and \ref{fig:resm210}.  
Again, the identical spin up and down DOS of the $ {\rm V_1} $ atoms 
reflect the vanishing magnetic moments of these atoms. In contrast, 
we observe spin-dependent shifts of the $ {\rm V_2} $ partial DOS in the 
energy interval between -1 and 1\,eV and, to a lesser degree, between 2 
and 4\,eV. While spin-splitting of the $ d_{xz} $ and $ d_{yz} $ states 
is rather small on these atoms the magnetic moment originates mainly 
from the local $ d_{x^2-y^2} $ states. 

\begin{figure}[htp]
\centering
\includegraphics[width=0.75\textwidth]{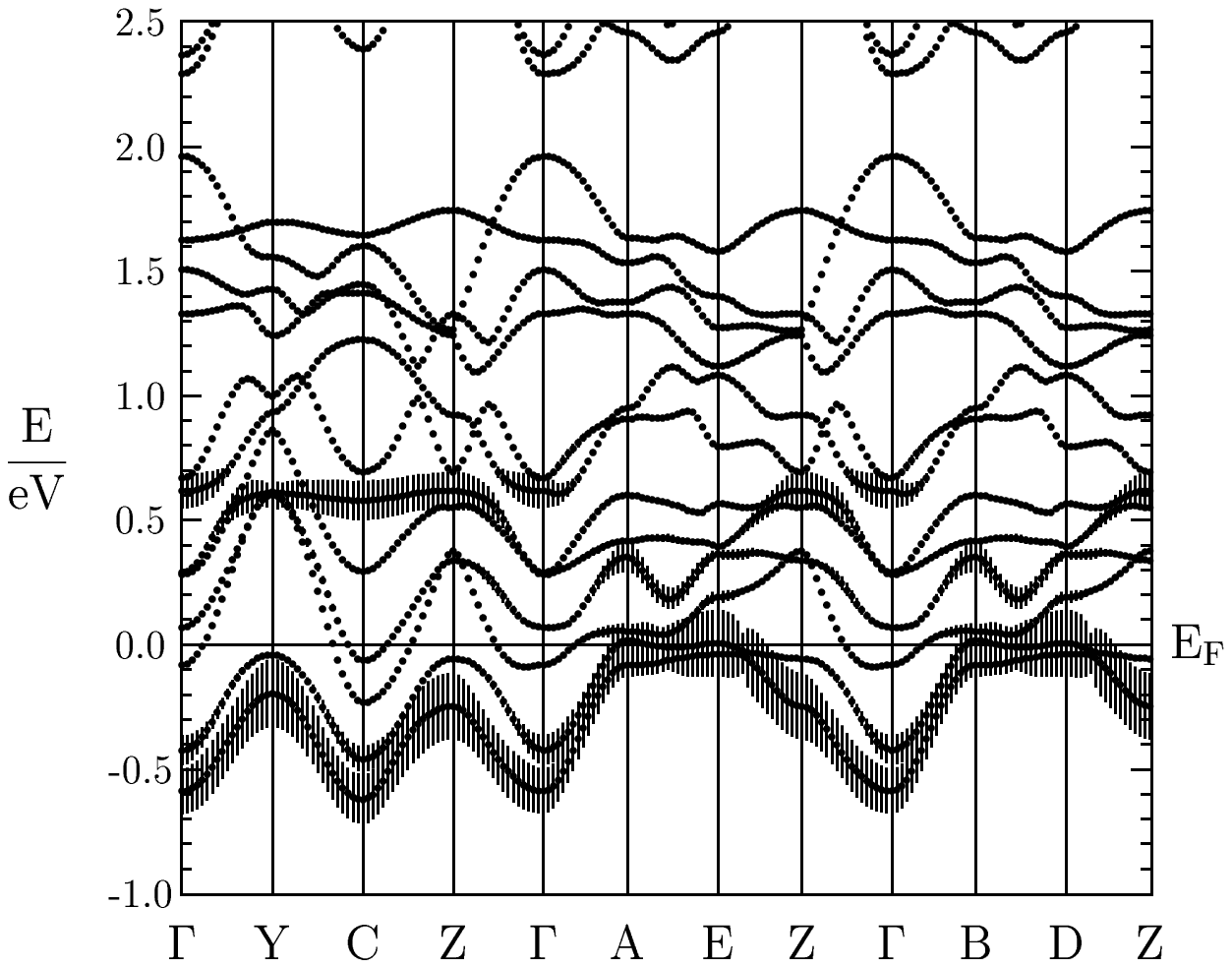}
\caption{Weighted electronic bands of antiferromagnetic monoclinic 
         $ {\rm M_2} $ $ {\rm VO_2} $. 
         The width of the bars given for each band indicates the contribution 
         due to the spin up $ 3d_{x^2-y^2} $ state of atom $ {\rm V_2} $, 
         which belongs to a zigzag chain, relative to the local rotated 
         reference frame.} 
\label{fig:resm211}
\end{figure}

\begin{figure}[htp]
\centering
\includegraphics[width=0.75\textwidth]{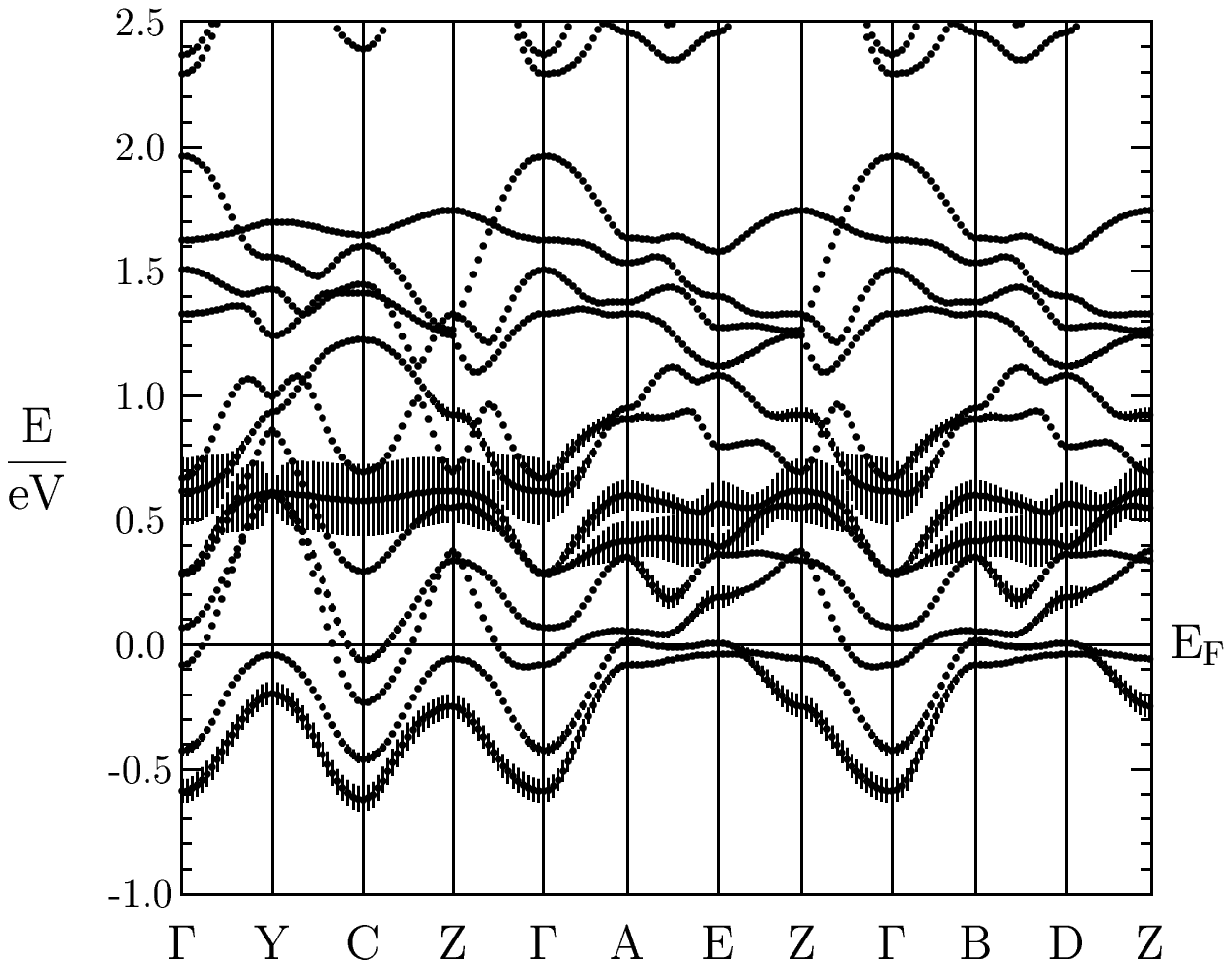}
\caption{Weighted electronic bands of antiferromagnetic monoclinic 
         $ {\rm M_2} $ $ {\rm VO_2} $. 
         The width of the bars given for each band indicates the contribution 
         due to the spin down $ 3d_{x^2-y^2} $ state of atom $ {\rm V_2} $, 
         which belongs to a zigzag chain, relative to the local rotated 
         reference frame.} 
\label{fig:resm212}
\end{figure}

Recalling the non-spinpolarized $ d_{x^2-y^2} $ partial DOS as shown in Fig.\ 
\ref{fig:resm24} with its prominent peak at $ \approx 0.5 $\,eV and the
broad shoulder below $ {\rm E_F} $ we can now attribute the peak at 
0.5\,eV mainly to the spin down states while the broad shoulder comprises 
almost exclusively the spin up states. This situation becomes clearer 
from the weighted electronic structures as shown in Figs.\ \ref{fig:resm211} 
and \ref{fig:resm212}. 
These band structures highlight the spin up and down $ {\rm V_2} $ 
$ 3d_{x^2-y^2} $ bands relative to the local reference frame of the 
single atom belonging to the spin up sublattice. While the occupied 
$ 3d_{x^2-y^2} $ bands are mainly of spin up character, bands at higher 
energies are almost exclusively of spin down character. The spin 
majority and minority bands thus play a similar role as the bonding 
and antibonding bands of the vanadium $ 3d_{x^2-y^2} $ states located 
at the dimerizing chains. The spin majority states are almost completely 
filled and show only small contributions to the bands at $ \approx 0.6 $ 
eV. In contrast, the spin minority states dominate the bands in the energy 
range between 0.4 and 0.6\,eV. As a consequence, antiferromagnetic ordering 
is obtained in the zigzag chains. At the same time, singlet formation 
arises in the dimerizing chains.  

As for the $ {\rm M_1} $ phase, the band structures and densities of states 
calculated with the $ {\rm M_2} $ structure lack the observed optical band 
gap. Yet, the situation is very similar to that found in Sec.\ \ref{resm1}. 
We find an overlap of the valence and conduction bands of $ \approx 0.2 $\,eV, 
which we attribute again to the shortcomings of the LDA. However, bands 
of different orbital character are clearly separated. For this reason, it 
is well justified to regard DFT-based calculations as being capable to 
describe also the metal-insulator rutile-to-$ {\rm M_2} $ transition. This 
holds even if the opening of the optical band gap is just missed due to 
the LDA. 

In summary, while the rutile phase is characterized by two types of 
rather weakly hybridized quasi-onedimensional $ d_{\parallel} $ and 
isotropically dispersing $ \pi^{\ast} $ bands the changes in crystal 
structure and magnetic order coming with the $ {\rm M_2} $ structure 
lead to distinct changes in the electronic structure. Both the zigzag-type  
displacement on the $ {\rm V_2} $ chains and the dimerization on the 
$ {\rm V_1} $ chains induce antiferroelectric shifts of the vanadium 
atoms inside the surrounding oxygen octahedra. The resulting increase 
of metal-oxygen bonding shifts the $ \pi^{\ast} $ bands to higher 
energies. In addition, dimerization of the $ {\rm V_1} $ atoms causes  
splitting of the corresponding $ d_{\parallel} $ bands into bonding 
and antibonding branches. In contrast, the $ d_{\parallel} $ bands 
belonging to the $ {\rm V_2} $ chains display spin-splitting and 
antiferromagnetic order in these chains. As for the $ {\rm M_1} $ 
phase coupling between the different types of bands is still weak 
and mainly via the common Fermi energy. The scenario of an embedded 
Peierls instability proposed for the transition from the rutile to 
the $ {\rm M_1} $ phase has to be extended by the possibility of 
antiferromagnetic ordering as already mentioned by Goodenough.

\section{Conclusion}
\label{concl}

In the present first principles study, augmented spherical wave calculations 
were used to investigate the structural, electronic, and magnetic properties 
as well as the metal-insulator transitions of vanadium dioxide. The essential 
features of the metallic rutile as well as the insulating $ {\rm M_1} $ 
and $ {\rm M_2} $ phases are well described within DFT and LDA. In metallic 
$ {\rm VO_2} $, the electronic structure consists of oxygen $ 2p $ bands 
well below the Fermi energy, as well as crystal field split vanadium $ 3d $ 
$ t_{2g} $ and $ e_g^{\sigma} $ bands. States near $ {\rm E_F} $ consist of 
two very weakly hybridizing types of $ t_{2g} $ bands, namely, the 
$ d_{\parallel} $ states with a quasi-onedimensional dispersion parallel 
to the characteristic V chains, and isotropically dispersing $ e_g^{\pi} $ 
bands. In the low-temperature $ {\rm M_1} $ structure of stoichiometric 
$ {\rm VO_2} $, metal-metal dimerization splits the $ d_{\parallel} $ band 
into bonding and antibonding branches, whereas the $ e_g^{\pi} $ states 
shift to higher energies due to reduced V--O distances. Although, due to 
the imperfections of the LDA, semimetallic behaviour with a band overlap 
of $ \approx 0.1 $\,eV is obtained rather than the observed optical band 
gap, the $ d_{\parallel} $ and $ e_g^{\pi} $ bands are completely separated. 
Since both types of bands are still coupled by charge conservation rather 
than hybridization, the insulating state is interpreted as due to a 
Peierls-like instability of the $ d_{\parallel} $ bands in an embedding 
reservoir of $ e_g^{\pi} $ electrons. The situation is thus identical to 
the one previously found for $ {\rm MoO_2} $ and $ {\rm NbO_2} $ 
\cite{moo2pap,nbo2pap}, and gives rise to a unified picture for the 
transition-metal dioxides at the beginning of the $ d $ series. This 
approach explains both 
(i) the destabilization of the rutile structure in terms of increased 
    metal-metal bonding and 
(ii) the metal-insulator transitions of the $ d^1 $ members. 

In the insulating $ {\rm M_2} $ phase, obtained on doping or applying uniaxial 
pressure, which generally is viewed as a metastable phase of stoichiometric 
$ {\rm VO_2} $, the metal-insulator transition results from different, 
complementary mechanisms. Still, depopulation of the $ e_g^{\pi} $ states 
is due to the antiferroelectric components of both the zigzag-like 
displacements and the dimerization, which increase the metal-oxygen bonding. 
In addition, the embedded Peierls-like instability proposed for the 
rutile-to-$ {\rm M_1} $ transition causes bonding-antibonding splitting of 
the $ d_{\parallel} $ bands connected with the dimerizing chains. In 
contrast, the $ d_{\parallel} $ states located on the chains with 
equidistant vanadium atoms experience spin-splitting due to antiferromagnetic 
order of the localized moments, which are carried predominantly by the 
$ d_{\parallel} $ electrons. As for the other phases, hybridization between 
the $ d_{\parallel} $ and $ e_g^{\pi} $ states is very weak, and the 
transition can be interpreted as a combined embedded 
Peierls-like/antiferromagnetic instability of rutile $ {\rm VO_2} $.

\vspace*{0.25cm} \baselineskip=10pt{\small \noindent 
I am grateful to U.\ Eckern, K.-H.\ H\"ock, S.\ Horn, and D.\ Vollhardt 
for their continuous support of this work as well as many helpful 
discussions. Thanks are also due to R.\ Horny, S.\ Klimm, D.\ Maurer, 
V.\ M\"uller, P.\ S.\ Riseborough, U.\ Schwingenschl\"ogl, and 
W.-D.\ Yang for many fruitful discussions as well as to E.\ Goering 
and O.\ M\"uller for supplying the experimental data. 
This work was supported by the Deutsche Forschungsgemeinschaft 
(Forschergruppe 241 and Sonderforschungsbereich 484, Augsburg).}


\end{document}